\documentclass[10pt]{article}
\begin{document}
\title{Issues in the Philosophy of Cosmology}
\author{George F\ R\ Ellis. \\
Mathematics Department and Applied Mathematics,\\
University of Cape Town,\\
Rondebosch, Cape Town 8001, South Africa.\\
email: ellis@maths.uct.ac.za}
\maketitle

\begin{abstract}
\noindent After a survey of the present state of cosmological theory and
observations, this article discusses a series of major themes underlying the
relation of philosophy to cosmology. These are:

A: The uniqueness of the universe;

B: The large scale of the universe in space and time;

C: The unbound energies in the early universe;

D: Explaining the universe --- the question of origins;

E: The universe as the background for existence;

F: The explicit philosophical basis;

G: The Anthropic question: fine tuning for life;

H: The possible existence of multiverses;

I: The natures of existence.

\noindent Each of these themes is explored and related to a series of Theses
that set out the major issues confronting cosmology in relation to
philosophy.
\end{abstract}

\section{Introduction}
Cosmology is the study of the large-scale structure of the Universe,
where `\textit{the Universe'} means all that exists in a physical
sense \cite{har81}. This is to be distinguished from \textit{the
Observable Universe}, namely that part of the Universe containing
matter accessible to our astronomical observations, which is a
subset of the Universe proper. Thus cosmology considers the vast
domain of galaxies, clusters of galaxies, quasi-stellar objects,
etc., observable in the sky by use of telescopes of all kinds,
examining their nature, distribution, origins, and relation to their
larger environment. \textit{Observational cosmology} \cite{hoy60,
krisac66, gunetal78, sanetal93, bot98} aims to determine the
large-scale geometry of the observable universe and the distribution
of matter in it from observations of radiation emitted by distant
objects, while \textit{physical cosmology}
\cite{pee71,sci71,wei72,sil01,per05,dod03} is the study of
interactions during the expansion of the universe in its early hot
big bang phase, and \textit{astrophysical cosmology}
\cite{sci71,pee93,pad93,ree95,dod03} studies the resulting later
development of large-scale structures such as galaxies and clusters
of galaxies. Various forms of quantum cosmology (see e.g.
\cite{haw93,gibetal03,coetal05}) and studies of particle physics
aspects of cosmology \cite{koltur90,pea99,all02,per05,dod03} attempt
to characterize the epochs before the hot big bang phase. These
studies function in a mainly symbiotic way, each informing and
supplementing the others to create an overall cosmological theory of
the origin and evolution of the physical universe
\cite{bon60,har81,sil97}.

A unique role of the universe is in creating the environment in
which galaxies, stars, and planets develop, thus providing a
setting in which local physics and chemistry can function in a way
that enables the evolution of life on planets such as the Earth.
If the cosmological environment were substantially different,
local conditions would be different and in most cases we would not
be here \cite{carree79,dav82,bartip84,teg93,ree99} --- indeed no
biological evolution at all would have taken place. Thus cosmology
is of substantial interest to the whole of the scientific
endeavor, for it sets the framework for the rest of science, and
indeed for the very existence of observers and scientists. It is
unique as the ultimate historical/geographical science.

Cosmology as a serious scientific study began with the discovery
of Einstein's static universe in 1917, followed by the key
observational discovery of the linear redshift-distance relation
by Hubble in 1929, indicating the expansion of the universe, and
the development of theoretical understanding of the geometry and
dynamics of the non-static Friedmann-Lema\^{\i}tre models with
their Robertson-Walker geometry
\cite{nor65,beretal76,smi82,ell89,kra96}. It has been transformed
in the past decades into a mainstream branch of physics
\cite{baretal96,nil91} by the linking of nuclear and particle
physics theories to observable features of the cosmos
\cite{wei72,koltur90,pea99,all02,dod03}, and into an important
part of astronomy because of the massive flow of new astronomical
data becoming available \cite{gunetal78,har84,bot98}, particularly
through new ground-based telescopes such as Keck and through
balloon and satellite observatories such as the Hubble Space
telescope (optical and ultra-violet), IRAS (infra-red), ROSAT
(x-ray), and COBE and WMAP (microwave). Thus the subject has
progressed from a mainly mathematical and even philosophical
exercise to an important part of mainstream science, with a
well-established standard model confirmed by various strands of
evidence \cite{wei72,peetal,sil97,pea99,dod03}. Nevertheless
because of its nature, it is different from any other branch of
the natural sciences, its unique features playing themselves out
in the ongoing interaction between speculation, theory, and
observation.

Cosmology's major difference from the other sciences is the
uniqueness of its object of study --- the Universe as a whole
\cite{mac53,mac60,mun86} --- together with its role as the
background for all the rest of physics and science, the resulting
problems being accentuated by the vast scale of the universe and
by the extreme energies occurring in the very early universe. We
are unable to manipulate in any way its originating conditions,
and there are limitations on our ability to observe both to very
distant regions and to very early times. Additionally, there are
limits to our ability to test the physics relevant at the earliest
epochs. Consequently it is inevitable that (as is also the case
for the other historical sciences) philosophical choices will to
some degree shape the nature of cosmological theory, particularly
when it moves beyond the purely descriptive to an explanatory role
\cite{matetal95} --- that move being central to its impressive
progress in recent decades. These philosophical choices will
strongly influence the resulting understanding, and even more so
if we pursue a theory with more ambitious explanatory aims.

After a substantial outline of present day cosmology in Section 2,
these issues will be explored in the subsequent sections, based on
a series of thirty-four \emph{Theses} clustered around nine key
aspects of the nature of cosmology, broadly speaking relating to
geometry, physics, and philosophy, that frame the context of the
philosophical issues facing cosmology and its relation to local
physics. I\ believe this formulation helps focus on specific
issues of importance in this relation. To those who believe
cosmology is simply about determining a number of physical
parameters, this will seem a vastly over-complicated approach; but
then a major purpose of this paper is precisely to counter such
simplistic visions of the nature of cosmology. For other reports
on the philosophy of cosmology, see
\cite{mac70,mun86,ell91,les94,les98}.

\section{Outline of cosmology}\label{sec:outline}
A series of basic features of present day cosmology are now well
established. Decades of painstaking work has established the
distances of galaxies and hence the huge scale of the universe, as
well as the basic feature that the universe is expanding and
evolving; the old dream of a static universe is unviable
\cite{ell90}. Cosmology proceeds by assuming \textit{the laws of
physics are the same everywhere, and underlie the evolution of the
universe}. The dominant role of gravity, despite its weakness,
then arises from the fact that it is the only known force acting
effectively on astronomical scales (the other known long-range
force is electromagnetism, but in this case negative charges
balance out positive charges, leaving no resultant large-scale
effect). Consequently, cosmological theory describing all but the
very earliest times is based on the classical relativistic theory
of gravitation, namely Einstein's General Theory of Relativity
\cite{mal05}, with the matter present determining space-time
curvature and hence the evolution of the universe. The way this
works out in any particular situation depends on the nature of the
matter/fields present, described by their effective equations of
state and interaction potentials.

The survey of cosmology in this section looks successively at the
basic models of cosmology; the hot big bang; cosmological
observations, including the Cosmic Background Radiation anisotropy
spectrum; causal and visual horizons, and their implications; recent
theoretical developments (including inflation); the very early
universe; and the present concordance model, which includes both
dark matter and dark energy.

\subsection{Basic Theory}
\label{sec:FL} Cosmology starts by assuming that \emph{the
large-scale evolution of spacetime can be determined by applying
Einstein's field equations of Gravitation} (`EFE')
\emph{everywhere}: global evolution will follow from local physics.
The standard models of cosmology \cite{rob33,ehl61,wei72,hawell73}
are based on the assumption that once one has averaged over a large
enough physical scale, \emph{isotropy is observed by all fundamental
observers} (the preferred family of observers associated with the
average motion of matter in the universe). When this isotropy is
exact, \emph{the universe is spatially homogeneous as well as
isotropic} \cite{wal44,ehl61,ell71}. The matter motion is then along
irrotational and shearfree geodesic curves with tangent vector
$u^{a}$, implying the existence of a canonical time-variable $t$
obeying $u_{a}=-t_{,a}$. The Robertson-Walker (`RW') geometries used
to describe the large-scale structure of the universe
\cite{rob35,wal36} embody these symmetries exactly. Consequently
they are conformally flat, that is, the Weyl tensor is zero:
\begin{equation}
C_{ijkl}:=R_{ijkl}+{\textstyle{\frac{1}{2}}}
(R_{ik}g_{jl}+R_{jl}g_{ik}-R_{il}g_{jk}-R_{jk}g_{il})-{\textstyle{\frac{1}{6}
}}R(g_{ik}g_{jl}-g_{il}g_{jk})=0;  \label{eq:weyl}
\end{equation}
this tensor represents the free gravitational field, enabling
non-local effects such as tidal forces and gravitational waves which
do not occur in the exact RW geometries.

Comoving coordinates can be chosen so that the metric takes the form:
\begin{equation}
ds^{2}=-dt^{2}+S^{2}(t)\,d\sigma ^{2},~~u^{a}=\delta
^{a}{}_{0}\;(a=0,1,2,3) \label{eq:rw}
\end{equation}
where $S(t)$ is the time-dependent scale factor, and the worldlines
with tangent vector $u^{a}=dx^{a}/dt$ represent the histories of
fundamental observers. The space sections $\{t=const\}$ are surfaces
of homogeneity and have maximal symmetry:\ they are 3-spaces of
constant curvature $ K=k/S^{2}(t) $ where $k$ is the sign of $K$.
The normalized metric $d\sigma^{2}$ characterizes a 3-space of
normalized constant curvature $k$; coordinates ($r,\theta ,\phi)$
can be chosen such that
\begin{equation}
d\sigma ^{2}=dr^{2}+f^{2}(r)\left({d\theta ^{2}+\sin ^{2}\theta
d\phi ^{2}} \right)  \label{three}
\end{equation}
where $f(r)=\{\sin r,\,r,\,\sinh r\}$ if $k=\{+1,\,0,\,-1\}$
respectively. The rate of expansion at any time $t$ is
characterised by the \emph{Hubble parameter} $H(t) = \dot{S}/S$.

To determine the metric's evolution in time, one applies the
Einstein Field Equations (`EFE'), showing the effect of matter on
space-time curvature, to the metric (\ref{eq:rw},\ref{three}).
Because of local isotropy, the matter tensor $T_{ab}$ necessarily
takes a perfect fluid form relative to the preferred worldlines with
tangent vector $u^{a}$:
\begin{equation}
T_{ab}=(\mu +p/c^2)u_{a}u_{b}+ (p/c^2)g_{ab}  \label{pf}
\end{equation}
($c$ is the speed of light). The energy density $\mu(t)$ and
pressure term $p(t)/c^2$ are the timelike and spacelike eigenvalues
of $T_{ab}$. The integrability conditions for the EFE are the
\textit{energy-density conservation equation}
\begin{equation}
T^{ab}_{\,\,\,\,\,;b}=0\,\Leftrightarrow\,\, \dot{\mu}+(\mu
+p/c^{2})3\dot{S}/S=0\,. \label{cons}
\end{equation}
This becomes determinate when a suitable equation of state
function $w: = pc^2/\mu$ relates the pressure $p$ to the energy
density $\mu$ and temperature $T$:  $p = w(\mu,T)\mu/c^2$ ($w$ may
or may not be constant). Baryons have \{$p_{bar}=0$
$\Leftrightarrow$ $w=0$\} and radiation has \{$
p_{rad}c^{2}=\mu_{rad}/3$ $\Leftrightarrow$ $w=1/3$, $\mu
_{rad}=aT_{rad}^{4}$\}, which by (\ref{cons}) imply
\begin{equation}
\mu _{bar}\propto S^{-3},\,\,\mu _{rad}\propto S^{-4},\,\,T_{rad}\propto
S^{-1}.  \label{scaling}
\end{equation}
The scale factor $S(t)$ obeys the \emph{Raychaudhuri equation}
\begin{equation}
3\ddot{S}/S=-\frac{1}{2}\kappa (\mu +3p/c^{2})+\Lambda ,  \label{Ray}
\end{equation}
where $\kappa $ is the gravitational constant and $\Lambda $ the
cosmological constant.\footnote{A cosmological constant can also
be regarded as a fluid with pressure $p$ related to the energy
density $\mu$ by \{$p=-\mu c^{2}$ $\Leftrightarrow$ $w=-1$\}. For
the history of the cosmological constant, see \cite{ear01,ear03}.}
This shows that the active gravitational mass density of the
matter and fields present is $\mu _{grav}:=\mu +3p/c^{2}$ . For
ordinary matter this will be positive:
\begin{equation}
\mu +3p/c^{2}>0\,\, \Leftrightarrow\,\, w > -1/3 \label{eq:ec}
\end{equation}
(the `Strong Energy Condition'), so ordinary matter will tend to
cause the universe to decelerate ($\ddot{S}<0).$ It is also apparent
that a positive cosmological constant on its own will cause an
accelerating expansion ($\ddot{S}>0).$ When matter and a
cosmological constant are both present, either result may occur
depending on which effect is dominant. The first integral of
equations (\ref{cons}, \ref{Ray}) when $\dot{S}\neq 0$ is the
\emph{Friedmann equation}
\begin{equation}
\frac{\dot{S}^{2}}{S^{2}}=\frac{\kappa \mu}{3}+\frac{\Lambda
}{3}-\frac{k}{ S^{2}}.  \label{Fried}
\end{equation}
This is just the Gauss equation relating the 3-space curvature to
the 4-space curvature, showing how matter directly causes a
curvature of 3-spaces \cite{ehl61,ell71}. Because of the spacetime
symmetries, the ten EFE are equivalent to the two equations
(\ref{Ray}, \ref{Fried}). Models of this kind, that is with a
Robertson-Walker (`RW') geometry with metric (\ref{eq:rw},
\ref{three}) and dynamics governed by equations (\ref{cons}, \ref
{Ray}, \ref{Fried}), are called \emph{Friedmann-Lema\^{\i}tre
universes} (`FL' for short). The Friedmann equation (\ref{Fried})
controls the expansion of the universe, and the conservation
equation (\ref{cons}) controls the density of matter as the universe
expands; when $\dot{S}\neq 0$ , equation (\ref{Ray}) will
necessarily hold if (\ref{cons}, \ref{Fried}) are both satisfied.

Given a determinate matter description (specifying the equation of
state $ w=w(\mu,T)$ explicitly or implicitly) for each matter
component, the existence and uniqueness of solutions follows both
for a single matter component and for a combination of different
kinds of matter, for example $\mu = \mu_{bar} + \mu_{rad} +
\mu_{cdm} + \mu_{\nu}$ where we include cold dark matter (cdm)\
and neutrinos ($\nu $). Initial data for such solutions at an
arbitrary time $t_{0}$ (eg. today) consists of,
\begin{itemize}
    \item The \emph{Hubble constant} $H_{0}: = (\dot{S}/S)_{0} = 100 h$
km/sec/Mpc;
    \item A dimensionless \emph{density parameter }$\Omega_{i0}: = \kappa
\mu_{i0}/3H_{0}^{2}$ for each type of matter present (labelled by
$i$);
    \item If $\Lambda \neq 0$, either $\Omega_{\Lambda 0}: = \Lambda
/3H_{0}^{2}$, or the dimensionless \emph{deceleration parameter}
$q_{0}: = -(\ddot{S}/S)_{0}H_{0}^{-2}$.
\end{itemize}
\noindent Given the equations of state for the matter, this data
then determines a unique solution \{$S(t)$, $\mu (t)$\}, i.e. a
unique corresponding universe history. The total matter density is
the sum of the terms $\Omega_{i0}$ for each type of matter present,
for example
\begin{equation}\label{omega}
    \Omega_{m0} = \Omega_{bar0} +\Omega_{rad0}+\Omega_{cdm0}+\Omega_{\nu
    0},
\end{equation}
and the total density parameter $\Omega_0$ is the sum of that for
matter and for the cosmological constant:
\begin{equation}\label{omega1}
    \Omega_{0} = \Omega_{m0} + \Omega_{\Lambda 0}.
\end{equation}
Evaluating the Raychaudhuri equation (\ref{Ray}) at the present
time gives an important relation between these parameters: when
the pressure term $p/c^{2}$ can be ignored relative to the matter
term $\mu $ (as is plausible at the present
time),\footnote{Assuming we represent `dark energy'
(Sec.\ref{sec:cdm}) as a cosmological
constant.}
\begin{equation}\label{q0}
q_{0}=\frac{1}{2}\,\Omega_{m0}-\Omega_{\Lambda 0}.
\end{equation}
 This shows that a
cosmological constant $\Lambda $ can cause an acceleration
(negative $q_{0}$); if it vanishes, the expression simplifies:
$\Lambda =0 \Rightarrow q=\frac{1}{2}\,\Omega_{m0}$, showing how
matter causes a deceleration of the universe. Evaluating the
Friedmann equation (\ref{Fried}) at the time $t_0$, the spatial
curvature is
\begin{equation}\label{K}
    K_0:=k/S_0^2=H_{0}^{2}\,(\Omega_{0} -1).
\end{equation}
The value $\Omega_{0}=1$ corresponds to spatially flat universes
($K_0=0$), separating models with positive spatial curvature
($\Omega_{0}
> 1 \Leftrightarrow K_0>0$) from those with negative spatial curvature
($\Omega_{0}<1\Leftrightarrow K_0<0$).

\emph{The FL models are the standard models of modern cosmology},
surprisingly effective in view of their extreme geometrical simplicity. One
of their great strengths is their explanatory role in terms of making
explicit the way the local gravitational effect of matter and radiation
determines the evolution of the universe as a whole, this in turn forming
the dynamic background for local physics (including the evolution of the
matter and radiation).

\subsubsection{The basic solutions}
\label{sec:basic}For baryons (pressure-free matter) and non-interacting
radiation, the Friedmann equation (\ref{Fried}) takes the form
\begin{equation}
\frac{3\dot{S}^{2}}{S^{2}}= \frac{A}{S^{3}}+\frac{B}{S^{4}}\ +
\frac{\Lambda }{3}-\frac{3k}{S^{2}} \label{Fried1}
\end{equation}
where $A: = \kappa\mu_{bar0}S_{0}^{3}$ and $B: = \kappa \mu_{rad0}
S_{0}^{4}.$ The behaviour depends on the cosmological constant
$\Lambda $ \cite{rob33,rin01}.

When $\Lambda =0$, the universe starts off at a very dense initial
state --- according to the classical theory, an initial singularity
where the density and curvature go infinite (see
Sec.\ref{sec:sing}). Its future fate depends on the value of the
spatial curvature, or equivalently the density parameter
$\Omega_{0}$. The universe expands forever if \{$k=0\Leftrightarrow
\Omega_{0}=1$\} or \{$k<0$ $\Leftrightarrow $ $ \Omega_{0}<1\},$ but
collapses to a future singularity if \{$k>0\Leftrightarrow
\Omega_{0}>1$\}. Thus $\Omega_{0} = 1$ \textit{corresponds to the
critical density} $\mu_{crit}$ \textit{separating} $\Lambda = 0$
\textit{FL models that recollapse in the future from those that
expand forever}, and $\Omega_{0}$ is just the ratio of the matter
density to this critical density:
\begin{equation}
\{\Omega_{crit}=1 \Leftrightarrow \kappa \mu_{crit} = 3H_{0}^{2}\}\;
\Rightarrow \;\Omega_{0}: = \kappa \mu_{0}/3H_{0}^{2} = \mu_{0}/\mu
_{crit}\;.  \label{crit}
\end{equation}
\indent When $\Lambda <0,$ all solutions start at a singularity
and recollapse.

When $\Lambda >0$, if $k=0$ or $k=-1$ all solutions start at a
singularity and expand forever$.$ If $k=+1$ there can again be
models with a singular start, either expanding forever or
collapsing to a future singularity. However in this case a static
solution (the Einstein static universe) is also possible, as well
as models asymptotic to this static state in either the future or
the past. Furthermore models with $k=+1$ can bounce (collapsing
from infinity to a minimum radius and re-expanding).

The dynamical behaviour of these models has been investigated in
depth: first for dust plus a cosmological constant
\cite{rob33,rin01}, followed by perfect fluids, fluids with bulk
viscosity, kinetic theory solutions, and scalar field solutions.
Current models employ a realistic mixture of matter components
(baryons, radiation, neutrinos, cold dark matter, a scalar field,
and perhaps a cosmological constant). Informative phase planes show
clearly the way higher symmetry (self-similar) models act as
attractors and saddle points for the other models
\cite{madell88,ehlrin89}.

The simplest expanding solutions are the following:

1. The \textit{Einstein-de Sitter} model, for which \{$p=0,$
$\Lambda =0$, $k=0$\} $\Rightarrow $ $\Omega_{0}=1$. This is the
simplest expanding non-empty solution:
\begin{equation}
S(t)= C\,t^{2/3}  \label{eds}
\end{equation}
starting from a singular state at time $t=0$ ($C$ is an arbitrary
constant). Its age (the proper time since the start of the universe)
when the Hubble constant takes the value $H_0$ is $\tau_{0} =
\frac{2}{3H_{0}}.$ This is a good model of the expansion of the
universe since radiation domination ended until the recent times
when a cosmological constant started to dominate the expansion. It
is also a good model of the far future universe if $k=0$ and
$\Lambda =0$.

2. The \textit{Milne} model, for which \{$\mu = p =0$, $\Lambda =
0$, $k=-1$\} $\Rightarrow$ $\Omega_{0} = 0$, giving a linearly
expanding empty solution:
\begin{equation}
S(t)=C\,t.  \label{milne}
\end{equation}
This is just flat space-time as seen by a uniformly expanding set of
observers (\cite{rin01}, pp. 360-363), singular at $t=0$. Its age is
$\tau_{0}=\frac{1}{H_{0}}.$ It is a good model of the far future
universe if $k<0$ and $\Lambda =0$.

3. The \textit{de Sitter} universe, for which \{$\mu = p = 0$,
$\Lambda \neq 0$, $k=0$\} $\Rightarrow$ $\Omega_{0} = 0,$ giving the
steady state expanding empty solution:\footnote{The Steady State
universe of Bondi, Hold and Hoyle \cite{bon60} utilised this metric,
but was non-empty as they abandoned the EFE.}
\begin{equation}
S(t) = C\,\exp\,(Ht),  \label{des}
\end{equation}
where $C$ and $H$ are constants. As the expansion rate is constant
forever, there is no start and its age is infinite.\footnote{It is
however singular in that it is geodesically incomplete; this metric
covers only half the de Sitter hyperboloid \cite{sch56,hawell73}.}
It is a good model of the far future universe for those cases which
expand forever with $\Lambda >0 $. It can alternatively be
understood as a solution with $ \Lambda = 0$ and containing matter
with the exceptional equation of state $\mu + p/c^{2}=0$. There are
other RW forms of the de Sitter Universe: a geodesically complete
form with ${k=+1}$, $S(t)=S_{0}\, \cosh{Ht}$ (a regular bounce), and
another geodesically incomplete form with ${k = -1}$ , $S(t) =
S_{0}\sinh{Ht}$ (a singular start). This lack of uniqueness is
possible because \emph{this is a spacetime of constant curvature},
with no preferred timelike directions or space sections
\cite{sch56,hawell73,rin01}.\footnote{There is also a static
(non-RW) form of the metric --- the first form of the metric
discovered.}

\subsubsection{An initial singularity?}
\label{sec:sing} The above are specific models: what can one say
generically? When the inequality (\ref{eq:ec}) is satisfied, one
obtains directly from the Raychaudhuri equation (\ref{Ray}) the

\begin{quotation}
\textbf{Friedmann-Lema\^{\i}tre Universe Singularity Theorem}
\cite{ehl61,ell71}: In a FL universe with $\Lambda \leq 0$ and $\mu
+ 3p/c^2 > 0$ at all times, at any instant $t_0$ when $H_{0}\equiv
(\dot{S}/S)_0 > 0$ there is a finite time $t_*$: $t_0 - (1/H_0) <
t_* < t_{0}$, such that $S(t) \rightarrow 0$ as $t \rightarrow t_*$;
the universe starts at a space-time singularity there, with $\mu
\rightarrow \infty$ and $T \rightarrow \infty$ if $\mu +p/c^2 > 0$.
\end{quotation}

\noindent This is not merely a start to matter --- it is a start to
space, to time, to physics itself. It is the most dramatic event in
the history of the universe: it is the start of existence of
everything. The underlying physical feature is the non-linear nature
of the EFE: going back into the past, the more the universe
contracts, the higher the active gravitational density, causing it
to contract even more. The pressure $p$ that one might have hoped
would help stave off the collapse makes it even worse because
(consequent on the form of the EFE) $p$ enters algebraically into
the Raychaudhuri equation (\ref{Ray}) with the same sign as the
energy density $\mu$. Note that the Hubble constant gives an
estimate of the age of the universe: the time $\tau_{0} = t_{0}-t_*$
since the start of the universe is less than $1/H_0$.

This conclusion can in principle be avoided by a cosmological
constant, but in practice this cannot work because we know the
universe has expanded by at least a ratio of $11$, as we have seen
objects at a redshift\footnote{The redshift $z$ for light emitted at
$t_{e}$ and observed at $t_{0}$ is related to the expansion by $1+z
= S(t_{0})/S(t_{e})$, see Sec.\ref{sec:obsrels}.} of $10$; from
(\ref{Fried1}), the cosmological constant would have to have an
effective magnitude at least $11^{3}=1331$ times the present matter
density to dominate and cause a turn-around then or at any earlier
time, and so would be much bigger than its observed present upper
limit (of the same order as the present matter density).
Accordingly, no turn around is possible while classical physics
holds \cite{ehlrin89}. However energy-violating matter components
such as a scalar field (Sec.\ref{sec:inflate}) can avoid this
conclusion, if they dominate at early enough times; but this can
only be when quantum fields are significant, when the universe was
at least $10^{12}$ smaller than at present.

Because $T_{rad} \propto S^{-1}$ (eqn.(\ref{scaling})), a major
conclusion is that \emph{a} Hot Big Bang \emph{must have occurred;
densities and temperatures must have risen at least to high enough
energies that quantum fields were significant}, at something like
the GUT energy. The universe must have reached those extreme
temperatures and energies at which classical theory breaks down.

\subsection{The hot big bang}
\label{sec:hbb} The matter and radiation in the universe gets hotter
and hotter as we go back in time towards the initial quantum state,
because it was compressed into a smaller volume. In this \emph{Hot
Big Bang} epoch in the early universe, we can use standard physical
laws to examine the processes going on in the expanding mixture of
matter and radiation \cite{wei72,per05}. A key feature is that about
300,000 years after the start of the Hot Big Bang epoch, nuclei and
electrons combined to form atoms. At earlier times when the
temperature was higher, atoms could not exist, as the radiation then
had so much energy it disrupted any atoms that tried to form into
their constituent parts (nuclei and electrons). Thus at earlier
times matter was ionized, consisting of negatively charged electrons
moving independently of positively charged atomic nuclei. Under
these conditions, the free electrons interact strongly with
radiation by Thomson scattering. Consequently matter and radiation
were tightly coupled in equilibrium at those times, and the Universe
was opaque to radiation. When the temperature dropped through the
ionization temperature of about 4000K, atoms formed from the nuclei
and electrons, and this scattering ceased: the Universe became very
transparent (today we are able to see galaxies at enormous distances
from us). The time when this transition took place is known as the
\emph{time of decoupling} --- it was the time when matter and
radiation ceased to be tightly coupled to each other, at a redshift
$z_{dec} \simeq 1100$ \cite{dod03}. By (\ref{scaling}), the universe
was radiation dominated ($\mu_{rad} \gg\mu_{mat}$) at early times
and matter dominated ($\mu_{rad}\ll\mu_{mat}$) at late
times;\footnote{The dynamically dominant Cold Dark Matter
(Sec.\ref{sec:cdm}) obeys the same density law (\ref{scaling}) as
baryons.} matter-radiation density equality occurred significantly
before decoupling (the temperature $T_{eq}$ when this equality
occurred was $T_{eq} \simeq 10^4$K; at that time the scale factor
was $S_{eq} \simeq 10^4 S_0$, where $S_0$ is the present-day value).
The dynamics of both the background model and of perturbations about
that model differ significantly before and after $S_{eq}$
\cite{dod03}.

\subsubsection{Cosmic Blackbody Radiation}
\label{sec:cbr0} Radiation was emitted by matter at the time of
decoupling, thereafter travelling freely to us through the
intervening space. When it was emitted, it had the form of blackbody
radiation, because this is a consequence of matter and radiation
being in thermodynamic equilibrium at earlier times. Thus \emph{the
matter at $z=z_{dec}$ forms the Last Scattering Surface} (LSS)
\emph{in the early universe, emitting Cosmic Blackbody Background
Radiation\footnote{This is often called ``Cosmic Microwave
Background", or CMB for short. However it is only microwave at the
present epoch.}} (`CBR') \emph{at 4000K, that since then has
travelled freely with its temperature $T$ scaling inversely with the
scale function of the universe}.\footnote{This scaling for freely
propagating radiation follows from the discussion in
Sec.\ref{sec:obsrels}.} As the radiation travelled towards us, the
universe expanded by a factor of about 1100; consequently by the
time it reaches us, it has cooled to 2.75 K (that is, about 3
degrees above absolute zero, with a spectrum peaking in the
microwave region), and so is extremely hard to observe. It was
however detected in 1965, and its spectrum has since been
intensively investigated, its blackbody nature being confirmed to
high accuracy \cite{par95}. Its existence is now taken as solid
proof both that the Universe has indeed expanded from a hot early
phase, and that standard physics applied unchanged at that era in
the early universe.

\subsubsection{Particle interactions and element formation}
\label{sec:nucleo} The thermal capacity of the radiation is hugely
greater than that of the matter. At very early times before
decoupling, the temperatures of the matter and radiation were the
same (because they were in equilibrium with each other), scaling as
$1/S(t)$ (eqn.(\ref{scaling})). The early universe exceeded any
temperature that can ever be attained on Earth or even in the centre
of the Sun; as it dropped towards its present value of 3 K,
successive physical reactions took place that determined the nature
of the matter we see around us today. At very early times and high
temperatures, only elementary particles can survive and even
neutrinos had a very small mean free path; as the universe cooled
down, neutrinos decoupled from the matter and streamed freely
through space. At these times the expansion of the universe was
radiation dominated, and we can approximate the universe then by
models with \{$k=0$, $w=1/3$, $\Lambda =0$\}, the resulting simple
solution of (\ref{Fried1}) uniquely relating time to temperature:
\begin{equation}
S(t)=S_{0}\,t^{1/2},\,\,\,  t=1.92\sec \left[
\frac{T}{10^{10}K}\right] ^{-2}.\label{edsradn}
\end{equation}
(There are no free constants in the latter equation).

At very early times, even neutrinos were tightly coupled and in
equilibrium with the radiation; they decoupled at about $10^{10}$K
(\cite{dod03}, pp. 44-46), resulting in a relic neutrino
background density in the universe today of about $\Omega_{\nu 0}
\simeq 10^{-5}$ if they are massless (but it could be higher
depending on their masses). Key events in the early universe are
associated with out of equilibrium phenomena (\cite{dod03}, p.
58). An important event was the era of \emph{nucleosynthesis}, the
time when the light elements were formed. Above about $10^9$K,
nuclei could not exist because the radiation was so energetic that
as fast as they formed, they were disrupted into their constituent
parts (protons and neutrons). However below this temperature, if
particles collided with each other with sufficient energy for
nuclear reactions to take place, the resultant nuclei remained
intact (the radiation being less energetic than their binding
energy and hence unable to disrupt them). Thus the nuclei of the
light elements --- deuterium, tritium, helium, and lithium ---
were created by neutron capture. This process ceased when the
temperature dropped below about $10^8$ K (the nuclear reaction
threshold). In this way, the proportions of these light elements
at the end of nucleosynthesis were determined; they have remained
virtually unchanged since. The rate of reaction was extremely
high; all this took place within the first three minutes of the
expansion of the Universe. One of the major triumphs of Big Bang
theory is that \emph{theory and observation are in excellent
agreement provided the density of baryons is low: $\Omega_{bar 0}
\simeq 0.044$. Then the predicted abundances of these elements}
(\emph{$25\%$ Helium by weight, $75\%$ Hydrogen, the others being
less than $1\%$})\emph{\ agrees very closely with the observed
abundances}. Thus the standard model explains the origin of the
light elements in terms of known nuclear reactions taking place in
the early Universe \cite{schtur98}. However heavier elements
cannot form in the time available (about 3 minutes).

In a similar way, physical processes in the very early Universe
(before nucleosynthesis) can be invoked to explain the ratio of
matter to anti-matter in the present-day Universe: a small excess of
matter over anti-matter must be created then in the process of
\emph{baryosynthesis}, without which we could not exist today (if
there were no such excess, matter and antimatter would have all
annihilated to give just radiation \cite{sil05}). However other
quantities (such as electric charge) are believed to have been
conserved even in the extreme conditions of the early Universe, so
their present values result from given initial conditions at the
origin of the Universe, rather than from physical processes taking
place as it evolved. In the case of electric charge, the total
conserved quantity appears to be zero: after quarks form protons and
neutrons at the time of baryosynthesis, there are equal numbers of
positively charged protons and negatively charged electrons, so that
at the time of decoupling there were just enough electrons to
combine with the nuclei and form uncharged atoms (it seems there is
no net electrical charge on astronomical bodies such as our galaxy;
were this not true, electromagnetic forces would dominate cosmology,
rather than gravity).

After decoupling, matter formed large scale structures through
gravitational instability (\cite{bot98}, pp. 183-222) which
eventually led to the formation of the first generation of stars
\cite{sil05} and is probably associated with the re-ionization of
matter (\cite{dod03}, p. 73). However at that time planets could not
form for a very important reason: there were no heavy elements
present in the Universe. The first stars aggregated matter together
by gravitational attraction, the matter heating up as it became more
and more concentrated, until its temperature exceeded the
thermonuclear ignition point and nuclear reactions started burning
hydrogen to form helium. Eventually more complex nuclear reactions
started in concentric spheres around the centre, leading to a
build-up of heavy elements (carbon, nitrogen, oxygen for example),
up to iron. These elements can form in stars because there is a long
time available (millions of years) for the reactions to take place.
Massive stars burn relatively rapidly, and eventually run out of
nuclear fuel. The star becomes unstable, and its core rapidly
collapses because of gravitational attraction. The consequent rise
in temperature blows it apart in a giant explosion, during which
time new reactions take place that generate elements heavier than
iron; this explosion is seen by us as a Supernova (\textquotedblleft
New Star") suddenly blazing in the sky, where previously there was
just an ordinary star. Such explosions blow into space the heavy
elements that had been accumulating in the star's interior, forming
vast filaments of dust around the remnant of the star. It is this
material that can later be accumulated, during the formation of
second generation stars, to form planetary systems around those
stars. Thus \emph{the elements of which we are made} (\emph{the
carbon, nitrogen, oxygen and iron nuclei for example}) \emph{were
created in the extreme heat of stellar interiors, and made available
for our use by supernova explosions}. Without these explosions, we
could not exist.

\subsection{Cosmological Observations}
\label{sec:obs} Cosmological models only become meaningful when
related to astronomical observations
\cite{hoy60,san61,ell71,wei72}. These are of two kinds:
astronomical observations of distant matter tells us what was
happening far away in the universe and (because of the finite
speed of light) a long time ago. On the other hand observations of
nearby objects (matter on Earth, the solar system, nearby stars
for example) when related to theories of origins tell us what was
happening very near our past world line a very long time ago. The
first set of observations may be characterized as ``null cone"
observations, the second as ``geological" observations, one of the
most important being the determination of local element
abundances, which are then related to nucleosynthesis calculations
(Sec.\ref{sec:nucleo}).

Observations are totally dependent on telescope and detector
technology \cite{har84,bot98}. After the initial establishment of
distance scales and the basic evidence of cosmic homogeneity and
expansion in the 1920s and 1930s, progress was slow until the 1960s
when observations were extended from the optical to the entire
electromagnetic spectrum. In recent decades cosmology has changed
from a data-poor to a data-rich subject. Massive new data sets are
now available because of the extraordinary improvement of telescope,
detector, and computer technology in recent decades, particularly
the advent of new detectors such as Charge Coupled Devices (CCD's)
and fibre optics (enabling simultaneous measurement of hundreds of
redshifts). We now have not only optical, ultraviolet, and infrared
observations of galaxies, determining luminosities and spectra with
unprecedented sensitivity, but also radio, X-ray, and gamma-ray sky
surveys. Galaxies have been detected up to a redshift of $6$ and we
have identified many quasi-stellar objects and gamma-ray bursters as
well as multiple images of very distant gravitationally-lensed
galaxies \cite{har84}. Large-scale structures (clusters of galaxies,
superclusters, walls, and voids) have been identified, with
associated large-scale velocity flows (\cite{bot98}, pp. 85-137).

In addition to large-scale number-count and redshift surveys, we
have measured the background radiation spectrum and anisotropies at
all wavelengths. We identify the radiation as `background' precisely
when it is constant on very small angular scales (as opposed to
discrete sources, which appear as isolated objects). There is a
complex relation of this radiation to the intergalactic matter
density and thermal history. The most important component of the
background radiation is the Cosmic Blackbody Radiation (`CBR')
mentioned above (Sec.\ref{sec:hbb}); detailed observations have
mapped its temperature over the whole sky at a sensitivity of better
than one part in $10^{5}$. However other components of the
background radiation (X-ray and radio in particular) convey
important information on the temperature and density of
intergalactic matter, and hence strongly restrict its possible
thermal history. For example hot matter emits X-rays, so the X-ray
background measurement restricts the amount of hot intergalactic
matter allowed; while neutral hydrogen strongly absorbs Ultra-Violet
radiation to give the Lyman alpha spectral absorption line, so
absence of such absorption gives strong limits on the amount of
neutral hydrogen and hence on the temperature of intergalactic
matter.

\subsubsection{Isotropy}
\label{sec:isotropy} The first important point about cosmological
observations is that \emph{when averaged on a large enough
physical scale} (\emph{clusters of galaxies and above}) \emph{they
are statistically isotropic about us}; there is no direction
apparently pointing to the centre of the universe. The \emph{high
degree of isotropy of the CBR} strongly supports this conclusion:
its temperature is the same in all directions about us to better
than one part in 10,000 after we have allowed for the motion of
the Earth relative to the cosmos (about 250 km/sec), which creates
a temperature dipole at one part in a thousand.\footnote{The CBR\
dipole that could be interpreted as due to a major cosmological
inhomogeneity is rather interpreted as being due to our motion
(`peculiar velocity') relative to a spatially homogeneous
universe.} Any inhomogeneities or anisotropies in the matter
distribution lead to anisotropies in this radiation, as recently
measured at only one part in $10^{5}$ by the extremely sensitive
detectors of the COBE and WMAP satellites. This high degree of
isotropy is the major reason we believe the Universe itself is
spatially homogeneous and isotropic to a good approximation (see
Sec.\ref{sec:homog}), providing good observational support for use
of the FL universe models as accurate models of the observed
region of the universe.

\subsubsection{Distance scale and ages}
\label{sec:ages} The underlying problem in all astronomy is
determining the distances of observed objects. This is done by a
`cosmic distance ladder' (\cite{bot98}, pp. 25-83) whereby nearest
objects have their distance determined by parallax (i.e.
essentially by local trigonometry); and more distant ones by a
series of consecutive distance indicators (Cepheid variables, RR
Lyrae variables, brightest red supergiants) until at a
cosmological distance, redshift $z$ is a primary distance
indicator, but is contaminated by local velocities of matter
relative to the rest-frame of the universe. Other distance
indicators (for example the Tully-Fisher method, the luminosity
function of planetary nebulae, the globular cluster luminosity
function, surface brightness fluctuations) serve to refine the
estimates \cite{bot98}.

Closely associated with the distance scale is determination of the
Hubble constant $H_0$ (the present rate of expansion of the
universe), because estimates of the size of the observable region of
the universe scale with the Hubble constant. But the Hubble constant
also determines the age of the universe, so its determination
underlies a crucial consistency condition for cosmology: \textit{the
age of objects in the universe} (\emph{rocks, planets, stars, star
clusters, galaxies}) \emph{must be less than the age of the
universe}. This condition has been a cause of concern ever since we
have had good estimates of ages and of the Hubble
constant.\footnote{ Indeed Hubble himself never fully accepted the
expanding universe theory because of age difficulties, preferring to
refer to a redshift-distance relation rather than a velocity
distance relation \cite{hub36}. However the problem has been eased
by a series of revisions of the value of the Hubble constant since
then, due to a better understanding of the primary distance
indicators.} It seems not to be violated by current observations of
low redshift objects given the current estimates of $H_{0}\simeq 70$
km/sec/Mpc, giving an age of the universe of about $15$ billion
years whereas the oldest star clusters seem to be about $14$ billion
years old. However it is very tight, perhaps even problematic, for
very distant (and so much younger) objects \cite{jaidev05}.

\subsubsection{Observational relations}
\label{sec:obsrels} Light travels on null geodesics $x^{a}(\lambda)$
in spacetime (the tangent vector $k^{a}:=dx^{a}/d\lambda $ is such
that $k_{\,\,;b}^{a}k^{b}=0$, $ k^{a}k_{a}=0$). In a RW\ geometry,
it suffices to consider only radial null geodesics (by the
symmetries of the model, these are equivalent to generic geodesics).
Then from (\ref{eq:rw}) we find that for light emitted at time
$t_{e}$ and received at time $t_{0}$, the comoving radial distance
$u(t_{0},t_{1}):=r_{0}-r_{1}$ between comoving emitters and
receivers is given by
\begin{equation}
\{ds^{2}=0,\;d\theta =0=d\phi\} \,\,\, \Rightarrow \,\,\,
u(t_{0},t_{1})=\int_{t_{1}}^{t_{0}}\frac{dt}{S(t)}
=\int_{S_{1}}^{S_{0}}\frac{dS}{S\dot{S}}  \label{geod}
\end{equation}
with $\dot{S}$ given by the Friedmann equation (\ref{Fried}). The
key quantities related to cosmological observations are redshift,
area distance (or apparent size), and local volume corresponding to
some increment in distance (determining number counts)
\cite{san61,ell71,wei72}. The redshift $ z$ measured for an object
emitting light of wavelength $\lambda _{e}$ that is observed with
wavelength $\lambda _{0}$ is given by
\begin{equation}
1+z:=\frac{\lambda _{0}}{\lambda _{e}}=(1+z_{c})(1+z_{v}),
\label{red}
\end{equation}
where $z_{v}$ is the redshift caused by the local peculiar motion of
the object observed ($z_{v}=0$ for comoving objects), and $z_{c}$ is
the cosmological redshift given by
\begin{equation}
1+z_{c}=\frac{S(t_{0})}{S(t_{e})}.  \label{redshift}
\end{equation}
From eqn.(\ref{red}), the same ratio of observed to emitted light
holds for all wavelengths:\ a key identifying property of redshift.
The problem in using redshifts of objects as a distance indicator is
to separate out the cosmological from the Doppler components, which
lead to redshift-space distortions (\cite{dod03}, pp. 275-282); this
can reasonably be done for a cluster of galaxies by appropriate
averaging over cluster members ($\langle z_{v}\rangle=0$ for a
comoving cluster). The area distance $r_{0}$ of an object at
redshift $z_c$ and of linear size $l$ which subtends angular size
$\alpha$ is given by\footnote{This depends only on $z_c$ because
apparent shapes and sizes are independent of the motion of the
source.}
\begin{equation}
r_{0}(z_c) :=\frac{l}{\alpha }=f(u)S(t_{e}). \label{area dist}
\end{equation}
Thus measures of apparent sizes will determine the area distance if the
source physical size is known. The flux of radiation $F$ measured from a
point source of luminosity $L$ emitting radiation isotropically is given by
the fraction of radiant energy emitted by the source in a unit of time that
is received by the telescope:
\begin{equation}
F=\frac{L}{4\pi }\frac{1}{f^{2}(u)S^{2}(t_{0})(1+z)^{2}} ;
\label{flux0}
\end{equation}
(the two redshift factors account firstly for time dilation observed
between observer and source, and secondly for loss of energy due to
redshifting of photons). The source's \textit{apparent magnitude}
$m$ is defined from the flux: $ m=-2.5log_{10}F+const$. On using
(\ref{redshift}, \ref{area dist}), equation (\ref{flux0}) becomes
\begin{equation}
F=\frac{L}{4\pi }\frac{1}{r_{0}^{2}(1+z)^{4}}.  \label{flux}
\end{equation}
showing that measures of magnitudes will determine the area distance
if the source's intrinsic luminosity is known. On using (\ref{area
dist}) it follows from (\ref{flux}) that \emph{the point-wise
surface brightness of extended objects (the flux received per unit
solid angle) depends only on redshift} \cite{krisac66,ell71} --- a
key feature in determining detection probabilities and in
gravitational lensing observations. It further follows from this
result that \emph{a blackbody spectrum emitted at temperature
$T_{e}$ when observed with a redshift $z$ remains a blackbody
spectrum but with observed temperature} $T_{0}=T_{e}/(1+z)$ --- a
crucial feature in analyzing the CBR observations.

Using the Friedmann equation and the relevant equation of state for
matter, the area distance can be determined as a function of
redshift $z_c$ in terms of the Hubble constant $H_{0}$, deceleration
parameter $q_{0}$, and cosmological constant $\Lambda $. In the case
of pressure-free matter with vanishing cosmological constant, one
obtains from (\ref{geod}), (\ref{Fried}), (\ref{redshift}), and
(\ref{area dist})\footnote{ Or, more elegantly, from the geodesic
deviation equation (see \cite{ellels99a}).} the Mattig relation
\cite{san61}
\begin{equation}
\ r_{0}(z_{c})=\frac{1}{H_{0}q_{0}^{2}}\frac{
(q_{0}-1)(1+2q_{0}z_{c})^{1/2}+(q_{0}(z_{c}-1)+1)}{(1+z_{c})^{2}}\,.
\label{mattig}
\end{equation}
Consequently measures of either apparent size of sources of known
physical size, or of radiant flux from sources of known intrinsic
luminosity, will determine the deceleration parameter $q_{0}$.
Generalizations of this relation hold if a cosmological constant or
radiation is present. An interesting aspect is that there is a
minimum apparent size for objects of fixed physical size at some
redshift $z_c=z_{\ast}$ depending on the density parameter and the
cosmological constant. The past light cone of the observer attains a
maximum area at $z_{\ast}$; the entire universe acts as a
gravitational lens for further off objects, magnifying their
apparent size so that very distant objects can appear to have the
same angular size as nearby ones \cite{hoy60,san61,ell71}. For the
Einstein-de Sitter universe, the minimum angular diameter is at
$z_{\ast }=1.25$; in low density universes, it occurs at higher
redshifts.

The number of objects seen in a solid angle $d\Omega $ for a
distance increment $du$ (characterized by increments $dz$, $dm$ in
the observable variables $z$ and $m$) is given by
\begin{equation}
dN = W(u)\,\rho (t_{e})S^{3}(t_{e})f(u)dud\Omega  \label{no}
\end{equation}
where the detection probability or `selection function' is $W(u)$
(\cite{dod03}, p. 263) and $\rho(t_{e})$ is the number density of
objects at the time of emission (spatial homogeneity is expressed
in the fact that this is independent of the spatial coordinates).
The observed total number $N$ of objects measured in a survey is
given by integrating from the observer to the survey limit: in
terms of the radial coordinate $r_{e}$ of the source (which can be
related to redshifts or magnitudes), $N=\int_{r_{0}}^{r_{e}}dN$.
If the number of objects is conserved (e.g. observing galaxies in
an epoch when they are neither created nor destroyed),
$\rho(t_{e}) = \rho (t_{0})(1+z)^{3}$ and we find from (\ref{no})
that in the idealized case when $W$ is independent of distance (a
reasonable assumption for relatively nearby objects),
\begin{equation}
N = W\rho (t_{0})d\Omega \int_{r_{0}}^{r_{e}}f(u)du.
\label{number}
\end{equation}
The simple integral has to be separately done for the cases
$k=+1,0,-1$ \cite{san61}.

The above equations enable one to determine observational relations
between observable variables, for example $(m,z),$ $(\alpha ,z)$ or
$(N,m)$ relations for objects with known intrinsic properties (known
size or luminosity, for example), and so to observationally
determine $q_{0}$. These relations have to be modified if there is
absorption by an intergalactic medium, gravitational lensing, or
anisotropic emission of radiation; and detailed comparisons with
observations have to take into account the spectrum of the source as
well as source detection and identification probabilities
\cite{har84}. Here we encounter \emph{the contrast between image and
reality} : there can be many objects out there that we either do not
detect, or do not recognize for what they are \cite{dis76}. An
``observational map" relating source properties to the nature of
their images gives a useful view of how this occurs
\cite{elletal84}.

One important feature here is that \emph{a specific object will look
completely different at different wavelengths} (optical, radio,
X-ray for example); indeed it may be detectable at one wavelength
and not at another. This shows very clearly how our images of
reality are dependent on the detectors we use. To get a full picture
of what is out there, we need to use multiple modes of investigation
--- imaging at all wavelengths together with intensity, spectral, and
polarization measurements \cite{har84}, as well as watching for
time variations. A second important feature is observational
selection effects such as the \textit{Malmquist bias} --- if we
have a population of objects with different luminosities, at large
distances we will only see the more luminous objects (the fainter
ones will not be detected); hence \textit{the average luminosity
will appear to increase with distance}, but this is just an
observational effect rather than the real state of affairs. Using
different detection thresholds controls this effect to some
degree.

\subsubsection{Number Counts and the visible matter density}
Number counts of galaxies as a function of redshift or luminosity
show approximate spatial homogeneity of the universe \cite{hub36}.
However counts of radio sources and quasi-stellar objects (qso's)
show that the universe has not been in a steady state as proposed
by Bondi, Gold, and Hoyle \cite{bon60}. Indeed \emph{number counts
are only compatible with a RW geometry if there has been evolution
of source numbers and/or luminosities} \cite{sci71}.

Number counts also give estimates of the density of visible
(luminous) matter in the universe: $\Omega_{vm0}\simeq 0.015$. This
is very low relative to the critical density ($\Omega_{0}=1)$ and is
also considerably less than the amount of baryons determined by
nucleosynthesis studies ($ \Omega_{bar 0}\simeq 0.044,$ see
Sec.\ref{sec:nucleo}). Thus \emph{much of the baryonic matter in the
universe is in some hidden} (\emph{non-luminous}) \emph{form}
(\cite{bot98}, pp. 223-272)), e.g. burnt out stars \cite{hog99}.

\subsubsection{Apparent Luminosities and sizes:\ Dark Energy}
\label{sec:super} Apparent sizes or luminosities as a function of
redshift can be used to determine the deceleration parameter $q_0$
(Sec.\ref{sec:FL}) if the intrinsic source sizes or luminosities are
known. The problem is that until recently there were not known
enough galaxies or other objects of standard size or luminosity to
use to determine $q_{0}$, and scatter in their properties leads to
biassing of observations by the Malmquist effect
(Sec.\ref{sec:obsrels}). However this dramatically changed with
recent observations of the decay curves of the luminosity of
supernovae in distant galaxies. It turns out that the peak
luminosity of type Ia supernovae is closely correlated with their
light curve decay time, for the first time giving reliable `standard
candles' for galaxies at large distances \cite{per98}. The
conclusion from these observations is that, rather than slowing down
as expected, \emph{the rate of expansion of the universe is speeding
up at a rate corresponding to a cosmological constant with}
$\Omega_{\Lambda 0}=0.7$. This evidence is concordant with that from
CBR observations (Sec.\ref{sec:cbr}) and number counts
\cite{dod03,sil05}.

The nature of the field or matter causing this acceleration is
unclear. Its equation of state $w :=pc^2/\mu$, is unknown, and
many physical and unphysical proposals are being made in this
regard. From (\ref{Ray}), it has to violate the strong energy
condition (\ref{eq:ec}) and so must have a large negative
pressure. It could indeed be due to a cosmological constant
($w=-1$), which would have dominated the expansion of the universe
since a redshift $z\simeq 0.33$, and would have been negligible
earlier (and is also negligible on small scales --- it does not
affect local astrophysics). However it could also be some other
form of matter or field with effective negative pressure so that
$w<-1/3$, as can happen in the case of a scalar field (see
eqn.(\ref{scalar1}) below). In that case it is called
`quintessence'. There are many speculations as to what this might
be, but there is no clarity on the matter. One should note here
that alternative explanations of the observations are possible,
for they can be exactly reproduced by a spherically symmetric
inhomogeneous universe model where we are near the centre
\cite{musetal98}, or could at least partly be due to the
back-reaction of inhomogeneities on the cosmic expansion
\cite{ellbuc05} or the effect of inhomogeneities on the effective
area distance \cite{kan95,kan98}. These alternatives are being
investigated, but the most probable cause remains some unknown
kind of matter or field with effective negative energies.

In summary, the standard gravitational equations together with the
supernovae observations imply \emph{presence of a cosmological
constant or some equivalent form of `dark energy' with a large
effective negative energy density $\mu_{grav}$ (due to negative
pressure) dominating the present expansion of the universe; its
physical nature is unknown}. There is no known physics reason why
this force should exist at this level where it is just detectable
--- quantum field theory relates the cosmological constant to the
zero-point energy of the vacuum, and suggests it should be
enormously larger than observed
\cite{wei89,wei00,rugzin02,zin02,sus05}. It is a major mystery why
it exists at the small (just detectable) level that observations
indicate \cite{sei03}. A key aspect of present day cosmology is
trying on the one hand to observationally determine the effective
equation of state of this `dark energy' (running the field
equations backwards to obtain $w(z)$ from the observations
\cite{saietal00}, and in particular determining whether $w$ is
constant or varying over time), and on the other attempting to
give a plausible theoretical explanation for its physical origin.

\subsubsection{Matter Distribution and Motion: Dark Matter}
\label{sec:cdm} Detailed studies have been made of the distribution
of galaxies and their motions. They occur in clusters, in turn
making up superclusters imbedded in vast walls surrounding
relatively sparsely populated intergalactic voids. The galaxy
\textit{luminosity function} characterizes the numbers of galaxies
occurring within each luminosity class; the \textit{covariance
function} characterizes their spatial clustering \cite{pee93,dod03}.
Large scale motions occur for galaxies in clusters, and for the
clusters themselves. It is easy to conceive of matter that is hard
to detect (for example, small rocks distributed through space);
studies of galactic rotation curves and of motions of galaxies in
clusters (\cite{bot98}, pp. 139-181) imply \emph{the existence of
huge amounts of unseen dark matter, dominating the dynamics of the
Universe}: its density is $\Omega_{dm 0} \simeq 0.3$, much greater
than both visible matter ($\Omega_{vm 0} = 0.015$) and baryons
($\Omega_{bar 0} = 0.044$), but significantly less than the critical
density $\Omega_{0} = 1$. Thus \emph{the dark matter is
non-baryonic}, meaning it has some kind of exotic nature rather than
being the protons and neutrons that are the substance of ordinary
matter \cite{sei03}. In contrast to the `dark energy' discussed in
the previous section, dark matter is dynamically effective on
astrophysical scales as well as on cosmological scales. Many
attempts have been made to identify its nature, for example it might
be axions, supersymmetric partners of known particles, quark
nuggets, or massive neutrinos \cite{griree91,per05}, but what it is
composed of is still unknown. Laboratory searches are under way to
try to detect this matter, so far without success. A key question is
whether its apparent existence is due to our using the wrong theory
of gravity on these scales. This is under investigation, with
various proposals for modified forms of the gravitational equations
that might explain the observations without the presence of large
quantities of dark matter. This is a theoretical possibility, but
the current consensus is that this dark matter does indeed exist.

An important distinction is whether dark matter consists of

(i) weakly interacting massive particles that cooled down quickly,
thereafter forming \emph{cold dark matter} (`CDM') moving slowly at
the time of structure formation (and resulting in a bottom-up
process with large scale structure forming from smaller scale
structures), or

(ii) particles that have a low mass and cooled slowly, therefore for
a long time forming \emph{hot dark matter}, moving very fast at the
time of structure formation (and resulting in a top-down galaxy
formation scenario).

\noindent Structure formation studies currently support the CDM
hypothesis, with hierarchical formation of gravitationally bound
objects taking place in a complex bottom up process involving
interactions of CDM, baryons, and radiation, with dwarf galaxies
forming initially \cite{sil05,moutan05} and then aggregating to form
larger structures. These studies are based on massive numerical
simulations, with initial conditions provided by the inflationary
scenario discussed below, see Sec.\ref{sec:inflate}. Unlike `dark
energy', CDM\ has an ordinary baryonic equation of state (it is a
perfect fluid (\ref{pf}) with $p_{cdm} = 0$ $\Leftrightarrow
w_{cdm}=0$).

Another way of detecting dark matter in clusters is by its
gravitational lensing effects \cite{schetal98}. The bending of light
by massive objects was one of the classic predictions of General
Relativity theory. Rich clusters of galaxies or galaxy cores can
cause strong lensing of more distant objects, where multiple images
of distance galaxies and qso's occur, sometimes forming rings or
arcs; and weaker lensing by closer masses results in characteristic
patterns of distortions of images of distant objects. Analysis of
multiple images can be used to reconstruct the lensing mass
distributions, and statistical analysis of weak lensing patterns of
image distortions are now giving us detailed information on the
matter distribution in distant galaxies and clusters. These studies
show that to get enough lenses in an almost flat cosmology
($\Omega_{0}\simeq 1)$ requires the presence of a cosmological
constant --- there cannot be a critical density of dark matter
present \cite{dod03,sil05}.

A key feature of present-day cosmology is attempts to identify the
nature of this dark matter, and if possible to detect it in a
laboratory situation. While observations favour the CDM\ scenario,
some residual problems as regards the emergence of fine-scale
structure still need resolution \cite{sil05}.

\subsubsection{The CBR Power spectrum}
\label{sec:cbr} The CBR angular anisotropies are characterized by an
angular power spectrum showing the amount of power in perturbations
at each physical scale on the LSS \cite{benetal03,sei03,dod03}. In
the time from the the end of inflation to the LSS, modes of
different wavelengths can complete a different number of
oscillations. This translates the characteristic periods of the
waves into characteristic lengths on the LSS, leading to a series of
maxima (`acoustic peaks') and minima in the inhomogeneities on the
LSS and consequently in the CBR angular anisotropy power spectrum
\cite{husug95,pea99,per05}. These inhomogeneities then form the
seeds for structure formation and so are related to the power
spectrum of physical scales for structures that form later. They are
characterised by a (3-dimensional) spatial power spectrum on the
LSS; because we receive the observed CBR radiation from the 2-sphere
$S_{2:LSS}$ where our past light cone intersects the LSS, this is
seen by us as a 2-dimensional power spectrum of anisotropies on the
sky (characterised by the unit sphere $S_2$ of all direction vectors
$e_a$: $e^{a}e_a=1$, $e^a u_a=0$).

The apparent angular size of the largest CBR peak (about $1^{o}$)
allows estimates of the area distance to the LSS and hence of the
density of matter in the universe for various values of the
cosmological constant, and determines the major cosmological
parameters \cite{spe03}:
\begin{quote}
   \emph{``By combining WMAP data with
other astronomical data sets, we constrain the geometry of the
universe: $\Omega_{tot}=1.02\pm 0.02$, the equation of state of the
dark energy, $w<-0.78$ (95\% confidence limit), and the energy
density in neutrinos, $\Omega_{\nu }h^{2}<0.0076$ (95\% confidence
limit). For 3 degenerate neutrino species, this limit implies that
their mass is less than 0.23 eV (95\% confidence limit). The WMAP
detection of early reionization rules out warm dark matter."}
\end{quote}
\noindent There is however a problem here: while the agreement of
theory and observations for all small angular scales is remarkable,
there is a divergence at the largest angular scales: the
observations show less power than expected. Specifically, the
quadrupole and octopole are much lower than theory predicts. Also
the axes of the quadrupole and octopole are very precisely aligned
with each other, and there are other angular anomalies
\cite{stasch05}. These effects might be due to (i) observational
contamination by the galaxy (which gets in the way of our view of
the LSS), (ii) a contingent (`chance') event (it represents `cosmic
variance', discussed below, see Sec.\ref{sec:unique}), (iii) our
living in a well-proportioned `small universe' which is spatially
closed so that there is a maximum size to possible fluctuations
\cite{weeetal04}, or (iv) some unexpected new physical effect or
deeper problem with our understanding of the early universe. The
jury is out as to which the case is; this could turn out to be a
crisis for the CBR analysis, but on the other hand one can always
just resort to saying it is a statistical fluke (the underlying
problem here being the uniqueness of the universe, as discussed in
Sec.\ref{sec:unique}).

There are similar expected peaks in the polarization spectrum of
this radiation, and polarization maps should have a mode associated
with gravitational waves predicted by inflation to exist in the very
early universe (Sec.\ref{sec:inflate}); detection of such modes will
be a crucial test of inflation \cite{dod03,sie05}. Studies of
polarization indicate that reionisation of the universe took place
as early as a redshift of 17, contrary to what is deduced from qso
studies. More detailed studies of anisotropies involve the
Sunyaev-Zel'dovich effect (changes in the observed temperature due
to scattering by hot matter in galaxy clusters) and gravitational
lensing.

There is a huge amount of information in the CBR\ maps, and their
more accurate measurement and interpretation is a central feature
of current cosmology \cite{stei95,pea99,dod03,per05}.

\subsection{Causal and visual horizons}
\label{sec:horizon} A fundamental feature affecting the formation of
structure and our observational situation is the limits arising
because causal influences cannot propagate at speeds greater than
the speed of light. Thus the region that can causally influence us
is bounded by our past null cone. Combined with the finite age of
the universe, this leads to the existence of particle horizons
limiting the part of the universe with which we can have had causal
connection.\footnote{There are also \emph{event horizons} and
\emph{apparent horizons} in some cosmological models (\cite{rin56};
\cite{tipetal80}; \cite{rin01}, pp. 376-383).}

A \textit{particle horizon} is by definition comprised by the
limiting worldlines of the furthest matter that ever intersects our
past null cone \cite{rin56,rin01}. This is the limit of matter that
we can have had any kind of causal contact with since the start of
the universe, characterized by the comoving radial coordinate value
\begin{equation}
u_{ph}=\int_{0}^{t_{0}}\frac{dt}{S(t)}\,.  \label{ph}
\end{equation}
The present physical distance to the matter comprising the horizon is
\begin{equation}
d_{ph}=S(t_{0})u_{ph}.
\end{equation}
The key question is whether the integral (\ref{ph}) converges or
diverges as we go to the limit of the initial singularity where
$S\rightarrow 0$. \emph{Horizons will exist in standard FL
cosmologies for all ordinary matter and radiation}, for $u_{ph}$
will be finite in those cases; for example in the Einstein-de Sitter
universe (see Sec.\ref{sec:basic}), $ u_{ph}=3t_{0}^{1/3} $,
$d_{ph}=3t_{0}$. We will then have seen only a fraction of what
exists, unless we live in a universe with spatially compact sections
so small that light has indeed had time to traverse the whole
universe since its start; this will not be true for universes with
the standard simply-connected topology. Penrose's powerful use of
conformal methods (see \cite{hawell73,tipetal80}) gives a very clear
geometrical picture of the nature of horizons \cite{ellwill}. They
may not exist in non-FL universes, for example Bianchi (anisotropic)
models \cite{mis69}. In universes with closed spatial sections, a
supplementary question arises: Is the scale of closure smaller than
the horizon scale? There may be a finite time when causal
connectivity is attained, and particle horizons cease to exist. In
standard $k=+1$ FL models, this occurs just as the universe reaches
the final singularity; if however there is a positive cosmological
constant or other effective positive energy density field, it will
occur earlier. The horizon always grows, because (\ref{ph}) shows
that $ u_{ph}$ is a monotonically increasing function of $t_{0}$.
Despite many contrary statements in the literature, \emph{it is not
possible that matter leave the horizon once it has entered}. In a
(perturbed) FL model, once causal contact has taken place, it
remains until the end of the universe.

The importance of horizons is two-fold: they underlie causal
limitations relevant in the origin of structure and uniformity
\cite{mis69,gut81}, and they represent absolute limits on what is
testable in the universe \cite{ell75,ell80}.

\subsubsection{Causal limitations}
\label{sec:puzzles} As to causal limitations, horizons are important
in regard both to the smoothness of the universe on large scales,
and the lumpiness of the universe on small scales. The issue of
smoothness is encapsulated in the \textit{horizon problem}: if we
measure the temperature of the CBR arriving here from opposite
directions in the sky in a standard FL\ model, it came from regions
of the surface of last scattering that can have had no causal
contact of any kind with each other since the start of the universe.
In a radiation-dominated early universe with scale factor
(\ref{edsradn}), the size of the particle horizon at the time of
last scattering appears as an angular scale of about $1^{o}$ in the
sky today, and corresponds to a comoving physical length of about
400,000 light years when evaluated today. Why then are conditions so
similar in these widely separated regions?
\cite{mis68,gut81,blagut87,koltur90}. Note that this question is of
a philosophical rather than physical nature, i.e. there is no
contradiction here with any experiment, but rather an unease with an
apparent fine tuning in initial conditions. This problem is claimed
to be solved by the inflationary universe scenario mentioned below,
see Sec.\ref{sec:inflate}.

Associated with the existence of horizons is the prediction that physical
fields in different regions in the universe should be uncorrelated after
symmetry breaking takes place, because they cannot have interacted causally.
Consequently, if grand unified theories are correct, topological defects
such as monopoles and cosmic strings may be expected as relics of the
expansion of the very early universe \cite{koltur90}. In a standard
cosmology, far too many monopoles are predicted. This is also solved by
inflation.

As to the lumpiness, the issue here is that if we believe there was
a state of the universe that was very smooth --- as indicated at the
time of decoupling, by the low degree of anisotropy of the CBR, and
represented by the RW geometry of the FL models --- then there are
limits to the sizes of structures that can have grown since then by
causal physical processes, and to the relative velocities of motion
that can have been caused by gravitational attraction in the
available time (for example, the peculiar motion of our own galaxy
relative to the CBR\ rest frame caused by the huge overdensity
called the `Great Attractor'). If there are larger scale structures
or higher velocities, these must have been imprinted in the
perturbations at the time of last scattering, for they cannot have
been generated in a causal way since that time. They are set into
the initial conditions, rather than having arisen by physical
causation from a more uniform situation. This is a key factor in the
theory of growth of perturbations in the early universe where the
expansion damps their growth. The quantity determining the relevant
physical scales for local causal influences in an expanding universe
is the \emph{comoving Hubble radius} $\lambda_{H}: =(SH)^{-1}$; the
way perturbations of wavelength $\lambda$ develop depends on whether
$\lambda
>\lambda_{H}$ or $\lambda<\lambda_{H}$ (\cite{dod03}, pp. 146-150).

Actually the domain of causal influence is even more tightly
constricted than indicated by the past light cone: the limits coming
from the horizon size are limits on what can be influenced by
particles and forces acting at the speed of light. However only
freely travelling photons, massless neutrinos, and gravitons can
move at that speed; and such particles coming from cosmological
distances have very little influence on our galaxy or the solar
system (indeed we need very delicate experiments to detect them).
Any massive particles, or massless particles that are interacting
with matter, will travel slower (for example before decoupling,
light has a very small mean free path and information will travel
only by sound waves and diffusion in the tightly coupled
matter-radiation fluid). The characteristics for pressure-free
scalar and vector perturbations are timelike curves, moving at zero
velocity relative to the matter; while density perturbations with
pressure can move at the speed of sound, only tensor perturbations
can travel at the speed of light. Thus the true domain that
influences us significantly is much less than indicated by the
particle horizon. It is the small region round our past world line
characterised after decoupling by the comoving scale from which
matter coalesced into our galaxy: a present distance of about 1 to
1.95 Mpc,\footnote{Dodelson \cite{dod03}, p. 283; W Stoeger, private
communication.} corresponding to an observed angle of about 0.64
arcminutes on the LSS. Before decoupling it would have been limited
by the sound horizon (\cite{dod03}, p. 257) rather than the particle
horizon.

\subsubsection{Observational limitations}
\label{sec:vizhorizon} Clearly we cannot obtain any observational
data on what is happening beyond the particle horizon; indeed we
cannot even see that far because the universe was opaque before
decoupling. \emph{Our view of the universe is limited by the} visual
horizon, \textit{comprised of the worldlines of furthest matter we
can observe --- namely, the matter that emitted the CBR at the time
of last scattering} \cite{ellsto88,rotell93}. This occurred at the
time of decoupling $t=t_{dec}$ ($z_{dec}\simeq 1100$), and so the
visual horizon is characterized by $r=u_{vh}$ where
\begin{equation}
u_{vh}=\int_{t_{dec}}^{t_{0}}\frac{dt}{S(t)}\,<\,u_{ph}.
\end{equation}
Indeed the LSS delineates our visual horizon in two ways: we are
unable to see to \emph{earlier times} than its occurrence (because
the early universe was opaque for $t<t_{dec}$), and we are unable to
detect matter at \emph{ larger distances} than that we see on the
LSS (we cannot receive radiation from matter at co-moving coordinate
values $r>u_{vh}$). The picture we obtain of the LSS by measuring
the CBR from satellites such as COBE\ and WMAP\ is just a view of
the matter comprising the visual horizon, viewed by us at the time
in the far distant past when it decoupled from radiation. The
position of the visual horizon is determined by the geometry since
decoupling. Visual horizons do indeed exist, unless we live in a
small universe, spatially closed with the closure scale so small
that we can have seen right around the universe since decoupling.
This is a possibility that will be discussed below
(Sec.\ref{sec:small}). There is no change in these visual horizons
if there was an early inflationary period, for inflation does not
affect the expansion or null geodesics during this later period. The
major consequence of the existence of visual horizons is that many
present-day speculations about the super-horizon structure of the
universe --- e.g. the chaotic inflationary theory
(Sec.\ref{sec:inflate}) --- are not observationally testable,
because one can obtain no definite information whatever about what
lies beyond the visual horizon \cite{ell75,ell80}. This is one of
the major limits to be taken into account in our attempts to test
the veracity of cosmological models (Sec.\ref{sec:horizon1}).

\subsection{Theoretical Developments}
The cosmological application of Einstein's Theory of Gravitation has also
progressed greatly in past decades, as regards exact solutions and generic
properties of the field equations; as regards approximate solutions; and in
terms of understanding the relationship between them.

\subsubsection{Exact solutions and generic properties}
Theory initially predicted there must have been a start to the universe, but
it was not clear for a long time if this was simply due to the very special
exactly isotropic and spatially homogeneous geometry of the standard
Friedmann-Lema\^{\i}tre models. It was possible that more realistic models
with rotation and acceleration might show the prediction was a mathematical
artefact resulting from the idealized models used. The singularity theorems
developed by Penrose and Hawking \cite{hawell73,tipetal80,ear99} showed this
was not the case: even for realistic geometries, classical gravitational
theory predicts a beginning to the universe at a space-time singularity,
provided the usual energy conditions were satisfied. This study has led
inter alia to a greatly increased understanding of causality and topology of
generic universe models \cite{tipetal80}, including the fact that
singularities may have a quite different nature than those in the
Robertson-Walker models, for example being anisotropic \cite{tipetal80} or
of a non-scalar nature \cite{ellki74}.

Various classes of exact cosmological solutions are known
(Kantowski-Sachs and Bianchi spatially homogeneous but anisotropic
models, Tolman-Bondi spherically symmetric inhomogeneous models, and
`Swiss-Cheese' non-analytic models) enabling understanding of
dynamical and observational behaviour of more general classes of
models than just the FL models \cite{ellels99}. Dynamical systems
studies \cite{waiell96,uggetal03} relate the behaviour of whole
classes of anisotropic models in suitable state spaces, enabling
identification of generic patterns of behaviour (fixed points,
saddle points, attractors, etc.) and hence the relationship between
dynamics of higher symmetry and lower symmetry universes. These
studies help understanding to what degree the FL models are generic
within the families of possible cosmological models, and which
models might give observations similar to those in the FL models. In
particular they are relevant in considering the possible geometry of
the universe at very early or very late times.

\subsubsection{Perturbation theory, the gauge issue, and back reaction}
\label{sec:perturb} Sophisticated perturbation theory has been
developed to underlie the theory of structure formation in the
expanding universe, examining \emph{the dynamics of perturbed FL
models}. The fluid flow in these models can have shear, vorticity,
and acceleration, and the Weyl tensor $C_{ijkl}$ (see
(\ref{eq:weyl})) is not zero, so that density variations, tidal
forces, peculiar velocities, and gravitational waves can be present.
Detailed studies use the kinetic theory approximation for matter
(electrons, protons, dark matter) and radiation (photons,
neutrinos), with their dynamics described by the Boltzmann equation
(\cite{dod03}, Ch.4; \cite{uff05}), interacting with space-time
inhomogeneities characterised by a perturbed FL metric. A key issue
here is the \emph{gauge problem} --- how to choose the background
model in the perturbed spacetime \cite{ellsto87}. If this is not
properly handled then one may attain apparent perturbation solutions
that are pure gauge (they are mathematical rather than physical), so
that one can alter the apparent growth rate simply by changing
coordinates. The key to handling this is either to keep careful
track at all stages of remaining gauge freedom and possible changes
of gauge, or (preferably, in my view) to use gauge invariant
variables (see \cite{bar80,ellbru89,chalas98}).

Most of the literature on perturbation theory deals with the linear
case, but some studies tackle the non-linear regime (e.g.
\cite{lanver05}), and some consider questions such as the origin of
magnetic fields and the causes of galactic rotation. A key problem
here is properly relating relativistic analyses of astrophysical
dynamics to the Newtonian approaches most often used by
astrophysicists (e.g. {\cite{bot98}, pp. 183-222); this is not
straightforward.\footnote{ Some exact General Relativity results,
which must necessarily apply in the Newtonian limit of General
Relativity, have no Newtonian analogue; an example is the shear-free
theorem applying to pressure-free matter \cite{ell67}. The
underlying issue is that there are 10 field equations to be
satisfied in General Relativity, with 20 integrability conditions
(the Bianchi identities), but only one field equation to be
satisfied in Newtonian theory (Poisson's equation) together with 4
conservation equations.} A further unresolved issue is the nature of
gravitational entropy \cite{pen89,ell02,pen04}. Many statements
about the nature of entropy in physics textbooks are wrong when
gravity is dominant, leading to the spontaneous formation of
structures such as stars and galaxies. There is as yet no agreed
definition of gravitational entropy that is generally applicable;
until there is, cosmological arguments relying on entropy concepts
are ill-founded.

The existence of inhomogeneities in the universe raises the issue of
fitting and back-reaction. To what degree does the nature of the
exactly smooth FL models reflect the geometrical and dynamical
nature of more realistic `lumpy' universe models? \cite{ellsto87}.
Inhomogeneities lead to extra terms appearing in the evolution
equations for the idealized background models, representing the
back-reaction of the perturbations on their dynamics \cite{ell84}.
These could possibly be dynamically significant \cite{ellbuc05}, but
this is a matter of dispute.

\subsection{Inflation} \label{sec:inflate} Particle physics
processes dominated the very early eras, when exotic processes took
place such as the condensation of a quark-gluon plasma to produce
baryons. Quantum field theory effects were significant then, and
this leads to an important possibility: scalar fields producing
repulsive gravitational effects could have dominated the dynamics of
the universe at those times. This leads to the theory of the
inflationary universe, proposed by Alan Guth \cite{gut81,gut97}: if
$\mu_{grav} = \mu +3p/c^{2}<0$, which can happen if a scalar field
dominates the dynamics of the early universe, an extremely short
period of accelerating expansion will precede the hot big bang era
\cite{blagut87}. This produces a very cold and smooth
vacuum-dominated state, and ends in `reheating': conversion of the
scalar field to radiation, initiating the hot big bang epoch. This
inflationary process is claimed to explain the puzzles mentioned
above (Sec.\ref{sec:puzzles}): why the universe is so special (with
spatially homogeneous and isotropic geometry and a very uniform
distribution of matter), and also why the space sections are so
close to being flat at present (we still do not know the sign of the
spatial curvature), which requires very fine tuning of initial
conditions at very early times. Inflationary expansion explains
these features because particle horizons in inflationary FL models
will be much larger than in the standard models with ordinary
matter, allowing causal connection of matter on scales larger than
the visual horizon, and inflation also will sweep topological
defects outside the visible domain.

In more detail: in the case of a single scalar field $\phi$ with
spacelike surfaces of constant density, on choosing $u^{a}$
orthogonal to these surfaces, the stress tensor has a perfect fluid
form with
\begin{equation}
\mu = \frac{1}{2}\dot{\phi}^{2} + V(\phi),\; \ p/c^2 =
\frac{1}{2}\dot{\phi} ^{2} - V(\phi),  \label{scalar}
\end{equation}
and so
\begin{equation}
\mu +3p/c^2 = 2\dot{\phi}^{2}-2V(\phi).  \label{scalar1}
\end{equation}
The slow-rolling case is $\dot{\phi}^{2}\ll V(\phi)$, leading to
$\mu +p/c^2=2 \dot{\phi}^{2}\simeq 0\Rightarrow \mu +3p/c^2\simeq
-2\mu <0$. This then enables a resolution of the horizon problem in
inflationary FL models: if sufficient inflation took place in the
early universe, then all the regions from which we receive CBR were
causally connected; indeed if the universe began in an inflationary
state, or was inflationary with compact spatial sections, there may
be no causal horizons at all. The inflationary models also cause
initial perturbations to die away, including velocity perturbations,
hence explaining the observed smoothness of the universe on large
scales. This process is expected to create a universe with very flat
spatial sections at late times:
\begin{equation}
\Omega_0  = \Omega_{dm\,0} + \Omega_{\Lambda\,0} \simeq 1
\,\Leftrightarrow\, \Omega_k \simeq 0.  \label{flatness}
\end{equation}

This theory led to a major bonus: a proposal that initial tiny
quantum fluctuations were expanded to such a large scale by
inflation that they provided seeds initiating growth by
gravitational attraction of large scale structures such as clusters
of galaxies. This theory makes clear observational predictions for
the spectrum of CBR anisotropies, which have since been
spectacularly verified by observations from balloons and satellites,
such as WMAP \cite{spe03}. Thus inflation has provided us with our
first coherent theory of structure formation. Inhomogeneities
started as quantum fluctuations in the inflationary epoch which are
then amplified in physical scale by the inflationary expansion but
remain constant in amplitude when larger than the contemporary
Hubble scale, leading to Gaussian scale-free perturbations at the
start of the Hot Big Bang era. Starting from these fluctuations,
Cold Dark Matter\ (`CDM') creates potential wells for baryons to
fall into, but the radiation (tightly coupled to the electrons and
baryons) resists collapse. Gravity wins if the wavelength $\lambda$
is greater than the \emph{Jean's length} $\lambda_J$ (which is
proportional to the speed of sound \cite{ree95,ellels99}). There are
acoustic oscillations (sound waves) when $\lambda < \lambda_J$;
these oscillations ceased at decoupling, which led to a dramatic
decrease in $\lambda_J$ and the growth of structure by gravitational
instability in a `bottom up' way (Sec.\ref{sec:cdm}).\footnote{This
is a highly simplified account; for more detailed versions, see e.g.
\cite{dod03,sil05}.}

A popular version of inflation is \textit{chaotic inflation}
\cite{lin90,gut01,sus05} where inflation ends at different times in
different places, so that one ends up with numerous `pocket
universe' (expanding universe domains like the one we see around us,
or perhaps very different) all imbedded in a still-inflating
universe region and starting at different times, the whole forming a
fractal-like structure. It is argued this is an inevitable
consequence of the nature of plausible scalar field potentials.

Inflation is not an inevitable conclusion, for there are some
alternatives proposed \cite{holwal02,khoetal01}, and the WMAP
results can be reproduced by any scenario where Gaussian scale-free
perturbations of suitable amplitude occur at the start of the Hot
Big Bang era. However inflation is regarded by most cosmologists as
the best proposal available for the era prior to the Hot Big Bang
epoch, leading to the presence of such perturbations. Nevertheless
one should note it is a generic proposal for what happened, rather
than a specific physical theory. While a great many possibilities
have been proposed (it could for example be an effective field due
to higher-order gravity effects, or it could involve multiple scalar
fields), at the present time the identity of the proposed
inflationary field (`the inflaton') has not been established or
linked to any known particle or field. The hoped-for link between
early universe dynamics and particle physics is potential rather
than real \cite{earmos99}. Detailed studies of the CBR\ anisotropies
and structure formation in conjunction with the observations hope to
distinguish between the various possibilities, for example testing
whether the spectral index $n$ takes the scale-free value: $ n=1,$
or whether rather there is a tilted power spectrum ($n\neq 1)$. A
unique spectrum of gravitational waves will also be produced at very
early times in an inflationary universe, and detection of these
waves either directly by proposed gravitational wave detectors or
indirectly by measuring the associated curl mode in the CBR
polarization will be an important test of inflation, for example
determining the ratio $r$ of scalar to tensor perturbations in the
early universe \cite{dod03}.

\subsection{The very early universe}
Quantum gravity processes are presumed to have dominated the very
earliest times, preceding inflation. There are many theories of
the quantum origin of the universe, but none has attained
dominance. The problem is that we do not have a good theory of
quantum gravity \cite{rov05}, so all these attempts are
essentially different proposals for extrapolating known physics
into the unknown. A key issue is whether quantum effects can
remove the initial singularity and make possible universes without
a beginning. Preliminary results suggest this may be so
\cite{boj01,rov04,mul05}.

\subsubsection{Is there a quantum gravity epoch?}
\label{sec:avoid} A preliminary issue is, can there be a
non-singular start to the inflationary era, thus avoiding the need
to contemplate a preceding quantum gravity epoch? In the
inflationary epoch the existence of an effective scalar field
leads to a violation of the strong energy condition (\ref{eq:ec}),
therefore at first sight it seems that a bounce may be possible
preceding the start of the expanding inflationary era and avoiding
the inevitability of a quantum gravity epoch.

However a series of theorems suggest that inflationary models cannot
bounce: they are stated to be future infinite but not past infinite
\cite{gut01}. This is an important issue, so it is worth looking at
it further. There are two major requirements to get a bounce. The
Friedmann equation (\ref{Fried}) relates the scale factor $S(t),$
curvature constant $k$, and the effective total energy density
$\mu(t)$, which is \textit{defined} by this equation whatever
dynamics may be involved (multiple scalar fields, higher order
gravity, higher dimensional theories leading to effective
4-dimensional theories, etc.).\footnote{See \cite{coetal05} for the
ways various quantum gravity theories result in modified Friedmann
equations.} The Raychaudhuri equation (\ref{Ray}) includes the
effective total pressure $p(t)$, which again is \textit{defined} by
this equation. In this section, a cosmological constant $\Lambda$ is
represented as perfect fluid with $\mu_{\Lambda}+p_{ \Lambda}/c^2 =
0$. To get a bounce, first one needs the curve $S(t)$ of the scale
factor as a function of time to bend up: that is,
\begin{equation}
\frac{\ddot{S}}{S}\geq 0\Leftrightarrow \mu+3p/c^2<0,
\end{equation}
which is just a violation of the strong energy condition
(\ref{eq:ec}). This is the case if $\mu +p/c^2=0$ (a vacuum); and
indeed by eqn.(\ref{scalar1}) it is possible for example for any
slow-rolling scalar field. Second, one also needs a time when the
scale factor is a minimum. Thus there must be a time $t_{\ast}$
such that $\dot{S}(t_{\ast}) = 0$. From the Friedmann equation
(\ref{Fried}),
\begin{equation}
\dot{S}^{2}(t_{\ast}) = 0 \Leftrightarrow \frac{\kappa \mu(t_{\ast})}{3 } =
\frac{k}{S^{2}(t_{\ast})}.
\end{equation}
With $k\leq 0$ this is possible only if $\mu(t_{\ast}) < 0.$ Even with a
scalar field (see eqn.(\ref{scalar})) this can only be achieved by having
negative potential energies, which appears to be an unphysical requirement.
With $k=+1$ this is possible with $\mu(t_{\ast}) > 0$ \cite{rob33}, which is
compatible with ordinary matter.

Thus if you want a bounce in an inflationary universe, it is
sensible to look to $k=+1$ inflationary models, which indeed will
turn around if a vacuum domain occurs for long enough (curvature
will eventually always win over a vacuum as we go back into the past
\cite{infplus1,infplus2}). The theorems mentioned above do
\emph{not} include this case (see \cite{gut01}); they only consider
inflationary universes with $k=0$ and $k=-1$. And one should note
here that although the scale-free $ k=0$ exponential case clearly is
the model underlying the way many people approach the problem, it is
highly exceptional --- it is of zero measure within the space of all
inflationary FL models.

Explicit non-singular models can be constructed, the simplest
being the de Sitter universe in the $k=+1$ frame
(Sec.\ref{sec:basic}), which is an exact eternal solution that
bounces at a minimum radius $S_{0}$. This model has the problem
that it does not exit inflation (it corresponds to an exactly
constant potential), but variants exist where exit is possible;
there are also viable non-singular models that start off
asymptotic to the Einstein Static universe in the distant past and
avoid the need for a quantum gravity epoch \cite{ellmaa04}. These
models start off in a very special state, precisely because they
are asymptotic to the Einstein static universe in the distant
past. This is a possible situation. It seems likely that the
options for the start of inflation are (i) avoiding the quantum
gravity era, but at the cost of having special (`fine tuned')
initial conditions, or (ii) having a quantum gravity epoch
preceding the inflationary era. Thus a key issue is whether the
start of the universe was very special or generic.

\subsubsection{Quantum gravity effects: The origin of the universe}
\label{sec:start} Contemporary efforts to explain the beginning of
the universe, and the particular initial conditions that have
shaped its evolution, usually adopt some approach or other to
applying quantum theory to the creation of the universe
\cite{lem31}. Many innovative attempts have been made here; as
this article focuses on General Relativity and its application to
cosmology, and it would be impossible to do justice to the various
approaches to quantum cosmology \cite{rov05} without a very much
longer article. I will just make a few comments on these
approaches.

The attempt to develop a fully adequate quantum gravity approach to
cosmology is of course hampered by the lack of a fully adequate
theory of quantum gravity, as well as by the problems at the
foundation of quantum theory (the measurement problem, collapse of
the wave function, etc. --- see \cite{ish97,dic05,lan05}) which can
be ignored in many laboratory situations but have to be faced in the
cosmological context \cite{peretal05}. The various attempts at
quantum cosmology each develop in depth some specific aspect of
quantum theory that may be expected to emerge from a successful
theory of quantum gravity applied to the universe as a whole, being
for example based on either (i) the Wheeler-deWitt equation and the
idea of the wave function of the universe, or (ii) on some version
of embedding in higher dimensional space time (inspired by string
theory), or (iii) an appropriate application of
loop quantum gravity. In effect they attempt either 

\textbf{(a)} to give a true theory of creation \textit{ex nihilo}
\cite{vil82}; such efforts however cannot truly ``solve'' the issue
of creation, for they rely on some structures or other (e.g. the
elaborate framework of quantum field theory and much of the standard
model of particle physics) pre-existing the origin of the universe,
and hence themselves requiring explanation; or 

\textbf{(b)} to describe a self-sustaining or self-referential
universe which by-passes the issue of creation, either by

\noindent \textbf{(b1)} originating from an eternally pre-existing
state, via the recurring idea of a Phoenix universe \cite{dicpee79}
(as in Veneziano's `pre-big bang theory' based on analogues of the
dualities of string theory, or self-repeating universes such as the
chaotic inflationary models of Linde); creation from fluctuations in
some quite different pre-existing structure (e.g. emergence from de
Sitter space time; or the `ekpyrotic universe' initiated by a
collision between pre-existing `branes' in a higher dimensional
spacetime); or emerging from an eternal static initial state; or

\noindent \textbf{(b2)} starting from a state with different
properties of time than usual (or with an an emergent notion of
time): as in the Hartle--Hawking no-boundary proposal
\cite{haw87,haw93}, and the Gott causal violation proposal
\cite{gotli97} where the universe `creates itself' and starts normal
expansion in the domain without closed timelike lines.

\noindent Any of these may be combined with one or other proposals
for

\textbf{(c)} an effective ensemble of universes \cite{teg03a},
realized either

\noindent \textbf{(c1)} in space-time regions that are part of
either a larger entangled quantum entity, or are part of a single
classical space-time, but are effectively disconnected from each
other, or

\noindent \textbf{(c2)} in truly disconnected form.

\noindent All of these proposals however are strongly speculative, none
being based solidly in well-founded and tested physics, and none being in
any serious sense supported by observational evidence. They are all vast
extrapolations from the known to the unknown. They may or may not be true.
One thing is certain: they can't all be true!

\subsection{The concordance model}
\label{sec:concord} Observational support for the idea of expansion
from a Hot Big Bang epoch is very strong, the linear
magnitude-redshift relation for galaxies demonstrating the
expansion,\footnote{The alternative interpretation as gravitational
redshifts in a static universe does not work because of the
linearity of the observed redshift-distance relation
\cite{elletal78}.} with source number counts and the existence of
the blackbody CBR being strong evidence that there was indeed
evolution from a hot early stage. Agreement between measured light
element abundances and the theory of nucleosynthesis in the early
universe confirms this interpretation. This basic theory is robust
to critical probing. Much present activity attempts to link particle
physics interactions during very early stages of the expansion of
the universe to the creation of structures by gravitational
instability much later, traces of the early seed fluctuations being
accessible to us through present day CBR anisotropy patterns. Thus
the present dominant cosmological paradigm is \emph{a quantum
gravity era of some kind followed by inflation; a hot big bang
epoch; decoupling of matter and radiation; and then gravitational
instability leading to formation of clusters of galaxies from the
seed density perturbations that occur on the LSS}.

Together with supernova data, analysis of the CBR angular
anisotropies and in particular their peaks gives a \emph{concordance
model} of this kind \cite{benetal03,teg03,teg04,dod03,sco05} that is
then confirmed by the statistics of matter clustering \cite{eis05}
together with observations of gravitational lensing and large-scale
motions of matter \cite{sil05}. This model is characterized by
specific values for a set of cosmological parameters \cite{lid04},
in particular
\begin{equation}
\Omega_{cdm 0} \simeq 0.3,\, \Omega_{\Lambda 0}\simeq 0.7,\,
T_{cbr0} = 2.75K, \,H_{0}\simeq 65\mathrm{km/sec/mpc},\,\,
t_{0}\simeq 1.4\times 10^{10} \mathrm{years}.  \label{concordance}
\end{equation}
Also $\Omega_{bar 0} \simeq 0.044$ is the density of baryons,
$\Omega_{vis 0} \simeq 0.015$ that of luminous matter, and
$\Omega_{\nu 0} \simeq 10^{-5}$ that of massless neutrinos, implying
$\Omega_{0}\simeq  0.3 + 0.7 \simeq 1$ in agreement with the
inflationary prediction (\ref{flatness}). The sign of $k$ is
uncertain, but if the combined evidence of all current observations
is taken at face value it is positive, with $\Omega_{0} = 1.02\pm
0.02$ \cite{spe03}. As noted above, there are some concerns firstly
over age issues (see Sec.\ref{sec:ages}); secondly concerning the
large angle CBR anisotropies (see Sec.\ref{sec:cbr}); and thirdly
regarding details of CDM\ structure formation at small scales (see
Sec.\ref{sec:cdm}); but none of these issues seems to be crucial at
present.

\subsubsection{Some misunderstandings}
Despite its simplicity, there are some common misconceptions about
the standard universe models (cf. \cite{lindav05}) that can lead to
philosophical misunderstandings.

\textbf{Misconception 1}: \textit{The universe is expanding into
something}. It is not, as it is all there is. It is just getting
bigger, while always remaining all that is. One should note here
that a RW universe can be represented as a 4-dimensional curved
spacetime expanding in a 5-dimensional flat embedding space time
\cite{rob33}; however there is no necessity to view the
5-dimensional spacetime in this representation as physically real.
Furthermore this embedding is no longer possible when we take
perturbations into account; a 10 dimensional flat spacetime is
needed for locally embedding a realistic (perturbed) universe model
(and to do so globally requires many more dimensions, in general).

\textbf{Misconception 2}: \textit{The universe expands from a specific
point, which is the centre of the expansion.} All spatial points are
equivalent in these universes, and the universe expands equally about all of
them. Every fundamental observer sees exactly the same thing in an exact RW
geometry. There is no centre to a FL universe.

\textbf{Misconception 3}: \textit{Matter cannot recede from us
faster than light}. It can, at an instant; two distantly separated
fundamental observers in a surface $\{t=const\}$ can have a relative
velocity greater than $c$ if their spatial separation is large
enough \cite{rotell93,davlin04}. No violation of special relativity
is implied, as this is not a local velocity difference, and no
information is transferred between distant galaxies moving apart at
these speeds. For example, there is presently a sphere around us\ of
matter receding from us at the speed of light;\footnote{This sphere
is not the same as the particle horizon, as is sometimes claimed
(see \cite{rotell93}).} matter beyond this sphere is moving away
from us at a speed greater than the speed of light. The matter that
emitted the CBR was moving away from us at a speed of about $61c$
when it did so \cite{rotell93}.

\textbf{Misconception 4}: \textit{The existence of a preferred FR
frame (that in which the universe appears isotropic) contradicts
relativity theory, which says all reference frames are equally
good}. But this equivalence of frames is true for the equations
rather than their solutions. Almost all particular solutions will
have preferred world lines and surfaces; this is just a particular
example of a \textit{broken symmetry} --- the occurrence of
solutions of equations with less symmetries than the equations
display. This feature is a key theme in modern physics
\cite{bracas05,har05}.

\textbf{Misconception 5}: \textit{The space sections are necessarily
infinite if }$\mathit{k=0}$\textit{\ or $-1$}. This is only true if they
have their `natural' simply connected topology. If their topology is more
complex (e.g. a 3-torus) they can be spatially finite \cite{ell71,lalu94}.
There are many ways this can happen; indeed if $k=-1$ there is an infinite
number of possibilities.

\textbf{Misconception 6}: \textit{Inflation implies} \textit{spatial
flatness } ($k=0\Leftrightarrow \Omega _{k}=1$) \textit{exactly}.
There is nothing in inflationary theory which determines the sign of
the spatial curvature. Inflationary universes are very nearly flat
at late times; this is very different from being exactly flat (a
condition which requires \textit{infinite} fine tuning of initial
conditions; if say the two millionth digit in the value of $\Omega
_{k}$ is non-zero at any time, then the universe is not spatially
flat). Inflationary theory does not have the theoretical teeth
required to imply that the universe has exactly flat spatial
sections; hence a key issue for cosmology is observationally
determining the sign of the spatial curvature, which is potentially
dynamically important in both the very early universe
\cite{infplus1,infplus2} and the late universe (it determines if
recollapse is possible, should the dark energy decay away).

\subsubsection{Overall}
Cosmology has changed from a speculative enterprize into a
data-driven science that is part of standard physical theory
\cite{baretal96}; a wealth of observations supports this dominant
theory \cite{peetal,sil97,per05}. Nevertheless some theoretical
proposals are being made for the very early stages that have no
observational support; and sometimes it may be impossible to ever
obtain such support, both as regards the proposed physics and the
geometry. Thus in some respects it remains a principle driven
enterprise, with observation subordinate to theory.

We now explore the relation between cosmology and philosophy in terms of a
series of \emph{Theses} clustered around a set of major \emph{Issues}. One
can obtain a synoptic overview of the overall argument by simply considering
the full set of \emph{Issues} and \emph{Theses}. They are summarized in the
Table at the end.

\section{Issue A: The uniqueness of the universe.}
\label{sec:unique} The first and most fundamental issue is that
there is only one Universe \cite{mun86,mac60,ell91}. This
essential uniqueness of its object of study sets cosmology apart
from all other sciences. In particular, the unique initial
conditions that led to the particular state of the universe we see
were somehow ``set" by the time that physical laws as we know them
started governing the evolution of both the universe and its
contents, whenever that time may be. We cannot alter these unique
initial conditions in any way --- they are given to us as absolute
and unchangeable, even though they are understood as contingent
rather than necessary; that is, they could have been different
while still being consistent with all known physical laws. The
implications are that

\textbf{Thesis A1: The universe itself cannot be subjected to physical
experimentation}. \textit{We cannot re-run the universe with the same or
altered conditions to see what would happen if they were different, so we
cannot carry out scientific experiments on the universe itself}. Furthermore,

\textbf{Thesis A2: The universe cannot be observationally compared
with other universes}. \textit{We cannot compare the universe with
any similar object, nor can we test our hypotheses about it by
observations determining statistical properties of a known class
of physically existing universes.}

\noindent Where this all becomes of observational relevance is in
the idea of \emph{cosmic variance} (\cite{dod03}, pp. 241, 343).
The theory of structure formation in the early universe makes
statistical predictions only (it cannot attempt to predict the
specific structures that will actually be formed). Testing the
theory compares our universe to a theoretical ensemble of
universes, and declares a variance between what is measured in the
actual universe and the expected properties based on the ensemble
of models. If this variance is small enough, a deviation from
expected values is pronounced as a statistical deviation, i.e. of
no physical significance --- we do not need to explain it any
further; if it is large, it needs explanation. This is a key issue
for example in the analysis of the CBR anisotropy observations
\cite{whietal93,kamloe97}. The power spectrum of the CBR as
measured by WMAP is less than expected at large angular scales
(Sec.\ref{sec:cbr}). One school of thought claims this is just a
statistical fluctuation; another that it needs explanation, and
might for example be evidence of a small universe
\cite{lumetal03}. This debate arises because there is just one
universe, and on large angular scales there are just a few
measurements that can possibly be made (on small angular scales we
can make many measurements and so this uncertainty becomes very
small).

Consequent on \textbf{A1} and \textbf{A2},

\textbf{Thesis A3: The concept of `Laws of Physics' that apply to only one
object is questionable}. \textit{We cannot scientifically establish `laws of
the universe' that might apply to the class of all such objects, for we
cannot test any such proposed law except in terms of being consistent with
one object (the observed universe).}

\noindent This is insufficient: one observational point cannot establish the
nature of a causal relation. Indeed the concept of a `law' becomes doubtful
when there is only one given object to which it applies \cite{mun86}. The
basic idea of a physical law is that it applies to a set of objects all of
which have the same invariant underlying behaviour (as defined by that law),
despite the apparent variation in properties in specific instances, this
variation resulting from varying initial conditions for the systems on which
the law acts. This understanding is tested by physical experiments in which
initial conditions for evolution of a set of similar systems are varied, and
observations by which the statistical nature of a set of objects of the same
broad kind is investigated. Neither is possible in the case of cosmology.

The laws of physics apply locally to the objects in the cosmos,
and determine the evolution of the cosmos as a whole when locally
applied everywhere with suitable initial/boundary conditions
imposed (in the case of the FL models, via the Friedmann equation
for example). Apart from this, we cannot establish higher-level
effective laws that apply to all universes and determine their
structure, as we can at all other levels of the hierarchy of
complexity. All that we can do at this level of structure is
observe and analyze the one unique object that exists. This is
expressed by McCrea as follows: ``When we speak of the other
solutions of the equations of stellar structure, besides the one
we are interested in at the moment, as representing systems that
could exist, we mean that they could exist in the universe as we
know it. Clearly no such attitude is possible towards the universe
itself" \cite{mac53}.

Since the restriction of a global solution to a local neighborhood
is also a solution, we have zillions of ``mini-universe" on which to
test the laws that control the local nature of the universe. But a
mini-universe is not the universe itself; it is a small part of the
whole. By examining these ``mini-universes" and seeing if they are
essentially the same everywhere, we can to some degree check firstly
\emph{that the laws of physics are the same everywhere in the
universe} (a key feature of all cosmological analysis, cf.
Sec.\ref{sec:laws}), and secondly \emph{that the universe is
spatially homogeneous} (this is discussed in depth below, see
Sec.\ref{sec:homog}). But the latter feature is what has to be
\emph{explained} by a `law of the universe'; verifying homogeneity
does not explain why it is the case; this comes about because of
specific initial conditions, which some suggest are due to
hypothesized `laws of the universe', applicable to the whole rather
than to its parts. Finally,

\textbf{Thesis A4: The concept of probability is problematic in
the context of existence of only one object}. \textit{Problems
arise in applying the idea of probability to cosmology as a whole
--- it is not clear that this makes much sense in the context of
the existence of a single object which cannot be compared with any
other existing object.}

\noindent But a concept of probability underlies much of modern
argumentation in cosmology. Talk of `fine tuning' for example is
based on the use of probability (it is a way of saying something is
improbable). This assumes both that things could have been
different, and that we can assign probabilities to the set of
unrealized possibilities in an invariant way. The issue here is to
explain in what sense they could have been different with
well-defined probabilities assigned to the different theoretical
possibilities, if there is indeed only one universe with one set of
initial conditions fixed somehow before physics came into being, or
perhaps as physics came into being. We cannot scientifically
establish laws of creation of the universe that might determine such
initial conditions and resulting probabilities. If we use a Bayesian
interpretation, which some suggest can be meaningfully applied to
only one object \cite{garcol93}, the results depend on our `prior
knowledge', which in this case can be varied by changing our initial
pre-physics assumptions. Related issues arise concerning the meaning
of `the wave function of the universe', at the heart of quantum
cosmology. This wave function gives no unique prediction for any
specific single universe.

Two comments on the above. First, \emph{it is useful to
distinguish between the experimental sciences --- physics,
chemistry, microbiology for example --- on the one hand, and the
historical and geographical sciences --- astronomy, geology,
evolutionary theory for example, on the other}. It is the former
that are usually in mind in discussions of the scientific method.
The understanding in these cases is that we observe and experiment
on a class of identical or almost identical objects and establish
their common behaviour. The problem then resides in just how
identical those objects are. Quarks, protons, electrons, are all
exactly identical to each other, and so have exactly the same
behaviour (indeed this feature underlies well-tested quantum
statistics). All DNA molecules, frogs, human beings, and
ecosystems are somewhat different from each other, but are similar
enough nevertheless that the same broad descriptions and laws
apply to them; if this were not so, then we would be wrong in
claiming they belonged to the same class of objects in the first
place. Water molecules, gases, solids, liquids are in an
intermediate category --- almost identical, certainly describable
reliably by specific physical and chemical laws.

As regards the geographical and historical sciences, here one
explicitly studies objects that are unique (the Rio Grande, the
continent of Antarctica, the Solar System, the Andromeda galaxy,
etc.) or events that have occurred only once (the origin of the
Solar System, the evolution of life on Earth, the explosion of
SN1987a, etc.). Because of this uniqueness, comment \textbf{A1}
above applies in these cases also: we can only observe rather than
experiment; the initial conditions that led to these unique objects
or events cannot be altered or experimented with. However comment
\textbf{A2} does not apply: at least in principle, there is a class
of similar objects out there (other rivers, continents, planetary
systems, galaxies, etc.) or similar events (the origin of other
galaxies, the evolution of other planetary systems, the explosion of
other supernovae, etc.) which we can observe and compare with our
specific exemplar, also carrying out statistical analyses on many
such cases to determine underlying patterns of regularity; and in
this respect these topics differ from cosmology.

If we truly cannot carry out such analyses --- that is, if
\textbf{A2} applies as well in some particular case --- then that
subject partakes in this respect of the nature of cosmology. One may
claim that \emph{the dividing line here is that if we convince
ourselves that some large-scale physical phenomenon essentially
occurs only once in the entire universe, then it should be regarded
as part of cosmology proper}; whereas if we are convinced it occurs
in many places or times, even if we cannot observationally access
them (e.g. we believe that planets evolved around many stars in
other galaxies), then study of that class of objects or events can
be distinguished from cosmology proper precisely because there is a
class of them to study.

The second comment is that some workers have tried to get around
this set of problems by essentially \emph{denying the uniqueness
of the universe}. This is done by proposing the physical existence
of `many universes' to which concepts of probability can be
properly applied (cf. Sec.\ref{sec:start}), envisaged either as
widely separated regions of a larger universe with very different
properties in each region (as in chaotic inflation for example),
as multiple realizations of quantum outcomes, or as an ensemble of
completely disconnected universes --- there is no physical
connection whatever between them --- in which all possibilities
are realized. We return to this in Sec.\ref{sec:multiverse}.

\section{Issue B: The large scale of the universe in space and time.}
\label{sec:scale} The problems arising from the uniqueness of the
universe are compounded by its vast scale in both space and time,
which poses major problems for observational cosmology. We therefore
need to adduce various Principles in addition to the observations,
in order to attain unique models: theory comes in as basis for
interpreting observations.

\subsection{Observations in a large scale universe}
The distance to the nearest galaxy is about $10^{6}$ light years,
that is about $10^{24}$cm., while the size of the earth is about
$10^{9}$cm. The present size of the visible universe is about
$10^{10}$ light years, that is about $10^{28}$cm. This huge size
relative to our own physical scale (about $10^{2}$cm) places major
constraints on our ability to observe distant regions (and certainly
prevents us experimenting with them). The uniqueness of cosmology in
this respect is that it deals with this scale: the largest with
which we can have causal or observational contact.

\textbf{Thesis B1: Astronomical observations are confined to the past null
cone, fading with distance}. \textit{We can effectively only observe the
universe, considered on a cosmological scale, from one space-time event}.
\textit{Visual observations are possible only on our past light cone, so we
are inevitably looking back into the past as we observe to greater
distances. Uncertainty grows with distance and time}.

\noindent The vast scale of the universe implies we can effectively
only view it from one spacetime event (`here and now')
\cite{ell71,ell75}. If we were to move away from this spatial
position at almost the speed of light for say $10,000$ years, we
would not succeed in leaving our own galaxy, much less in reaching
another one; and if we were to start a long-term astronomical
experiment that would store data for say $20,000$ years and then
analyze it, the time at which we observe the universe would be
essentially unchanged (because its age is of the order of $10^{10}$
years: the extra time would make a negligible difference). This is
quite unlike other geographic sciences: we can travel everywhere on
earth and see what is there. The situation would be quite different
if the universe were much smaller. Given its actual scale, such that
we are now seeing galaxies whose present distance from us is about
$10^{9}$ light years, the effect is as if we were only able to
observe the earth from the top of one mountain, and had to deduce
its nature from those observations alone \cite{ell75}.

Because we can only observe by means of particles --- photons,
massless neutrinos, gravitons --- travelling to us at the speed of
light, astronomical observations of distant sources and background
radiation by telescopes operating at all wavelengths (optical,
infrared, ultraviolet, radio, X-ray) are constrained to rays lying
in our past light cone. These allow detailed observations
(including visual pictures, spectral information, and polarization
measurements) of matter as it intersects our past light cone. In
observing distant regions, we can also aspire to use neutrino and
gravitational wave telescopes, and perhaps cosmic rays, also
representing information coming to us at the speed of light or
less. However all our detailed data about distant regions is
gathered along our past light cone.

As a consequence, three interrelated problems occur in
interpreting the astronomical observations. The first is that
(because we can only view the universe from one point) \emph{we
only obtain a 2-dimensional projection on the sky of the
3-dimensional distribution of matter in the universe}. To
reconstruct the real distribution, we need reliable distance
measurements to the objects we see. However because of variation
in the properties of sources, most are not reliable standard
candles or standard size objects to use in calibrating distances,
and in these cases we have to study statistical properties of
classes of sources to estimate distances.

Second, \emph{we necessarily see distant galaxies and other
objects at earlier times in their history} (where their world
lines intersect this past light cone).\footnote{For example we see
the Andromeda galaxy as it was two million years ago, long before
humans existed on Earth \cite{sil05}.} Thus cosmology is both a
geographic and a historical science combined into one: we see
distant sources at an earlier epoch, when their properties may
have been different. As we are looking back in the past, source
evolution must be taken into account; their properties at the time
they emitted the light may be quite different from their
properties now. We can only determine the distances of objects if
we understand this evolution; but in practice it is one of the
unknowns we have to try to determine (cf. Sec.\ref{sec:param}).

Third, distant sources appear very small and very faint, both
because of their physical distance, and because their light is
highly redshifted (due to the expansion of the universe). Simply
detecting them, let alone determining their characteristics,
becomes rapidly more difficult with distance. Furthermore
absorption by intervening matter can interfere with light from
distant objects. The further back we look, the worse these
problems become; thus our reliable knowledge of the universe
decreases rapidly with distance \cite{ell75}.

The situation is however improved by the availability of
geological-type data \cite{hoy60}; that is, the present-day status
of rocks, planets, star clusters, galaxies, and so on, which
contains much information on the past history of the matter
comprising those objects. Thus we can obtain detailed information on
conditions near our past world-line in spacetime \cite{ell71,ell75}
at very early times if we can interpret this data reliably, for
example by relating theories of structure formation to statistical
studies of source properties.

\textbf{Thesis B2: `Geological' type observations can probe the
region near our past world line in the very distant past}.
\textit{Physical and astrophysical observations tell us about
conditions near matter world-lines in the far distant past. They
can be used also to investigate the far distant past of more
distant objects.}

\noindent This involves us in physical cosmology: namely the study
of the evolution of structures in the universe, tested by
comparison with astronomical observation. Particularly useful are
measurements of the abundances of elements which resulted from
nucleosynthesis in the Hot Big Bang, giving us data about
conditions long before decoupling (Sec.\ref{sec:nucleo}). If we
can obtain adequate quality data of this kind for objects at high
redshifts, we can use this to probe conditions very early on in
their histories at some distance from our past worldline.
Encouraging in this regard is the possibility of determination of
element abundances at high redshift (\cite{dod03}, pp. 11-12;
\cite{petetal05}).

\subsection{Determining Spacetime Geometry: Observational Limits.}
The unique core business of observational cosmology is determining the
large-scale geometry of everything there is, or at least of everything we
can observe.

\subsubsection{Direct determination versus theory based approaches}
One can go about this in a direct manner: trying to determine the
geometry of the universe directly from observations (assuming one
has some understanding of the sources observed). The way this can
be done (curiously known as the `inverse approach') has been fully
characterized \cite{krisac66,elletal85}; indeed there is an
interesting result here, namely

\textbf{Observational Cosmology Theorem}:\ \textit{The data in
principle available on our past null cone from astronomical
observations is just necessary and sufficient to determine the
space-time geometry on that null cone} \cite{elletal85}.
\textit{From this data one can in principle determine the space time
in the past of the null cone and, if a no-interference conditions is
assumed, to its future.}

\noindent However this is difficult to carry out both because of the
problem of estimating distances for all observed sources, requiring
a knowledge of the nature of the sources
(Sec.\ref{sec:param}),\footnote{The link between observations and
models always requires some theory, and is never direct.} and
because of the serious difficulty in obtaining some of the needed
data (which include apparent distortions of all distant objects, and
the transverse velocities of all observed matter). The further we
observe down the past light cone, the larger the uncertainty
becomes. This direct observational approach, where no prior model is
assumed for the space-time geometry, has been pursued to some degree
(and in essence underlies for example the observational studies that
discovered large-scale structure such as the great walls and voids).
Nevertheless it is not widely adopted as an overall approach to
cosmology, both because of these observational difficulties, but
also because it has little explanatory value; it just tells us what
the geometry and matter distribution is, but not why it is of that
nature.

The usual option in cosmology proper is rather to use a theory-based
approach: we \emph{a priori} assume a model based on a space-time
geometry with high symmetry (usually a FL model, see
Sec.\ref{sec:FL}), and then determine its essential free parameters
from comparison of theoretical relations with astronomical
observations (Sec.\ref{sec:obsrels}). Detailed observations of the
matter distribution and large-scale velocities as well as CBR
anisotropies then help us determine deviations from the exact model,
both statistically (an astrophysical description \cite{dod03}) and
in detail (an astronomical description \cite{ellsto87}).

\subsubsection{Indirect determination: justifying a Friedmann-Lema\^{\i}tre
geometry} \label{sec:homog} The standard models of cosmology are the
Friedmann-Lema\^{\i}tre (FL) family of universe models that are
exactly spatially homogeneous and isotropic everywhere
(Sec.\ref{sec:FL}). They are easy to understand, and have tremendous
explanatory power; furthermore their major physical predictions (the
existence of blackbody CBR and specific light element production in
the early universe) seem confirmed. The issue is, to what degree
does observational data uniquely indicate these universe models for
the expanding universe geometry?\ Here one is assuming a large
enough averaging scale for spatial homogeneity to be valid; this
scale should be explicitly indicated \cite{ell84} (it is about 100
Mpc at present \cite{dod03}).\footnote{There exist
\emph{hierarchical models} where neither the fluid approximation nor
homogeneity is ever attained at any scale because of their fractal
nature \cite{dev70}. The regularity of the observed galactic
motions, as evidenced by the $(m,z)$ relations, speaks against these
models, as do large-scale observations of the matter distribution
\cite{pee80}.} These are the background models for cosmology;
perturbed FL models then characterize the nature of deviations from
the exact FL geometry that are expected on smaller scales
(Sec.\ref{sec:perturb}).

The key feature here is the observed \emph{isotropy about our
location} (Sec.\ref{sec:isotropy}). Considered on a large enough
angular scale, astronomical observations are very nearly isotropic
about us, both as regards source observations and background
radiation; indeed the latter is spectacularly isotropic, better than
one part in $10^{4}$ after a dipole anisotropy, understood as
resulting from our motion relative to the rest frame of the
universe, has been removed \cite{par95}. Because this applies to all
observations (in particular, there are not major observed matter
concentrations in some other universe region), this establishes that
in the observable region of the universe, to high accuracy
\emph{both the space-time structure and the matter distribution are
isotropic about us}. We can easily construct exact spherically
symmetric universe models \cite{bon47,ellels99}, as indicated by
these observations. In general they will be spatially inhomogeneous,
with our Galaxy located at or near the centre; this is currently a
philosophically unpopular proposal, but is certainly possible. The
question is whether we can give convincing observational evidence
for spatial homogeneity in addition to the spherical symmetry.
Various arguments are used for this purpose.

(a) \emph{The cosmological principle} \cite{bon60,wei72}: Just
assume spatial homogeneity because it is the simplest case and you
don't need anything more complex on the basis of current data. We
simply adopt a philosophical principle as the basis of argument.
This is essentially an \emph{a priori prescription for initial
conditions for the universe} (a universe that initially has a RW
geometry will have that geometry at later times, because symmetries
of the initial data are preserved by the Einstein equations
\cite{hawell73}); but it is not usually expressed that way.

(b) \emph{FL observational relations}: If we could show that the
source observational relations had the unique FL form (\ref{mattig},
\ref{number}) as a function of distance, this would establish
spatial homogeneity in addition to the isotropy, and hence a RW
geometry \cite{elletal85}. This is essentially what is done for
example in using number counts towards establishing spatial
homogeneity \cite{hub36}. However because of Thesis \textbf{B1}
above, the observational problems mentioned earlier
--- specifically, unknown source evolution --- prevent us from
carrying this through: we cannot measure distances reliably
enough. Astrophysical cosmology could resolve this in principle,
but is unable to do so in practice. Indeed the actual situation is
the inverse: \emph{taking radio-source number-count data at its
face value, without allowing for source evolution, contradicts a
RW geometry}.

In the face of this, the usual procedure is to assume that spatial
homogeneity is known in some other way, and deduce the source
evolution required to make the observations compatible with this
geometric assumption \cite{ell75}. It is always possible to find a
source evolution that will achieve this \cite{musetal98}. Thus
attempts to observationally prove spatial homogeneity this way fail;
indeed an alternative interpretation would be that this data is
evidence of spatial inhomogeneity, i.e. that we live in a
spherically symmetric inhomogeneous universe where we are situated
somewhere near the centre, with the cosmological redshift being
partly gravitational, cf. \cite{elletal78} (and conceivably with a
contribution to the CBR dipole from this inhomogeneity if we are a
bit off-centre). Similarly the supernova data usually understood as
implying the existence of a cosmological constant
(Sec.\ref{sec:super}) could also be interpreted in this way as
evidence of inhomogeneity, without the need for `dark energy'. Most
people regard such proposals as very unappealing --- but that does
not prove they are incorrect.

(c) \emph{Physical arguments}: One can claim that physical processes
such as inflation (Sec.\ref{sec:inflate}) make the existence of
almost-RW regions highly likely, indeed much more probable than
spherically symmetric inhomogeneous regions. This is a viable
argument, but we must be clear what is happening here --- we are
replacing an observational test by a theoretical argument based on a
physical process that may or may not have happened (for there is no
definitive observational proof that inflation indeed took place). It
is strongly bolstered because predictions for the detailed pattern
of CBR anisotropy on small scales \cite{husug95}, based on the
inflationary universe theory, have been confirmed \cite{per05}; but
that argument will only become rigorous if it is shown that
spherically symmetric inhomogeneous models (with or without
inflation) cannot produce similar patterns of anisotropy. But they
probably can, because the acoustic oscillations that lead to the
characteristic predicted anisotropy patterns in fact take place
after inflation, and can equally happen if suitable initial
conditions occur without a previous inflationary phase.

What about alternative observational routes? Another proposal is,

(d) \emph{Uniform thermal histories}: the idea is to use the
uniformity in the nature of the objects we see in the sky (we see
the same types of galaxy at large distances, for example) to deduce
they must have all undergone essentially the same thermal history,
and then to prove from this homogeneity of thermal histories that
the universe must be spatially homogeneous. For example,
observations showing that element abundances at high redshift in
many directions are the same as locally, are very useful in
constraining inhomogeneity by showing that conditions in the very
early universe at the time of nucleosynthesis must have been the
same at distant locations in these directions [82]. However turning
this idea into a proper test of homogeneity has not succeeded so
far: indeed it is not clear if this can be done, because some
(rather special) counter-examples to this conjecture have been found
\cite{bonell86}. Nevertheless the approach could be used to give
evidence against spatial homogeneity: for example, if element
abundances were measured to be different at high redshifts in any
direction \cite{petetal05,sigfur05}, or if ages of distant objects
were incompatible with local age estimates \cite{jaidev05}.

Finally the argument for spatial homogenity that is most generally
accepted:

(e) \emph{Isotropy everywhere:} If all observers see an isotropic
universe, then spatial homogeneity follows \cite{wal44,ehl61,ell71};
indeed homogeneity follows if only three spatially separated
observers see isotropy. Now we cannot observe the universe from any
other point, so we cannot observationally establish that far distant
observers see an isotropic universe. Hence the standard argument is
to assume a \textit{Copernican Principle:} that we are not
privileged observers. This is plausible in that all observable
regions of the universe look alike: we see no major changes in
conditions anywhere we look. Combined with the isotropy we see about
ourselves, this implies that \textit{all observers see an isotropic
universe}, and this establishes a RW geometry
\cite{wal44,ell71,hawell73}. This result holds if we assume isotropy
of \emph{all} observations; a powerful enhancement was proved by
Ehlers, Geren, and Sachs \cite{ehletal86,hawell73}, who showed that
if one simply assumes isotropy of freely-propagating radiation about
each observer in an expanding universe domain,\footnote{This result
does not hold in a static universe, for then the radiation
temperature depends only on the potential difference between the
emitter and observer, hence the radiation is isotropic everywhere
even if the universe inhomogeneous, cf. \cite{elletal78}.} the
result follows from the Einstein and Liouville equations; that is,

\textbf{EGS Theorem:}\ \textit{Exact isotropy of the CBR for every
geodesically moving fundamental observer at each point in an
expanding universe domain U implies an exact RW geometry in U.}

\noindent Thus we may establish spatial homogeneity by assuming a
weak Copernican principle: we are not in a privileged position where
the CBR just happens to be highly isotropic by chance; hence all
comoving observers may be assumed to measure highly isotropic CBR,
and the result follows. This is currently the most persuasive
observationally-based argument we have for spatial homogeneity.

A problem is that it is an exact result, assuming exact isotropy of
the CBR; is the result stable? Indeed it is: \textit{\
almost-isotropy of freely-propagating CBR for an expanding family of
geodesically-moving fundamental observers everywhere in some region
proves the universe geometry is almost-RW in that region}
\cite{stoetal95}. Thus the result applies to the real universe ---
provided we make the Copernican assumption that all other observers,
like us, see almost isotropic CBR. And that is the best we can do at
present. Weak tests of the isotropy of the CBR at other spacetime
points come from the Sunyaev-Zel'dovich effect \cite{goo95} and from
CBR polarization measurements \cite{kamloe97}, giving broad support
to this line of argument but not enough to give good limits on
spatial inhomogeneity.

The observational situation is clear:

\textbf{Thesis B3: Establishing a Robertson-Walker geometry for the universe
relies on plausible philosophical assumptions}. \textit{The deduction of
spatial homogeneity follows not directly from astronomical data, but because
we add to the observations a philosophical principle that is plausible but
untestable.}

\noindent The purpose of the above analysis is not to seriously support the
view that the universe is spherically symmetric and inhomogeneous, as is
allowed by the observations, but rather to show clearly the nature of the
best observationally-based argument by which we can (quite reasonably)
justify the assumption of spatial homogeneity.

Accepting this argument, the further question is, \textit{in which
spacetime regions does it establish a RW-like geometry}? The CBR we
detect probes the state of the universe from the time of decoupling
of matter and radiation (at a redshift of about $1100$) to the
present day, within the visual horizon. The argument from CBR
isotropy can legitimately be applied for that epoch. However, it
does not necessarily imply isotropy of the universe at much earlier
or much later times, because there are spatially homogeneous
anisotropic perturbation modes that are unstable in both directions
of time; and they will occur in a generic situation. Indeed, if one
examines the Bianchi (spatially homogeneous but anisotropic)
universes, using the powerful tools of dynamical systems theory, one
can show that \emph{intermediate isotropisation} can occur
\cite{waiell96,waietal98}: \emph{despite being highly anisotropic at
very early and very late times, such models can mimic a RW geometry
arbitrarily closely for an arbitrarily long time}, and hence can
reproduce within the errors any set of FL-like observations. We can
obtain strong limits on the present-day strengths of these
anisotropic modes from CBR anisotropy measurements and from data on
element abundances, the latter being a powerful probe because (being
of the `geological' kind) they can test conditions at the time of
element formation, long before decoupling. But however low these
observational limits, anisotropic modes can dominate at even earlier
times as well as at late times (long after the present). If
inflation took place, this conclusion is reinforced: it washes out
any information about very early universe anisotropies and
inhomogeneities in a very efficient way.

As well as this time limitation on when we can regard homogeneity as
established, there are major spatial limitations. The above argument does
not apply far outside the visual horizon, for we have no reason to believe
the CBR is highly isotropic there. Indeed if chaotic inflation is correct,
conditions there are not the same.

\subsubsection{Determining the model parameters}
\label{sec:param} Given that a RW geometry is a good description
of the observable universe on a large scale, the further issue is
what are the best-fit parameters that characterize it, selecting
the specific universe we observe from the family of all FL models
(Sec.\ref{sec:FL}). Important observational issues are:
\begin{itemize}
\item Determining the Hubble parameter $H_0$, which sets the
overall scale of the observed universe region. \item Determining
the trio of the density parameter $\Omega_0$, deceleration
parameter $q_0$, and cosmological constant $\Lambda$ (or
equivalently the density parameter $\Omega_\Lambda$), which are
the major defining characteristics of a specific FL model. The CBR
data, supernova observations, deep number counts, source
covariance functions, velocity measurements, and gravitational
lensing observations can determine these quantities. \item
Determining the sign of the curvature $k$, showing whether the
universe has closed spatial sections and also whether it is
possible for it to recollapse in the future or not. Analyses of
the observations should always attempt to determine this sign, and
not assume that $k=0$ (as is often done) \cite{wri06}. \item
Various parameters are used to characterize the nature of dark
matter (Sec.\ref{sec:cdm}) and dark energy (Sec.\ref{sec:super}).
As their dynamics is unknown, these too have to be determined
observationally.
\end{itemize}
\noindent We only obtain good estimates of these quantities by the
observational relationships characterized above
(Sec.\ref{sec:obsrels}) using statistical analysis of the classes
of objects we observe. Problems arise because of our lack of
adequate theories of their historical development.

\textbf{Thesis B4: Interpreting cosmological observations depends on
astrophysical understanding}. \textit{Observational analysis depends on
assessing a variety of auxiliary functions characterizing the sources
observed and the observations made. These introduce further parameters that
have to be observationally or theoretically determined, allowing
considerable freedom in fitting specific models to the observations.
Physical cosmology aims to characterize perturbed FL models (which account
for structure formation) rather than just the background exactly smooth FL
models; this introduces further parameters to be determined.}

\noindent It is useful here to distinguish between methods aimed at
determining the properties of the background (zeroth order) FL model
directly, and those aimed at determining properties of the
perturbations of these models \cite{teg03}. Methods for determining
the parameters of the background model (Sec.\ref{sec:FL}) depend on
assuming properties of the distance indicators used (galaxies, radio
sources, etc.). They will have their own properties (brightness
profiles, luminosities, physical sizes, spectra, etc.) and dynamical
evolution; but these are often not well understood, and will have to
be represented in a parametric way (e.g. by parameters describing
luminosity evolution). In each case we end up assuming important
aspects of the astrophysics and evolutionary histories of the
objects observed, which are not part of the cosmological model
proper. The statistical properties of the sources observed are also
characterized by parametrized functions (e.g. the luminosity
function characterizing the numbers of galaxies in each luminosity
class) that have to be known in order to analyze the observations.
This situation is an exemple of Lakatos' view of how scientific
programmes work, with a belt of auxiliary hypotheses interposing
between the core theoretical proposal and the data used to test it
\cite{lak80}. This makes the analysis rather model-dependent, where
the models are only indirectly related to the background model ---
their explanation is the aim of astrophysics rather than cosmology.
Thus if observational results disagree with a particular
cosmological model, one can always claim it is the understanding of
the auxiliary hypotheses that is at fault rather than the model
being proposed \cite{lak80}.

By contrast, many of the methods of estimating $\Omega_{0}$ (and to
some degree $\Lambda$) depend on studying the growth and nature of
inhomogeneities in the universe, that is they investigate perturbed
FL models (Sec.\ref{sec:perturb}), whose properties of course depend
on the background model, but introduce a whole set of further
functions and parameters describing the perturbations \cite{dod03},
for example the angular correlation function for matter (or its
Fourier transform, the 2-dimensional power spectrum), the power
spectrum of density fluctuations \cite{teg03}, red-shift space
correlation functions \cite{pee80,eis05}, and correlation function
for velocities \cite{dod03}. Associated parameters include a scalar
\emph{spectral index} (characterizing the spectrum of physical sizes
of inhomogeneities), the \emph{bias parameter} $b$ (expressing how
galaxy formation is biassed towards density peaks in the
inhomogeneities (\cite{dod03}, p. 280)) and the \emph{initial
fluctuation magnitudes} $Q$ (the seeds for structure formation).
Determining these parameters is part of the task of cosmology
proper: to fully characterize the perturbed cosmological model,
\emph{we aim to determine both the background parameters and the
quantities describing the perturbations}. Model selection then
depends on the parameters used to describe them --- what is assumed
known, and what is to be determined \cite{lid04,sco05}. For example,
standard inflationary theory predicts a scale-invariant spectrum of
Gaussian perturbations; do we test that assumption, or take it for
granted? This comes up in the issue of what `priors' are assumed
when conducting statistical tests.

\subsubsection{Consistency tests}
A key question for cosmology is \emph{what kinds of observations
provide critical tests of the standard FL models}. If there were no
observations that could disprove them, the subject would be of
questionable scientific status. An important such test is obtaining
\emph{estimates of the age of the universe} $t_0$, which is
dependent on $H_{0}$, $\Omega_{0}$, and $\Lambda$, and comparing
them with estimates of the ages of objects in the universe
(determined on astrophysical grounds):

 \textbf{Thesis B5: A crucial observational test for cosmology is
 that the age of the universe must be greater than the ages of stars}.
 \textit{The tension between the age of the universe and ages of stars
 is one area where the standard models are vulnerable to being shown to
 be inconsistent, hence the vital need to establish reliable distance
scales, basic to estimates of both} $H_{0}$ \textit{and the ages
of stars, and good limits on $\Lambda $. Other consistency tests
help confirm the standard model and consolidate cosmology's
standing as an empirical science.}

\noindent At present this age issue is acceptable for local objects,
because of a recent revision of our distance scale estimates
\cite{harretal98}, assisted by data that $\Lambda $ is positive
\cite{per98}; but continued vigilance is needed on this front,
particularly as there are indications of problems for high redshift
objects \cite{jaidev05}. If this ever became serious we might have
to resort to spherically symmetric inhomogeneous models rather than
spatially homogeneous models, with the `bang time' (characterizing
the start of the universe) dependent on distance from us
\cite{musetal98}.

Note that this issue is crucially unlike the case of the large angle
CBR anisotropies (Sec.\ref{sec:cbr}): \emph{the low CBR anisotropies
at large angular scales can as a last resort be dismissed as a
statistical fluke; the age issue cannot}. It is to do with the
internal consistency of individual cosmological models, not with
probabilities. Thus it is a plus for cosmology that the age issue
exists. Other consistency tests include
\begin{itemize}
\item Showing that the CBR temperature $T_{cbr}$ varies with redshift according
to $T_{cbr}=2.75\,(1+z)$ \cite{mey94};
\item Confirming that helium abundances are consistent with a
primordial value of 25\% at large distances (high redshifts) in all
directions (\cite{dod03}, pp. 11-12; also
\cite{petetal05,sigfur05}); and
\item Checking that there is a $2\%$ number count dipole parallel to the CBR
dipole for all cosmological sources \cite{ellbal84}.
\end{itemize}

\subsection{The hidden universe}
\label{sec:horizon1} If we do not live in a small universe
(Sec.\ref{sec:small}), the further essential point is that the
region of the universe we can observe is restricted, firstly because
we cannot see to earlier times than the LSS (the universe was opaque
before then (see Sec.\ref{sec:hbb})), and secondly because a finite
time has elapsed since the universe became transparent to radiation,
and light can only have travelled a finite distance in that time. As
no signal can travel to us faster than light, we cannot receive any
information from galaxies more distant than our visual horizon
\cite{ellsto88}. The most distant matter we can observe is that
which emitted the CBR (Sec.\ref{sec:vizhorizon}).

\textbf{Thesis B6: Observational horizons limit our ability to
observationally determine the very large scale geometry of the
universe}. \textit{We can only see back to the time of decoupling of
matter and radiation, and so have no direct information about
earlier times; and unless we live in a `small universe', most of the
matter in the universe is hidden behind the visual horizon.
Conjectures as to its geometry on larger scales cannot be
observationally tested. The situation is completely different in the
small universe case: then we can see everything there is in the
universe, including our own galaxy at earlier times.}

\noindent The key point here is that unless we live in a small
universe, \emph{the universe itself is much bigger than the
observable universe}. There are many galaxies --- perhaps an
infinite number --- at a greater distance than the horizon, that we
cannot observe by any electromagnetic radiation. Furthermore no
causal influence can reach us from matter more distant than our
particle horizon --- the distance light can have travelled since the
creation of the universe, so this is the furthest matter with which
we can have had any causal connection
\cite{rin56,hawell73,tipetal80}. We can hope to obtain information
on matter lying between the visual horizon and the particle horizon
by neutrino or gravitational radiation observatories; but we can
obtain no reliable information whatever about what lies beyond the
particle horizon. We can in principle feel the gravitational effect
of matter beyond the horizon because of the force it exerts (for
example, matter beyond the horizon may influence velocities of
matter within the horizon, even though we cannot see it). This is
possible because of the constraint equations of general relativity
theory, which are in effect instantaneous equations valid on
spacelike surfaces.\footnote{They are valid at any late time in a
solution of the EFE because they were valid initially --- the
initial data must satisfy constraint equations --- and once they are
satisfied, the constraints are preserved by the dynamic field
equations.} However we cannot uniquely decode that signal to
determine what matter distribution outside the horizon caused it: a
particular velocity field might be caused by a relatively small mass
near the horizon, or a much larger mass much further away
\cite{ellsch72}. Claims about what conditions are like on very large
scales --- that is, much bigger than the Hubble scale --- are
unverifiable \cite{ell75}, for we have no observational evidence as
to what conditions are like far beyond the visual horizon. The
situation is like that of an ant surveying the world from the top of
a sand dune in the Sahara desert. Her world model will be a world
composed only of sand dunes
--- despite the existence of cities, oceans, forests, tundra,
mountains, and so on beyond her horizon.

It is commonly stated that if we live in a low-density universe
and the cosmological constant vanishes, the universe has infinite
spatial sections. However this deduction only applies if firstly
the RW-like nature of the universe within the past light cone
continues to be true indefinitely far outside it, and secondly the
space sections have their `natural' simply-connected topology ---
and there is no way we can obtain observational evidence that
these conditions are both true. In contrast to this, in chaotic
inflationary models (Sec.\ref{sec:inflate}), it is a definite
prediction that the universe will not be like a RW geometry on a
very large scale --- rather it will consist of many RW-like
domains, each with different parameter values, separated from each
other by highly inhomogeneous regions outside our visual horizon
\cite{lin90}, the whole forming a fractal-like structure. This
prediction is just as untestable as the previously prevalent
assumption (based on a Cosmological Principle) that the universe
is RW-like on such scales \cite{bon60,wei72}. Neither can be
observationally confirmed or denied. The same issue arises in an
even more extreme form in relation to the idea of a multiverse. We
return to this below, see Sec.\ref{sec:multiverse}.

\subsubsection{Small universes}
\label{sec:small} There is one case where this kind of spatial
observational limit does not obtain. This is when a \emph{Small
Universe} occurs, that is, a universe which closes up on itself
spatially for topological reasons \cite{ell71a}, and does so on such
a small scale that we have seen right round the universe since the
time of decoupling. Then we can see all the matter that exists, with
multiple images of many objects occurring \cite{ellsch86}. This
possibility is observationally testable by examining source
statistics, by observation of low power in the large angle CBR
anisotropies, and by detecting identical temperature variation on
various circles in the CBR sky \cite{lalu94}. There are weak hints
in the observed CBR anisotropies (the lack of power on large angular
scales) that this could actually be the case \cite{lumetal03}, but
this is not solidly confirmed. Checking if the universe is a small
universe or not is an important task; the nature of our
observational relationship to the universe is fundamentally
different if it is true \cite{ellsch86}.

\subsection{The observed universe}
The observable part of the universe (i.e. back to the visual
horizon) is strictly limited, and we have already seen most of it.
We can only observe distant objects by electromagnetic radiation at
all wavelengths, by neutrinos, and by gravitational waves. We
already have very complete broad coverage of the entire sky by
electromagnetic observations at all wavelengths right back to the
surface of last scattering, which is the limit of what will ever be
observable by electromagnetic radiation. Detailed observations (such
as the Hubble Deep Field) are available for restricted domains in
angle and depth. Detailed observations at suitable wavelengths are
beginning to discern what lies behind the Milky Way, which tends to
obscure a substantial fraction of the sky. It is unlikely there are
many new astronomical phenomena undiscovered in this observable
region, although it will be crucial determining more detailed
features of the phenomena we have already discovered (e.g. the
nature of dark matter and dark energy).

\textbf{Thesis B7: We have made great progress towards
observational completeness}. \textit{We have already seen most of
the part of the universe that is observable by electromagnetic
radiation. It is plausible that not many new astronomical
phenomena remain to be discovered by us observationally; we will
determine more details (so understanding more about what we have
seen) and see more objects, but not discover many new kinds of
things}.

\noindent Indeed Harwit \cite{har84} has used the multiplicity of discovery
of specific astronomical phenomena to estimate how many new essentially
different such phenomena there are still waiting to be discovered.

Neutrinos and gravitational waves will in principle allow us to peer back to
much earlier times (the time of neutrino decoupling and the quantum gravity
era respectively), but are much harder to observe at all, let alone in
useful directional detail. Nevertheless the latter has the potential to open
up to us access to eras quite unobservable in any other way. Maybe they will
give us unexpected information on processes in the very early universe which
would count as new features of physical cosmology.

\section{Issue C: The unbound energies in the early universe}
\label{sec:energies} The analogous problems for physical cosmology
arise because energies occurring in the Hot Big Bang early universe
phase (Sec.\ref{sec:hbb}) are essentially unbounded, so the highest
energies we can attain in particle accelerators cannot reach the
levels relevant to very early times. The uniqueness of cosmology in
this regard is that it is the only science contemplating spacetime
regions that have experienced such high energies, and with which we
are in intimate causal contact despite the huge timescales involved
--- indeed events at those early times determined much of what we
see around us today. The nuclear reactions underlying
nucleosynthesis are well understood, and their cross-sections
reasonably well-known; the processes of baryosynthesis and
quark-gluon recombination are reasonably understood and are on the
border of being testable; but physical processes relevant at earlier
times are inaccessible to testing by laboratory or accelerator-based
experiment. The \emph{Physics Horizon} by definition separates those
aspects of physics we can hope to test by high-energy experiments on
Earth or in the Solar System, from those where it is reasonable to
expect no such test will ever be possible:

\textbf{Thesis C1: The Physics Horizon limits our knowledge of physics
relevant to the very early universe}. \textit{We cannot experimentally test
much of the physics that is important in the very early universe because we
cannot attain the required energies in accelerators on Earth. We have to
extrapolate from known physics to the unknown and then test the
implications; to do this, we assume some specific features of known lower
energy physics are the true key to how things are at higher energies. We
cannot experimentally test if we have got it right.}

\noindent Note that this is independent of the issue of setting of
initial conditions for the universe, considered below, see
Sec.\ref{sec:initial}: the problem arises after the initial
conditions have been set and the universe is running according to
invariable physical laws. We cannot be confident of the validity
of the physics we presuppose then. Rather than using known physics
to predict the evolution of the universe, \emph{we end up testing
proposals for this physics by exploring their implications in the
early universe}, which is the only `laboratory' where we can test
some of our ideas regarding fundamental physics at the highest
energies \cite{yos78}; this is particularly true in the case of
quantum gravity proposals. The problem is we cannot simultaneously
do this and also carry out the aim of physical cosmology, namely
predicting the evolution of the early universe from known physical
theory.

Our understanding of physics at those times has of necessity to be
based on extrapolation of known physics way beyond the circumstances
in which it can be tested. The trick is to identify which features
are the key to use in that extrapolation: for example, variational
principles, broken symmetries and phase changes, duality invariance,
entropy limits are candidates. If we confirm our guesses for the
relevant physics by their satisfactory implications for the early
universe, tested in some suitable way, then this is impressive
progress; but if this is the \emph{only} way we can test the
proposed physics, the situation is problematic. If the hypothesis
solves only the specific issues it was designed to solve in the
early universe and nothing else, then in fact it has little
explanatory power, rather it is just an alternative (perhaps
theoretically preferable) description of the known situation. One
obtains positive observational support for a particular proposal for
the relevant physics only if it predicts multiple confirmed outcomes
(rather than just one), for example predicting particles that are
then confirmed to exist in a laboratory, so that a single hypothesis
simultaneously solves several different observational issues. Some
of the options may be preferred to others on various theoretical
grounds; but one must distinguish this from their having
observational support. They lack physical power if they have no
other testable consequences. A particular example is the
inflationary universe proposal (Sec.\ref{sec:inflate}): the supposed
inflaton field underlying an inflationary era of rapid expansion in
the early universe \cite{gut81,bibetal83,koltur90,gut97} has not
been identified, much less shown to exist by any laboratory
experiment. Because this field $\phi$ is unknown, one can assign it
an arbitrary potential $V(\phi)$, this arbitrariness reflecting our
inability to experimentally determine the relevant behaviour. It can
be shown that virtually any desired scale evolution $S(t)$ of the
universe can be attained by suitable choice of this potential
\cite{ellmad91}; and also almost any desired perturbation spectrum
can be obtained by a (possibly different) suitable choice
\cite{lidetal97}. Indeed in each case one can run the mathematics
backwards to determine the required potential $V(\phi)$ from the
desired outcome (Sec.\ref{sec:darken} below). The mathematical
existence of such a theoretical potential of the desired form for
cosmological purposes does not by itself prove a particle or field
exists with that effective potential.

\textbf{Thesis C2: The unknown nature of the inflaton means that
inflationary universe proposals are incomplete}. \textit{The promise
of inflationary theory in terms of relating cosmology to particle
physics has not been realized. This will only be the case when the
nature of the inflaton has been pinned down to a specific field that
experiment confirms or particle physics requires to exist.}

\noindent The very impressive achievement of inflation is that the
predicted CBR anisotropy spectrum is verified and agrees with the
matter power spectrum \cite{eis05}; but that prediction depends only
on the physics from the era of tight coupling of matter and
radiation to the present day, given a suitable initial fluctuation
spectrum in the early universe, rather than on the specific
hypothesis of an inflationary origin for that spectrum. The true
clincher would be if properties of an inflationary field were
predicted from the cosmology side and then confirmed in the
laboratory; indeed that would count as one of the great feats of
theoretical physics. This may not happen however because of the
experimental problems focused on here, arising because we cannot
reproduce on Earth all the conditions relevant to very early
cosmology.

One key application where this issue becomes significant is in
respect of the chaotic inflation theory (Sec.\ref{sec:inflate}).
As remarked above, see Sec.\ref{sec:horizon1}, its geometric
predictions are observationally unverifiable. It would
nevertheless be a good physical prediction if it was a more or
less inevitable outcome of known and tested underlying physics.
However this is not the case: the proposed underlying physics is
not experimentally tested, indeed it is not even uniquely defined
or associated with any specific known physical particle or field.
The claim that it inevitably follows from string theory
\cite{sus05} suffers from the problem that string theory is not a
well-defined or tested part of physics.

\section{Issue D: Explaining the universe --- the question of origins.}
\label{sec:origins} This is the unique core business of physical
cosmology: explaining both why the universe has come into existence
and evolved to the present very high-symmetry FL geometry on large
scales, and how structures come into existence on smaller scales.

\subsection{Start to the universe}
\label{sec:start1} Did a start to the universe happen? If so, what
was its nature? This has been discussed above (Sec.\ref{sec:start}),
and the issue is unresolved. The major related question is whether
the process of expansion only happens once in the life of the
Universe, or occurs repeatedly. The first option is the standard
model, where the entire evolution of the Universe is a once-off
affair, with all the objects we see, and indeed the Universe itself,
being transient objects that will burn out like dead fireworks after
a firework display. In this case everything that ever happens occurs
during one expansion phase of the Universe (possibly followed by one
collapse phase, which could occur if $k=+1$ and the present `dark
energy' field dies away in the future). This evolution might have a
singular start at a space-time singularity; a beginning where the
nature of time changes character; a non-singular bounce from a
single previous collapse phase; or a start from a non-singular
static initial state \cite{mul05}. An alternative is that many such
phases have occurred in the past, and many more will occur in the
future; the Universe is a \emph{Phoenix Universe} \cite{dicpee79},
new expansion phases repeatedly arising from the ashes of the old.
While the idea of one or more bounces is an old one \cite{tol34},
actual mechanisms that might allow this bounce behaviour have not
yet been elucidated in a fully satisfactory way. A variant is the
chaotic inflation idea of new expanding universe regions arising
from vacuum fluctuations in old expanding regions, leading to a
universe that has a fractal-like structure at the largest scales,
with many expanding regions with different properties emerging out
of each other in a universe that lasts forever
(Sec.\ref{sec:inflate}).

As discussed above, see Sec.\ref{sec:avoid}, it is possible (if the
universe has positive spatial curvature) that the quantum gravity
domain can be avoided and there was no start to the universe;
however this probably requires special initial conditions
\cite{ellmaa04}. If a quantum gravity domain indeed occurred, we
cannot come to a definite conclusion about whether there was a
creation event or not because we do not know the nature of quantum
gravity, nor how to reliably apply it in the cosmological context
where the issue of initial conditions arises. Loop quantum gravity
suggests the universe may be singularity-free \cite{boj01}, with
bounces or a non-singular start, but that theory is unconfirmed.
Tested physics cannot give a decisive answer; it is possible that
\emph{testable} physics also cannot do so.

\textbf{Thesis D1: An initial singularity may or may not have occurred}.
\textit{A start to the universe may have occurred a finite time ago, but a
variety of alternatives are conceivable: eternal universes, or universes
where time as we know it came into existence in one or another way. We do
not know which actually happened, although quantum gravity ideas suggest a
singularity might be avoided.}

\noindent This is a key issue in terms of the nature of the
universe: a space-time singularity is a dramatic affair, where the
universe (space, time, matter) has a beginning and all of physics
breaks down and so the ability to understand what happens on a
scientific basis comes to an end. However eternal existence is also
problematic, leading for instance to the idea of Poincar\'{e}'s
eternal return: everything that ever happened will recur an infinite
number of times in the future and has already occurred an infinite
number of times in the past \cite{bartip84}. This is typical of the
problems associated with the idea of infinity (discussed further
below, see Sec.\ref{sec:infinity}). \emph{It is not clear in the end
which is philosophically preferable: a singularity or eternal
existence}. That decision will depend on what criteria of
desirability one uses (such criteria are discussed below, see
Sec.\ref{sec:criteria}).

\subsection{The issue of initial conditions}
\label{sec:initial} While occurrence of an initial singularity is striking
in that it is a start to physics and spacetime as well as matter, whether it
occurred or not is in a sense irrelevant to the key issue of what determined
the nature of the universe:

\textbf{Thesis D2: Testable physics cannot explain the initial state and
hence specific nature of the universe}. \textit{A choice between different
contingent possibilities has somehow occurred; the fundamental issue is what
underlies this choice. Why does the universe have one specific form rather
than another, when other forms consistent with physical laws seem perfectly
possible? The reasons underlying the choice between different contingent
possibilities for the universe (why one occurred rather than another) cannot
be explored scientifically. It is an issue to be examined through philosophy
or metaphysics. }

\noindent Even if a literal creation does not take place, as is the
case in various of the present proposals, this does not resolve the
underlying issue of what determined why the universe is the way it
is, given that it could presumably have been otherwise. If the
proposal is evolution from a previous eternal state --- Minkowski
space for example --- then why did that come into existence, and why
did the universe expansion as a bubble from that vacuum start when
it did, rather than at some previous time in the pre-existent
eternity? Whenever it started, it could have started before! Some
attempts involve avoiding a true beginning by going back to some
form of eternal or cyclic initial state, for example Tolman's series
of expansion and collapse cycles \cite{tol34}, proposals for
creation of the universe as a bubble formed in a flat space-time
\cite{try73}, Linde's eternal chaotic inflation \cite{lin90},
Veneziano's re-expansion from a previous collapse phase
\cite{ghoetal98}, the ekpyrotic universe proposal \cite{khoetal01},
and theories involving foundational limits on information through a
``holographic principle" \cite{suslin04}. These do not avoid the
ultimate problem; it can be claimed they simply postpone facing it,
for one now has to ask all the same questions of origins and
uniqueness about the supposed prior state to the Hot Big Bang
expansion phase. The Hartle-Hawking `no-boundary' proposal
\cite{haw93} avoids the initial singularity because of a change of
space-time signature, and so gets round the issue of a time of
creation in an ingenious way; and Gott's causality violation in the
early universe \cite{gotli97} does the same kind of thing in a
different way. Such proposals cannot overcome the ultimate
existential question: \emph{Why has one specific state occurred
rather than any of the other possibilities? How was it decided that
this particular kind of universe would be the one actually
instantiated?} This question cannot be solved by physics alone,
unless one can show that only one form of physics is
self-consistent; but the variety of proposals made is evidence
against that suggestion.

The explanation of initial conditions has been the aim of the family
of theories one can label collectively as `quantum cosmology'
\cite{haw93,gotli97,gibetal03}; however as discussed earlier, here
we inevitably reach the limits to what the scientific study of the
cosmos can ever say --- if we assume that such studies must of
necessity involve an ability to observationally or experimentally
check our theories. No physical experiment at all can help here
because of the uniqueness of the universe, and the feature that no
spacetime exists prior to (in a causal sense) such a beginning; so
brave attempts to define a `physics of creation' stretch the meaning
of `physics'. Prior to the start (if there was a start) physics as
we know it is not applicable and our ordinary language fails us
because time did not exist, so our natural tendency to contemplate
what existed or happened `before the beginning' is highly misleading
--- there was no `before' then, indeed there was no `then' then!
Talking as if there was is commonplace, but quite misleading in
trying to understand a scientific concept of `creation'
\cite{grun89}. We run full tilt into the impossibility of testing
the causal mechanisms involved, when physics did not exist. No
experimental test can determine the nature of any mechanisms that
may be in operation in circumstances where even the concepts of
cause and effect are suspect. This comes particularly to the fore in
proposing `laws of initial conditions for the universe' --- for here
we are apparently proposing a theory with only one object. Physics
laws are by their nature supposed to cover more than one event, and
are untestable if they do not do so (Sec.\ref{sec:unique}).

\subsection{Special or general}
The present state of the universe is very special. Explanation of
the present large-scale isotropy and homogeneity of the universe
means determining the dynamical evolutionary trajectories relating
initial to final conditions, and then essentially either
\emph{explaining initial conditions}, where we run into
difficulties (Sec.\ref{sec:initial}), or \emph{showing they are
irrelevant}. The issue raised is whether the universe started off
in a very special geometrical state:

\textbf{Thesis D3: The initial state of the universe may have been special
or general}. \textit{Whether there was generality or speciality of
geometrical initial conditions for the universe is a key question. It seems
likely that the initial state of the observed part of the universe was not
generic.}

The assumption that the universe is geometrically special was
encoded in the Cosmological Principle, taken as a founding principle
in cosmology until the 1960's, i.e. as an `explanation' of special
initial conditions \cite{bon60,wei72}. Then Misner introduced the
chaotic cosmology programme \cite{mis68}, based on the idea of a
universe with generic initial conditions being isotropised at later
times by physical processes such as viscosity, making initial
conditions irrelevant. This concept of isotropisation then became
central to the inflationary family of theories
(Sec.\ref{sec:inflate}), with the underlying assumption being that
`fine tuning' of initial conditions is unphysical and to be avoided.
Both programmes are however only partially successful: one can
explain a considerable degree of isotropisation and homogenization
of the physical universe by either process, but this will not work
in all circumstances. Inflation can get rid of much anisotropy
\cite{wal83} but inhomogeneity must be restricted if inflation is to
succeed in producing a universe like that we see today, and the
success of inflation in solving the horizon issue for FL models ---
where exact homogeneity exists to start with --- will not
necessarily be replicated in anisotropic models. Universes that are
initially too anisotropic may never inflate, and the horizon problem
may not be solved in such models if they do;\footnote{Most
inflationary studies show only that the \emph{geometric} horizon
problem is solved in the very special RW geometries; but there is no
\emph{physical} horizon problem in those geometries, for they are by
by assumption spatially homogeneous and isotropic \emph{ab initio}.}
and only rather special states lead to ordinary thermodynamics
\cite{pen89a,pen04,wal05,carche05}, which is taken to be true in
inflationary physics.

Inflation can only be guaranteed to succeed if initial conditions
are somewhat restricted; some degree of geometric speciality must
have occurred at the start of the observed region of the universe.
This special domain might possibly occur within the context of a
much larger universe domain where conditions vary randomly, and
only isolated regions lead to inflation and eventually domains
such as that we see around us; attractive as this may be, it is an
untestable hypothesis (essentially a version of the multiverse
proposal, see Sec.\ref{sec:multiverse}).

Special initial conditions (which inflation proposes to explain)
might have just occurred that way. The ultimate issue is that
\emph{we have no proof as to whether initial conditions for the
universe were special or general; either could have occurred.} If we
state these conditions must have been general, we are making a
philosophical claim, for it is not a provable physical statement.
Part of the problem is that we have no agreed measure on the space
of possible universes; what seems special or general depends on the
choice of such a measure.

\section{Issue E: The Universe as the Background for Existence}
\label{sec:background} The universe provides the environment for all
of science, by determining the initial conditions within which all
physical laws are constrained to operate, thus setting boundary
conditions for all local physics. Together with suitable equations
of state for the matter or structural equations for complex systems,
these determine the nature of physical outcomes. The uniqueness of
cosmology lies in that it considers the origin of such conditions.

\subsection{Laws and boundary conditions}
\label{sec:laws} A fundamental assumption underlying physical
cosmology is the idea that \emph{the laws of physics are the same
everywhere in the physical universe}: those we determine in a
laboratory here and now will be the same as apply at very distant
places (e.g. determining the astrophysics of qso's at redshift
$z=6$), at very early times (e.g. at the time of nucleosynthesis),
and at very late times. Without this assumption, explanatory
theories have no solid foundation. However because of the
uniqueness of the universe discussed above (see
Sec.\ref{sec:unique}), unlike the rest of physics where the
distinction is clear and fundamental, in the cosmological context
the distinction between laws and boundary conditions becomes
blurred.

\textbf{Thesis E1: Physical laws may depend on the nature of the universe}.
\textit{We have an essential difficulty in distinguishing between laws of
physics and boundary conditions in the cosmological context of the origin of
the universe. Effective physical laws may depend on the boundary conditions
of the universe, and may even vary in different spatial and/or temporal
locations in the cosmos}.

\noindent Because we cannot vary the initial conditions in any way,
as far as we are concerned they are necessary rather than contingent
--- so the essential distinction between initial conditions and laws
is missing. The distinction is clear once the cosmos has come into
existence --- but we are concerned with `prior' conditions
associated with the creation of the cosmos and the very existence of
physical laws. Certainly any proposal for distinguishing between
laws of nature and boundary conditions governing solutions to those
laws is untestable in this context. Given the feature that the
universe is the unique background for all physics, it is therefore
not far-fetched to suggest that it is possible the cosmos influences
the \emph{nature} of local physical laws, rather than just their
initial conditions \cite{ellsch72,ell02}. This has been examined
over many decades in three specific cases.

(a) \emph{Varying `constants'}: It might be that there is a time
variation in physical constants of nature \cite{bar03} related to
the expansion of the universe, as proposed in the case of the
gravitational constant $G$ by Dirac \cite{dir38}, developed in depth
by Jordan and then Brans and Dicke \cite{bradic61}. Such proposals
must be consistently developed in relation to the rest of physics
and should be related to dimensionless constants, as otherwise they
may simply be disguised variations in the units of measurements
used, rather than being a genuine physical change (some claims that
the speed of light `c' may vary fall into this category
\cite{elluza05}). This proposal has received impetus in recent times
from ideas based in quantum field theory and string theory,
suggesting that many of the `constants of nature' are in fact
contingent, depending on the nature of the vacuum state
\cite{sus03}. This kind of proposal is to some degree open to
observational test \cite{cowson95,wil79}, and in the cases where it
has been investigated it seems that it does not occur in the visible
region of the universe --- the constants of nature are indeed
invariant, with one possible exception: the fine structure constant,
where there is claimed to be evidence of a very small change over
astronomical timescales \cite{bar03}. That issue is still under
investigation. Testing such invariance is fundamentally important,
precisely because cosmology usually assumes as a ground rule that
physics is the same everywhere in the universe. If this were not
true, local physics would not guide us adequately as to the
behaviour of matter elsewhere or at other times, and cosmology would
become an arbitrary guessing game. In order to proceed in a
scientific manner when such variation is proposed, one needs then to
hypothesize the manner of such variation. Thus the old laws where
$G$ was constant are replaced by new laws governing its time
variation \cite{bradic61}; the principle of nature being governed by
invariant (unchanging) physical laws and associated constants
remains.\footnote{``Despite the incessant change and dynamic of the
visible world, there are aspects of the fabric of the universe which
are mysterious in their unshakeable constancy. It is these
mysterious unchanging things that make our universe what it is and
distinguish it from other worlds we might imagine" (Barrow
\cite{bar03}, p. 3).} Thus in the end the proposal is to replace
simpler old laws by new more complex ones. These must then be
assumed invariant, or we cannot proceed scientifically.

(b) \emph{Inertia and Mach's Principle}: It might be that the local
inertial properties of matter are determined by the distant
distribution of matter in the universe, so that if the universe were
different, inertia would be different. This is the complex of ideas
referred to as Mach's principle \cite{barpfi95}, which served as a
major impetus for Einstein's cosmological ideas. The precise meaning
and implications of this idea remain controversial.

(c) \emph{The arrow of time}: The existence and direction of the
macroscopic arrow of time in physics --- and hence in chemistry,
biology, psychology, and society --- is related to boundary
conditions in the past and future of the universe. The fundamental
physical laws by themselves are time symmetric, and so unable to
explain this feature \cite{dav74,ellsch72,zeh92,uff05}. A recent
argument of this kind is Penrose's claim that the existence of the
arrow of time is crucially based in the universe having had rather
special initial conditions \cite{pen89,pen89a,wal05}. Thus what
appears in ordinary physics as an immutable law of nature (viz. the
Second Law of Thermodynamics with a given arrow of time) may well be
the result of specific boundary conditions at the start and end of
the universe. It might not be true in all universes, even if the
underlying fundamental physical laws are the same.

\noindent In each case proposals have been made as to the possible
nature of the deeper underlying unchanging laws, and the relations
between the state of the universe and the resultant effective laws
in that context. This is also proposed in `landscape' of
possibilities of string theory \cite{sus05}. These proposals are
however intrinsically untestable, for the reasons explained above
(Sec.\ref{sec:unique}): we cannot change the boundary conditions of
the universe and see what happens; but they do serve as a continuing
fertile source of ideas.

\subsection{Alternative physics}
\label{sec:alternate} In any case, the important conclusion is that
it is certainly appropriate for cosmology to consider what would
have happened if, not only the boundary conditions at the beginning
of the universe, but also the laws of physics had been different
\cite{sus05}:

\textbf{Thesis E2: We cannot take the nature of the laws of
physics for granted}. \textit{Cosmology is interested in
investigating hypothetical universes where the laws of physics are
different from those that obtain in the real universe in which we
live --- for this may help us understand why the laws of physics
are as they are (a fundamental feature of the real physical
universe).}

\noindent One cannot take the existence and nature of the laws of
physics (and hence of chemistry) as unquestionable in cosmology ---
which seems to be the usual habit in biological discussions on the
origin and evolution of life. This is in stark contrast to the rest
of science, where we are content to take the existence and nature of
the laws describing the fundamental behaviour of matter as given and
unchangeable. Cosmological investigation is interested in the
properties of hypothetical universes with different physical
behaviour. Consideration of `what might have been' is a useful
cosmological speculation that may help throw light on what actually
is; this is a statement of the usefulness of `Gedanken experiments'
in cosmology.

Indeed if one wants to investigate issues such as why life exists in
the universe, consideration of this larger framework --- in essence,
a hypothetical ensemble of universes with many varied properties ---
is essential (this is of course not the same as assuming an ensemble
of such universes actually exists, cf. the discussion below in
Sec.\ref{sec:multiverse}). However we need to be very cautious about
using any claimed statistics of universes in such a hypothetical
ensemble of all possible or all conceivable universes. This is
usually not well defined, and in any case is only relevant to
physical processes if either the ensemble actually exists, rather
than being a hypothetical one, or if it is the outcome of processes
that produce well-defined probabilities --- an untestable proposal.
We can learn from such considerations the nature of possible
alternatives, but not necessarily the probability with which they
might occur (if that concept has any real meaning).

\subsection{Emergence of complexity}\label{sec:new}
As the universe evolves an increase of complexity takes place in
local systems as new kinds of objects come into being that did not
exist before --- nuclei, atoms, stars and galaxies, planets, life,
consciousness, and products of the mind such as books and computers
\cite{mor02}. New kinds of physical states come into being at late
times such as Bose-Einstein condensates, that plausibly cannot exist
without the intervention of intelligent beings.

\textbf{Thesis E3: Physical novelty emerges in the expanding
universe}. \textit{New kinds of physical existence come into being
in the universe as it evolves, that did not exist previously. Their
existence is allowed by the boundary conditions provided by the
universe for local systems, together with the possibility space
generated by the underlying physics. While their physical existence
is novel, every new thing that comes into being is foreshadowed in
possibility structures that precede their existence.}

\noindent Physical existence is new as the universe evolves, but
there had to be precursors of the novel in the possibility space
allowed by physics, so that they could come into being. In this
sense the truly novel does not emerge \emph{ex nihilo} but rather
is discovered. The universe is the environment that allows this to
happen. The nature of the features leading to the existence of
life, and their possible causes, is discussed in
Sec.\ref{sec:anthropic}.

\section{Issue F: The Explicit Philosophical Basis}
\label{sec:philos}

Consequent on the discussion above, and particularly items
\textbf{B6}, \textbf{C2}, and \textbf{D2}, it follows that

\textbf{Thesis F1: Philosophical choices necessarily underly cosmological
theory}. \textit{Unavoidable metaphysical issues inevitably arise in both
observational and physical cosmology. Philosophical choices are needed in
order to shape the theory.}

\noindent There is of course always a philosophical basis to any
scientific analysis, namely adoption of the basic scientific method
and a commitment to the attempt to explain what we see as far as
possible simply in terms of causal laws, ultimately based in
physics. This will clearly be true also in cosmology. However we
need further explicit philosophical input in order to attain
specific geometric models --- for example a Copernican principle, as
explained above, see Sec.\ref{sec:homog} --- and to determine what
form physical cosmology should take in the very early universe, for
example deciding which physical principle to use as the core of
one's extrapolation of known physics to the unknown
(Sec.\ref{sec:energies}). Underlying both sets of choices are
criteria for satisfactoriness of a cosmological model, which help
decide which feature to focus on in formulating a theory. Of
particular importance is the scope chosen for our cosmological
theory; together with the choice of criteria for a good theory, this
is a philosophical decision that will shape the rest of the
analysis. Some cosmologists tend to ignore the philosophical choices
underlying their theories; but simplistic or unexamined
philosophical standpoints are still philosophical standpoints!

\subsection{Criteria for theories}
\label{sec:criteria} As regards criteria for a good scientific
theory \cite{kuh77}, typical would be the following four areas of
assessment:

\textbf{1}. \emph{Satisfactory structure}: \textbf{(a)} internal
consistency, \textbf{(b)} simplicity (Ockham's razor), and \textbf{(c)}
aesthetic appeal (`beauty' or `elegance').

\textbf{2}. \emph{Intrinsic explanatory power}: \textbf{(a)} logical
tightness, \textbf{(b)} scope of the theory --- the ability to unify
otherwise separate phenomena, and \textbf{(c)} probability of the
theory or model with respect to some well-defined measure;

\textbf{3}. \emph{Extrinsic explanatory power, or relatedness}:
\textbf{(a)} connectedness to the rest of science, \textbf{(b)}
extendability --- providing a basis for further development;

\textbf{4}. \emph{Observational and experimental support}, in terms
of \textbf{(a)} testability: the ability to make quantitative as
well as qualitative predictions that can be tested; and \textbf{(b)}
confirmation: the extent to which the theory is supported by such
tests as have been made.

\noindent It is particularly the latter that characterizes a
scientific theory, in contrast to other types of theories claiming
to explain features of the universe and why things happen as they
do. It should be noted that \emph{these criteria are philosophical
in nature in that they themselves cannot be proven to be correct by
any experiment}. Rather their choice is based on past experience
combined with philosophical reflection. One could attempt to
formulate criteria for good criteria for scientific theories, but of
course these too would need to be philosophically justified. The
enterprise will end in infinite regress unless it is ended at some
stage by a simple acceptance of a specific set of criteria.

\textbf{Thesis F2: Criteria of satisfactoriness for theories cannot
be scientifically chosen or validated}. \textit{Criteria of
satisfactoriness are necessary for choosing good cosmological
theories; these criteria have to be chosen on the basis of
philosophical considerations. They should include criteria for
satisfactory structure of the theory, intrinsic explanatory power,
extrinsic explanatory power, and observational and experimental
support.}

\noindent The suggestion here is that the above proposed criteria
are a good set to use in investigating cosmology; they include those
most typically used (\cite{kuh77}; and see \cite{pen04,sus05} for
comments on such criteria).

\subsubsection{Conflicts between criteria.}
These criteria are all acknowledged as desirable. The point then is that
generally in pursuing historical sciences, and in particular in the
cosmological context, they will not all be satisfied to the same degree, and
may even lead to opposing conclusions:

\textbf{Thesis F3: Conflicts will inevitably arise in applying
criteria for satisfactory cosmological theories}.
\textit{Philosophical criteria for satisfactory cosmological
theories will in general come into conflict with each other, so that
one will have to choose between them to some degree; this choice
will shape the resulting theory.} \cite{ell91}.

\noindent The thrust of much recent development has been away from
observational tests towards strongly theoretically based proposals,
indeed sometimes almost discounting observational tests. At present
this is being corrected by a healthy move to detailed observational
analysis of the consequences of the proposed theories, marking a
maturity of the subject. However because of all the limitations in
terms of observations and testing [criteria \textbf{(4)}], in the
cosmological context we still have to rely heavily on other
criteria, and some criteria that are important in most of science
may not really make sense. This is true of \textbf{2(c)} in
particular, as discussed above, see Sec.\ref{sec:unique};
nevertheless many approaches still give the idea of probability
great weight. At a minimum, the ways this can make sense needs
exploration and explication. Furthermore the meaning of some of the
criteria may come into dispute. \textbf{1(b)} is clearly a case in
point : for example, is the idea of an existent ensemble of
universes displaying all possible behaviours simple (because it is a
single idea that can be briefly stated), or immensely complex
(because that statement hides all the complexities and ambiguities
involved in the idea of an infinity of possibilities)? \textbf{1(c)}
is also controversial(`beauty is in the eye of the beholder'), see
\cite{sus05} for a discussion.

The tenor of scientific understanding may change, altering the
balance of what is considered a good explanation and what is not. An
example \cite{ell90} is the way cosmologists strongly resisted the
idea of an evolving universe in the 1920's, at a time when
biological evolution was very well established but the idea of
continental drift was also being strongly resisted. The change to an
appreciation of the explanatory power of an evolving model came
later in both cases; but even then in the cosmological case, for
either aesthetic or metaphysical reasons, some still sought for a
steady state description, resisting the implication of a beginning
to the universe. That tendency is still with us today, in the form
of models that are eternal in one way or another (e.g. some forms of
chaotic inflation). Another example is the change from supposition
of underlying order, expressed in the idea of a Cosmological
Principle, to a broad supposition of generic disordered conditions,
embodied in the ideas of inflation. Associated with this is a shift
from making geometric assumptions to providing physical explanatory
models. It is this shift that underlies the major present support
for inflation:

\textbf{Thesis F4:\ The physical reason for believing in inflation
is its explanatory power as regards structure growth in the
universe}. \textit{Inflation predicts the existence of Gaussian
scale-free perturbations in the early universe thereby (given the
presence of cold dark matter) explaining bottom-up structure
formation in a satisfactory way.  This theory has been vindicated
spectacularly through observations of the CBR\ and matter power
spectra. It is this explanatory power that makes it so acceptable to
physicists, even though the underlying physics is neither
well-defined nor tested, and its major large-scale observational
predictions are untestable.}

The physical explanatory power of inflation in terms of structure
formation, supported by the observational data on the fluctuation
spectra, is spectacular. For most physicists, this trumps the lack
of identification and experimental verification of the underlying
physics (Sec.\ref{sec:energies}). Inflation provides a causal model
that brings a wider range of phenomena into what can be explained by
cosmology (Criterion \textbf{2(b)}), rather than just assuming the
initial data had a specific restricted form. Explaining flatness
($\Omega_{0}\simeq 1$ as predicted by inflation)\ and homogeneity
reinforces the case, even though these are philosophical rather than
physical problems (they do not contradict any physical law; things
could just have been that way). However claims on the basis of this
model as to what happens very far outside the visual horizon (as in
the chaotic inflationary theory) results from prioritizing theory
over the possibility of observational and experimental testing
\cite{earmos99}. It will never be possible to \textit{prove} these
claims are correct.

\subsection{The scope of cosmology}
\label{sec:scope} To sensibly choose priorities for the criteria just
discussed, we need an answer to the question, How much should we try to
explain?

\textbf{Thesis F5: Cosmological theory can have a wide or narrow
scope of enquiry}. \textit{The scope we envisage for our
cosmological theory shapes the questions we seek to answer. The
cosmological philosophical base becomes more or less dominant in
shaping our theory according to the degree that we pursue a theory
with more or less ambitious explanatory aims in terms of all of
physics, geometry, and underlying fundamental causation.}

\noindent This is a choice one has to make, as regards both
foundations and outcomes. Given a decision on this, one can
sensibly debate what is the appropriate philosophical position to
adopt in studying a cosmological theory with that scope. The study
of expansion of the universe and structure formation from
nucleosynthesis to the present day is essential and well-informed.
The philosophical stance adapted is minimal and highly plausible.
The understanding of physical processes at earlier times, back to
quantum gravity, is less well founded. The philosophical stance is
more significant and more debatable. Developments in the quantum
gravity era are highly speculative; the philosophical position
adopted is dominant because experimental and observational limits
on the theory are lacking.

One can choose the degree to which one will pursue the study of
origins \cite{fab89} back to earlier and earlier times and to more
fundamental causal issues, and hence the degree to which specific
philosophical choices are dominant in one's theory. The basic
underlying cosmological questions are \cite{ell91}:

(1) \emph{Why do the laws of physics have the form they do?}
Issues arise such as what makes particular laws work? For example,
what guarantees the behaviour of a proton, the pull of gravity?
What makes one set of physical laws `fly' rather than another? If
for example one bases a theory of cosmology on string theory
\cite{sus05}, then who or what decided that quantum gravity would
have a nature well described by string theory? If one considers
all possibilities, considering string theory alone amounts to a
considerable restriction.

(2) \emph{Why do boundary conditions have the form they do ?} The
key point here (Sec.\ref{sec:initial}), is how are specific
contingent choices made between the various possibilities, for
example whether there was an origin to the universe or not.

(3) \emph{Why do any laws of physics at all exist ?} This relates
to unsolved issues concerning the nature of the laws of physics:
are they descriptive or prescriptive? (Sec.\ref{sec:laws1}). Is
the nature of matter really mathematically based in some sense, or
does it just happen that its behaviour can be described in a
mathematical way?

(4) \emph{Why does anything exist ?} This profound existential
question is a mystery whatever approach we take.\footnote{But see
Grunbaum \cite{grun04} for a dissenting view.}

\noindent Finally the adventurous also include in these questions
the more profound forms of the contentious Anthropic question
\cite{carree79,dav82,bartip84,teg93,sus05}:

(5) \emph{Why does the universe allow the existence of intelligent
life?} This is of somewhat different character than the others and
largely rests on them but is important enough to generate
considerable debate in its own right.

\noindent The status of all these questions is philosophical
rather than scientific, for they cannot be resolved purely
scientifically. How many of them --- if any --- should we consider
in our construction of and assessments of cosmological theories?

One option is \emph{to decide to treat cosmology in a strictly
scientific way}, excluding all the above questions, because they
cannot be solved scientifically. One ends up with a solid
technical subject that by definition excludes such philosophical
issues. This is a consistent and logically viable option. This
logically unassailable position however has little explanatory
power; thus most tend to reject it because of criteria
\textbf{2(b)} and \textbf{3} above.

The second option is to decide that \emph{these questions are of
such interest and importance that one will tackle some or all of
them, even if that leads one outside the strictly scientific
arena}. It is here that criteria \textbf{2} and \textbf{3} above
are to some degree in conflict with criterion \textbf{4}. Thus if
we try to explain the origin of the universe itself, these
philosophical choices become dominant precisely because the
experimental and observational limits on the theory are weak; this
can be seen by viewing the variety of such proposals that are at
present on the market.

\subsection{Limits of Representation and Knowledge of Reality}
\label{sec:epist} It follows from the above discussion that there
are limits to what the scientific method can achieve in
explanatory terms. We need to respect these limits and acknowledge
clearly when arguments and conclusions are based on some
philosophical stance rather than purely on testable scientific
argument. If we acknowledge this and make that stance explicit,
then the bases for different viewpoints are clear and alternatives
can be argued about rationally.

A crucial underlying feature here is relating the nature of
epistemology to ontology:\ how do we relate evidence to our theories
of existence?\ A further key issue is the relation of models to
reality:

\textbf{Thesis F6: Reality is not fully reflected in either
observations or theoretical models}. \textit{Problems arise from
confusion of epistemology (the theory of knowledge) with ontology
(the nature of existence): existence is not always manifest
clearly in the available evidence. The theories and models of
reality we use as our basis for understanding are necessarily
partial and incomplete reflections of the true nature of reality,
helpful in many ways but also inevitably misleading in others.
They should not be confused with reality itself! }

\noindent The confusion of epistemology with ontology occurs all
the time, underlying for example the errors of both logical
positivism and extreme relativism. In particular, it is erroneous
to assume that lack of evidence for the existence of some entity
is proof of its non-existence. In cosmology it is clear for
example that regions may exist from which we can obtain no
evidence (because of the existence of horizons); so we can
sometimes reasonably deduce the existence of unseen matter or
regions from a sound extrapolation of available evidence (no one
believes matter ends at or just beyond the visual horizon).
However one must be cautious about the other extreme, assuming
existence can always be assumed because some theory says so,
regardless of whether there is any evidence of existence or not.
This happens in present day cosmology, for example in
presentations of the case for multiverses, even though the
underlying physics has not been experimentally confirmed. It may
be suggested that arguments ignoring the need for
experimental/observational verification of theories ultimately
arise because these theories are being confused with reality, or
at least are being taken as completely reliable total
representations of reality. This occurs in
\begin{itemize}
\item Confusing computer simulations of reality with reality
itself, when they can in fact represent only a highly simplified
and stylized version of what actually is; \item Confusing the laws
of physics themselves with their abstract mathematical
representation (if indeed they are ontologically real, c.f.
Sec.\ref{sec:lawscosm}), or confusing a construction of the human
mind (`Laws of Physics') with the reliable behaviour of ponderable
matter (if they are not ontologically real);
\item Confusing
theoretically based outcomes of models with proven observational
results (e.g. claiming the universe necessarily has flat spatial
sections: $\Omega_0=1$, and so this can be taken for granted, when
the value of $\Omega_0$ can and should be observationally
determined precisely because this then tests that prediction).
\end{itemize}
\noindent No model (literary, intuitive, or scientific) can give a
perfect reflection of reality. Such models are always selective in
what they represent and partial in the completeness with which
they do so. The only model that would reflect reality fully is a
perfect fully detailed replica of reality itself! This
understanding of the limits of models and theories does not
diminish the utility of these models; rather it helps us use them
in the proper way. This is particularly relevant when we consider
how laws of nature may relate to the origins of the universe
itself, and to the existence and nature of life in the expanding
universe. The tendency to rely completely on our theories, even
when untested, seems sometimes to arise because we believe they
are the same as reality --- when at most they are
\emph{descriptions} of reality.

\section{Key Issues}
\label{sec:issues}

There are some interrelated key issues where the features identified above
either are at the heart of current debates, or are likely to be at the heart
of future debates. They are: the reason cosmological conditions allow the
existence of life (anthropic issues), the closely related issue of the
possible existence of multiverses; and the natures of existence, including
the questions of the existence of infinities and the nature of the laws of
physics. We look at them in turn in this section. To some degree they have
already been considered above,but they are specifically featured here
because of the important role they will probably play in discussion in the
future.

\subsection{Issue G: The anthropic question:\ Fine tuning for life}
\label{sec:anthropic} One of the most profound fundamental issues in
cosmology is the Anthropic question, see
\cite{dav82,bartip84,ear87,fab89,dav87,bal91,ree99,ree02,bar03}:
\textit{why does the Universe have the very special nature required
in order that life can exist?}. The point is that a great deal of
``fine tuning" is required in order that life be possible. There are
many relationships embedded in physical laws that are not explained
by physics, but are required for life to be possible; in particular
various fundamental constants are highly constrained in their values
if life as we know it is to exist:

\noindent \textit{``A universe hospitable to life --- what we
might call a biophilic universe --- has to be special in many ways
... Many recipes would lead to stillborn universes with no atoms,
no chemistry, and no planets; or to universes too short lived or
too empty to evolve beyond sterile uniformity"} \cite{ree02}.

\noindent How has it come about that the Universe permits the
evolution and existence of intelligent beings at any time or
place? ``What features of the universe were essential for
creatures such as ourselves, and is it through coincidence or for
some deeper reason that our universe has these features?"
\cite{griree91}. Whether one regards this as an appropriate issue
for cosmology to discuss depends, as discussed above
(Sec.\ref{sec:scope}), on the scope one envisages for cosmology.
The viewpoint taken here will be that this is one of the major
issues one might wish to explain, and indeed a substantial
literature considers this. Here we explore the nature of this fine
tuning, and then consider possible answers as to how it arises.
There are three aspects that we consider in turn (cf.
\cite{sus05}).

\subsubsection{Laws of physics and the existence of complexity}
The laws of physics and chemistry are such as to allow the functioning of
living cells, individuals, and ecosystems of incredible complexity and
variety, and it is this that has made evolution possible. What requires
explanation, \emph{is why the laws of physics are such as to allow this
complex functionality to work}, without which no evolution whatever would
occur. We can conceive of universes where the laws of physics (and so of
chemistry) were different than in ours. Almost any change in these laws will
prevent life as know it from functioning.

The first requirement is \emph{the existence of laws of physics that
guarantee the kind of regularities that can underlie the existence
of life}. These laws as we know them are based on variational and
symmetry principles; we do not know if other kinds of laws could
produce complexity. If the laws are in broad terms what we presently
take them to be, the following \emph{inter alia} need to be right,
for life of the general kind we know to exist \cite{dav82,griree91}:
\begin{itemize}
\item Quantization that stabilizes matter and allows chemistry to
exist through the Pauli exclusion principle. \item The
neutron-proton mass differential must be highly constrained. If
the neutron mass were just a little less than it is, proton decay
could have taken place so that by now no atoms would be left at
all \cite{dav82}. \item Electron-proton charge equality is
required to prevent massive electrostatic forces overwhelming the
weaker electromagnetic forces that govern chemistry. \item The
strong nuclear force must be strong enough that stable nuclei
exist \cite{dav82}; indeed complex matter exists only if the
properties of the nuclear strong force lies in a tightly
constrained domain relative to the electromagnetic force
\cite{teg03a}. \item The chemistry on which the human body depends
involves intricate folding and bonding patterns that would be
destroyed if the fine structure constant (which controls the
nature of chemical bonding) were a little bit different. \item The
number $D$ of large spatial dimensions must be just 3 for
complexity to exist \cite{teg03a,ree02}.
\end{itemize}
\noindent Hogan has examined the freedom in the parameters of the
standard model of particle physics and concluded that 5 of the 17
free parameters of the standard model must lie in a highly
constrained domain if complex structures are to exist \cite{hog03}.
This is of course taking the basic nature of the standard model of
particle physics for granted. If this were not so, it is difficult
to determine what the constraints would be. However his study is
sufficient to show that whatever the nature of fundamental physics,
and in particular of particle physics, may be, only a small subset
of all possible laws of physics will be compatible with the
existence of complexity.

\subsubsection{Laws of physics and the existence of congenial environments}
The creation through astrophysical processes of suitable habitats
for life to exist (the existence of planets circling stable stars,
for example) depends to some degree on the nature of the
fundamental physical laws. If the laws are in broad terms what we
presently take them to be, the requirements for such habitats to
exist include:
\begin{itemize}
\item The gravitational force must create large stable structures
(planets and stars) that can be the habitat for life and their
energy source respectively. This requires the gravitational force
to be very weak relative to electrical forces. The ratio
$\mathcal{N}$ of the strength of the electromagnetic force to the
gravitational force must be close to the observed value:
$\mathcal{N} \simeq 10^{36}$ (\cite{ree99}, Chapter 3). \item The
weak force must allow helium production that leaves sufficient
hydrogen over; it is related to gravity through a numerical factor
of $ 10^{-11}$, which cannot be much different. And for this to
work, the neutron-proton mass difference must be close to the mass
of the electron \cite{dav82}. \item A stellar balance should allow
a long lifetime for stars like the sun, so allowing the
transmutation of the light elements into heavy elements. This
requires that the nuclear fusion efficiency $\mathcal{E}$ be close
to the observed value: $\mathcal{E}\simeq 0.007$ (\cite{ree99},
Chapter 4). \item One needs to overcome the beryllium
``bottleneck" in the making of heavy elements through nuclear
reactions in stars \cite{griree91,sus05}. The production of carbon
and oxygen in stars requires the careful setting of two different
nuclear energy levels to provide a resonance; if these levels were
just a little different, the elements we need for life would not
exist \cite{fab89}. Indeed it was on this basis that Hoyle
famously predicted a carbon-12 energy level that has since been
experimentally confirmed. \item One needs something like the
existence of neutrinos and the weak interaction with its specific
coupling constant in order to underly supernovae explosions that
spread heavy elements through space, as seeds for planetary
formation \cite{griree91}. \item The nuclear force must be weak
enough that di-protons do not exist, otherwise no protons will be
left over to enable heavier elements to exist \cite{dav82}. \item
The neutrino mass must not be too high, or the universe will not
last long enough \cite{dav82}.
\end{itemize}

\subsubsection{Cosmological boundary/initial conditions and congenial
environments} Finally, given laws of physics that are suitable in
terms of satisfying the requirements of both the previous
sections, the universe itself must also be suitable, in terms of
its initial or boundary conditions, for life to exist. If the laws
of physics are basically the same as we now believe them to be,
these cosmological requirements include
\begin{itemize}
\item The size of the universe and its age must be large enough.
There could be universes that expanded and then recollapsed with a
total lifetime of only 100,000 years; we need a sufficiently old
universe for second generation stars to come into existence and
then for planets to have a stable life for long enough that
evolution could lead to the emergence of intelligent life. Thus
the universe must be at about 15 billion years old for life to
exist \cite{griree91}, hence we must have $\Omega_{matter} \simeq
0.3$ (\cite{ree99}, Chapter 6). \item The size of the cosmological
constant must not be too large, or galaxies will not form; we need
$|\Omega_{\Lambda }|<1$ for galaxies to exist (\cite{ree99},
Chapter 7; \cite{sus05}). \item The seeds in the early universe
for fluctuations that will later grow into galaxies must be of the
right size that structures form without collapsing into black
holes: the number $Q$ characterizing the size of primordial
ripples on the LSS (and hence the geometry of the perturbed
cosmological model, see Sec.\ref{sec:perturb}) must therefore be
of the order $Q\simeq 10^{-5}$ (\cite{ree99}, Chapter 8).
\end{itemize}

The complex of interacting systems in a human body could not
possibly work if a series of delicate conditions were not
maintained. For example, the background radiation might never drop
below 3000 K, so that matter was always ionized (electrons and
nuclei always remaining separate from each other); the molecules
of life could then never form. Black holes might be so common that
they rapidly attracted all the matter in the universe, and there
never was a stable environment in which life could develop. Cosmic
rays could always be so abundant that any tentative organic
structures were destroyed before they could replicate. Overall,
\begin{itemize}
\item There must be non-interference with local systems. The
concept of locality is fundamental, allowing local systems to
function effectively independently of the detailed structure of the
rest of the Universe. We need the universe and the galaxies in it to
be largely empty, and gravitational waves and tidal forces to be
weak enough,\footnote{Thus the Weyl tensor $C_{abcd}$ must be
suitably small everywhere, presumably implying an almost-RW
geometry, cf. \cite{stoetal95}.} so that local systems can function
in a largely isolated way \cite{ell02}. \item The fact that the
night sky is dark (`Olbers' paradox' \cite{bon60,har81}) is a
consequence of the expansion of the universe together with the
photon to baryon ratio. This feature is a necessary condition for
the existence of life: the biosphere on Earth functions by disposing
of waste energy to the heat sink of the dark night sky \cite{pen89}.
Thus one way of explaining why the sky is observed to be dark at
night is that if this were not so, we would not be here to observe
it. \item The existence of the arrow of time, and hence of laws like
the second law of thermodynamics, are probably necessary for
evolution and for consciousness. This depends on boundary conditions
at the beginning and end of the Universe (Sec.\ref{sec:laws}). \item
Presumably the emergence of a classical era out of a quantum state
is required. The very early universe would be a domain where quantum
physics would dominate, leading to complete uncertainty and an
inability to predict the consequence of any initial situation; we
need this to evolve to a state where classical physics leads to the
properties of regularity and predictability that allow order to
emerge.\item Physical conditions on planets must be in a
quasi-equilibrium state for long enough to allow the delicate
balances that enable our existence, through the very slow process of
evolution, to be fulfilled.
\end{itemize}
\noindent Thus the existence of suitable local systems to be a
habitat for life depends critically on the large-scale properties of
very distant matter. These provides a stable local environment
within which life can develop.

\subsubsection{Fine tuning overall}
Thus there are many ways that conditions in a universe could
prevent life occurring. Life will occur only if: there exist heavy
elements; there is sufficient time for evolution of advanced life
forms to take place; there are regions in the universe that are
neither too hot nor too cold; there are precisely restricted
values of the fundamental constants that control chemistry and
local physics; and so on. These conditions will not be true in a
generic universe. In summary,

\textbf{Thesis G1: Life is possible because both the laws of physics and the
boundary conditions for the universe have a very special nature}. \textit{\
Only particular laws of physics, and particular initial conditions in the
Universe, allow the existence of intelligent life of the kind we know. No
evolutionary process whatever is possible for any kind of life if these laws
and conditions do not have this restricted form.}

\noindent Why is this so? One should note that we can only
meaningfully refer here to `life as we know it'. One of the
recurring issues is whether there could be some other quite
different basis for life. You can if you wish speculate that life
might exist in some immaterial form, or based only on light
elements, or existent deep in empty space without the need for
stars or planets to provide a viable habitat. The anthropic
literature is based on assuming this is not viable, but we cannot
\textit{prove} anything in this regard. We have no idea of any
basis by which life might come into existence other than the broad
principles we see in the life around us. The basic principles of
life as we understand it require a great degree of complex
organization enabling it to fulfil a complex variety of functions
that can only, as far as we know, be based in material existence
with information storage, energy usage, sensing of the external
world, etc., which requires at a minimum heavy elements (carbon,
nitrogen, oxygen, phosphorus for example), a long-term energy
source (such as the flow of energy from the sun), and a stable
environment (such as the surface of a planet). When we abandon
this basis for understanding --- saying `yes but some other form
of life might exist' without providing any proposal for its
possible structure --- one enters the unprofitable realm of
speculation. It does not seem to provide any useful way forward.

\paragraph{The Weak Anthropic Principle.}
There are two purely scientific approaches to the Anthropic
issue.\footnote{I omit the so-called \emph{Final Anthropic
Principle} (FAP for short), which maintains that intelligent life
must necessarily evolve and then remain in existence until the end
of the universe, for I do not believe it merits serious discussion
as a scientific proposal; indeed it led to a famous book review
referring to the \emph{Completely Ridiculous Anthropic Principle}
(CRAP for short) \cite{gar86}.} The first is the \emph{Weak
Anthropic Principle} (WAP), based on the comment: it is not
surprising the observed Universe admits the existence of life, for
the Universe cannot be observed unless there are observers in it
\cite{bartip84,bal91}. This seemingly empty statement gains
content when we turn it round and ask, at what times and places in
the Universe can life exist, and what are the inter-connections
that are critical for its existence? It could not for example
exist too early in the present expansion phase, for the night sky
would then have been too hot. Furthermore one can deduce various
necessary relations between fundamental quantities in order that
the observers should exist (e.g. those mentioned above), so that
if for example the fundamental constants vary with time or place
in the Universe, life will only be possible in restricted regions
where they take appropriate Anthropic values.

Hence this view basically interprets the Anthropic principle as a
selection principle: the necessary conditions for observers to
exist restricts the times and places from which the Universe can
be observed. Because it is quite possible that conditions would
not be right for life to exist anywhere in an arbitrarily selected
universe, it is also usually conjoined with the idea of the
existence of a multiverse, as discussed below, see
Sec.\ref{sec:multiverse}. This is an interesting and often
illuminating viewpoint. For example, neither the Chaotic
Inflationary Universe idea (Sec.\ref{sec:inflate}) nor any other
multiverse proposal works unless we add such an Anthropic
component into their interpretation to explain why we observe the
Universe from a viewpoint where life exists. It is now used by
some physicists to explain the low value of the cosmological
constant (which quantum field theory predicts should have a very
much larger value than observed, see Sec.\ref{sec:inftest}), and
occurs in the context of the possibility landscape of string
theory \cite{sus05}.

\paragraph{The Strong Anthropic Principle.}
By contrast, the \emph{Strong Anthropic Principle} (SAP)
\cite{bartip84,bal91} claims that it is necessary that intelligent
life exist in the Universe; the presence of life is required in
order that a universe model make sense. This is clearly a very
controversial claim, for it is hard to provide scientific reasons
to support this view. One can suggest that the most solid
justification attempted is through the claim that the existence of
an observer is necessary in order that quantum theory can make
sense. However, this justification is based on one of a number of
different interpretations of quantum theory; the nature of these
quantum foundations is controversial, and not resolved
\cite{ish97,dic05,lan05}.

Furthermore if we were to suppose this justification correct, then
the next step is to ask: Why does the Universe need quantum
mechanics anyway? The argument would be complete only if we could
prove that quantum mechanics was absolutely necessary for every
self-consistent Universe; but that line of reasoning cannot be
completed at present, not least because quantum mechanics itself
is not a fully self-consistent theory. Apart from the conceptual
problems at its foundation due to the unresolved measurement issue
\cite{ish97}, it suffers from divergences that so far have proved
irremediable in the sense that we can work our way round them to
calculate what we need, but cannot remove them. The SAP proposal
has no accepted physical foundation, and also raises problematic
philosophical issues \cite{ear87}. I will not pursue it further
here.

\subsubsection{The relation to fundamental physical theories}
Many physicists go further, rejecting any Anthropic form of
reasoning. They regard it as a cop-out resorted to when physical
theories fail to give the needed answers, and seek to obtain a
full answer from physics alone \cite{sco05,sus05}. One possibility
is that there is a fundamental theory of everything that
determines the nature of physics completely, with no arbitrary
parameters left, and this still to be discovered theory just
happens to be of such a nature as to admit life.

However in this case the Anthropic issue returns with a vengeance:
\emph{How could it be that such a theory, based for example on
variational principles and the specific invariance groups of
particle physics, could just happen to lead to biophilic parameter
values}? There is no clear way to answer such a question.
Uniqueness of fundamental physics resolves the parameter freedom
only at the expense of creating an even deeper mystery, with no
way of resolution apparent. In effect, the nature of the unified
fundamental force would be pre-ordained to allow, or even
encourage, the existence of life; but there would be no apparent
reason why this should be so.

A second possibility is that physics allows many effective
theories with varying parameters --- some form of multiverse, as
for example may be implied by string theory \cite{sus03,sus05}. If
these varying options are all equally real, life can occur because
in some cases the parameters will lie in the restricted biophilic
regime. Thus from this viewpoint the Anthropic idea is intimately
linked with the existence of multiverses, which provide a
legitimate domain for their application. We will turn to an
examination of multiverses in the next section, but before doing
so we will consider the range of metaphysical options for
resolving the anthropic question.

\subsubsection{The metaphysical options}
\label{sec:options} To make progress on the Anthropic issue, we
have to seriously consider the nature of ultimate causation: What
is the fundamental cause for the phenomena we see? If we pursue
the chain of physical cause and effect to its conclusion, we are
still left with the question: \emph{Why did this occur, and not
something else}? Whatever the reason is, it is the ultimate cause
we are seeking. Note that we are here leaving the terrain of
science itself, and starting to probe the domain of metaphysics
--- the foundations of science and indeed of existence. As noted
above, one can simply decide not to pursue such issues. If we do
continue to question, there appear to be basically six approaches
to the issue of ultimate causation: namely Random Chance,
Necessity, High Probability, Universality, Cosmological Natural
Selection, and Design. We briefly consider these in turn.

\textbf{Option 1}: \emph{Random Chance, signifying nothing}. The initial
conditions in the Universe just happened, and led to things being the way
they are now, by pure chance. Probability does not apply. There is no
further level of explanation that applies; searching for `ultimate causes'
has no meaning.

This is certainly logically possible, but not satisfying as an explanation,
as we obtain no unification of ideas or predictive power from this approach.
Nevertheless some implicitly or explicitly hold this view.

\textbf{Option 2}: \emph{Necessity}. Things have to be the way they
are; there is no other option. The features we see and the laws
underlying them are demanded by the unity of the Universe: coherence
and consistency require that things must be the way they are; the
apparent alternatives are illusory. Only one kind of physics is
self-consistent: all logically possible universes must obey the same
physics.

To really prove this would be a very powerful argument,
potentially leading to a self-consistent and complete scientific
view. But we can imagine alternative universes! --- why are they
excluded? Furthermore we run here into the problem that we have
not succeeded in devising a fully self-consistent view of physics:
neither the foundations of quantum physics nor of mathematics are
on a really solid consistent basis. Until these issues are
resolved, this line cannot be pursued to a successful conclusion.

\textbf{Option 3}: \emph{High probability}. Although the structure
of the Universe appears very improbable, for physical reasons it is
in fact highly probable.

These arguments are only partially successful, even in their own
terms. They run into problems if we consider the full set of
possibilities: discussions proposing this kind of view actually
implicitly or explicitly restrict the considered possibilities
\emph{a priori}, for otherwise it is not very likely the Universe
will be as we see it. Besides, we do not have a proper measure to
apply to the set of initial conditions, enabling us to assess
these probabilities. Furthermore, as discussed above, see
Sec.\ref{sec:unique}, application of probability arguments to the
Universe itself is dubious, because the Universe is unique.
Despite these problems, this approach has considerable support in
the scientific community, for example it underlies the chaotic
inflationary proposal (Sec.\ref{sec:inflate}). It attains its
greatest power in the context of the assumption of universality:

\textbf{Option 4}: \emph{Universality}. This is the stand that ``All that is
possible, happens": an ensemble of universes or of disjoint expanding
universe domains is realized in reality, in which all possibilities occur
\cite{ree99,ree02,teg03a}. In its full version, the anthropic principle is
realized in both its strong form (if all that is possible happens, then life
must happen) and its weak form (life will only occur in some of the
possibilities that are realized; these are picked out from the others by the
WAP, viewed as a selection principle). There are four ways this has been
pursued.

\textbf{1:} \emph{Spatial variation}. The variety of expanding
universe domains is realised in space through random initial
conditions, as in chaotic inflation (Sec.\ref{sec:inflate}). While
this provides a legitimate framework for application of
probability, from the viewpoint of ultimate explanation it does
not really succeed, for there is still then one unique Universe
whose (random) initial conditions need explanation. Initial
conditions might be globally statistically homogeneous, but also
there could be global gradients in some physical quantities so
that the Universe is not statistically homogeneous; and these
conditions might be restricted to some domain that does not allow
life. It is a partial implementation of the ensemble idea; insofar
as it works, it is really a variant of the ``high probability"
idea mentioned above. If it was the more or less unique outcome of
proven physics, then that would provide a good justification; but
the physics underlying such proposals is not even uniquely
defined, much less tested. Simply claiming a particular scalar
field with some specific stated potential exists does not prove
that it exists!

\textbf{2}: \emph{Time variation}. The variety of expanding universe
domains could be realised across time, in a universe that has many
expansion phases (a Phoenix universe), whether this occurs globally
or locally. Much the same comments apply as in the previous case.

\textbf{3}: \emph{Quantum Mechanical}. It could occur through the
existence of the Everett-Wheeler ``many worlds" of quantum
cosmology, where all possibilities occur through quantum branching
\cite{deu98}. This is one of the few genuine alternatives proposed
to the Copenhagen interpretation of quantum mechanics, which leads
to the necessity of an observer, and so potentially to the Strong
Anthropic interpretation considered above (see
Sec.\ref{sec:anthropic}). The many-worlds proposal is
controversial: it occurs in a variety of competing formulations
\cite{ish97}, none of which has attained universal acceptance. The
proposal does not provide a causal explanation for the particular
events that actually occur: if we hold to it, we then have to
still explain the properties of the particular history we observe
(for example, why does our macroscopic universe have high
symmetries when almost all the branchings will not?). And above
all it is apparently untestable: there is no way to experimentally
prove the existence of all those other branching universes,
precisely because the theory gives the same observable predictions
as the standard theory.

\textbf{4}: \emph{Completely disconnected}. They could occur as
completely disconnected universes: there really is an ensemble of
universes in which all possibilities occur, without any connection
with each other \cite{lew86,ree02,teg03a}. A problem that arises
then is, What determines what is possible? For example, what about
the laws of logic themselves? Are they inviolable in considering
all possibilities? We cannot answer, for we have no access to this
multitude of postulated worlds. We explore this further below
(Sec.\ref{sec:multiverse}).

In all these cases, major problems arise in relating this view to
testability and so we have to query the meaningfulness of the
proposals as scientific explanations. They all contradict Ockham's
razor: we ``solve" one issue at the expense of envisaging an
enormously more complex existential reality. Furthermore, they do
not solve the ultimate question: \emph{Why does this ensemble of
universes exist}? One might suggest that ultimate explanation of
such a reality is even more problematic than in the case of single
universe. Nevertheless this approach has an internal logic of its
own which some find compelling. We consider this approach further
below, see Sec.\ref{sec:multiverse}.

\textbf{Option 5}: \emph{Cosmological Natural Selection}. If a process of
re-expansion after collapse to a black hole were properly established, it
opens the way to the concept not merely of evolution of the Universe in the
sense that its structure and contents develop in time, but in the sense that
the Darwinian selection of expanding universe regions could take place, as
proposed by Smolin \cite{smo92}. The idea is that there could be collapse to
black holes followed by re-expansion, but with an alteration of the
constants of physics through each transition, so that each time there is an
expansion phase, the action of physics is a bit different. The crucial point
then is that some values of the constants will lead to production of more
black holes, while some will result in less. This allows for evolutionary
selection favouring the expanding universe regions that produce more black
holes (because of the favourable values of physical constants operative in
those regions), for they will have more ``daughter" expanding universe
regions. Thus one can envisage natural selection favouring those physical
constants that produce the maximum number of black holes.

The problem here is twofold. First, the supposed `bounce'
mechanism has never been fully explicated. Second, it is not clear
--- assuming this proposed process can be explicated in detail -
that the physics which maximizes black hole production is
necessarily also the physics that favours the existence of life.
If this argument could be made water-tight, this would become
probably the most powerful of the multiverse proposals.

\textbf{Option 6}: \emph{Purpose or Design}. The symmetries and
delicate balances we observe require an extraordinary coherence of
conditions and cooperation of causes and effects, suggesting that
in some sense they have been purposefully designed. That is, they
give evidence of intention, both in the setting of the laws of
physics and in the choice of boundary conditions for the Universe.
This is the sort of view that underlies Judaeo-Christian theology.
Unlike all the others, it introduces an element of meaning, of
signifying something. In all the other options, life exists by
accident; as a chance by-product of processes blindly at work.

The prime disadvantage of this view, from the scientific
viewpoint, is its lack of testable scientific consequences
(``Because God exists, I predict that the density of matter in the
Universe should be x and the fine structure constant should be
y"). This is one of the reasons scientists generally try to avoid
this approach. There will be some who will reject this possibility
out of hand, as meaningless or as unworthy of consideration.
However it is certainly logically possible. The modern version,
consistent with all the scientific discussion preceding, would see
some kind of purpose underlying the existence and specific nature
of the laws of physics and the boundary conditions for the
Universe, in such a way that life (and eventually humanity) would
then come into existence through the operation of those laws, then
leading to the development of specific classes of animals through
the process of evolution as evidenced in the historical record.
Given an acceptance of evolutionary development, it is precisely
in the choice and implementation of particular physical laws and
initial conditions, allowing such development, that the profound
creative activity takes place; and this is where one might
conceive of design taking place.\footnote{This is not the same as
the view proposed by the `Intelligent Design' movement. It does
not propose that God tweaks the outcome of evolutionary
processes.}

However from the viewpoint of the physical sciences \emph{per se},
there is no reason to accept this argument. Indeed from this
viewpoint there is really no difference between design and chance,
for they have not been shown to lead to different physical
predictions.

\subsubsection{Metaphysical Uncertainty}\label{sec:uncertain}
In considering ultimate causation underlying the anthropic question, in the
end we are faced with a choice between one of the options above. As pointed
out already by Kant and Hume, although we may be able to argue strongly for
one or other of them, we cannot \emph{prove} any of the options are correct
\cite{hum93}.

\textbf{Thesis G2: Metaphysical uncertainty remains about ultimate causation
in cosmology}. \textit{We cannot attain certainty on the underlying
metaphysical cosmological issues through either science or philosophy.}

\noindent If we look at the anthropic question from a purely
scientific basis, we end up without any resolution, basically
because science attains reasonable certainty by limiting its
considerations to restricted aspects of reality; even if it
occasionally strays into the area of ultimate causation, it is not
designed to deal with it. By itself, it cannot make a choice
between these options; there is no relevant experiment or set of
observations that can conclusively solve the issue. Thus a broader
viewpoint is required to make progress, taking into account both
the scientific and broader considerations. The issue is of a
philosophical rather than scientific nature. One important issue
that then arises is what kind of data is relevant to these
philosophical choices, in addition to that which can be
characterized as purely scientific data (Sec.\ref{sec:ultimate}).

\subsection{Issue H:\ The possible existence of multiverses}
\label{sec:multiverse} If there is a large enough ensemble of
numerous universes with varying properties, it may be claimed that
it becomes virtually certain that some of them will just happen to
get things right, so that life can exist; and this can help
explain the fine-tuned nature of many parameters whose values are
otherwise unconstrained by physics \cite{ree99,ree02}. As
discussed in the previous section, there are a number of ways in
which, theoretically, multiverses could be realized
\cite{lew86,teg03a}. They provide a way of applying probability to
the universe \cite{sci71,bos02} (because they deny the uniqueness
of the universe). However, there are number of problems with this
concept. Besides, this proposal is observationally and
experimentally untestable; thus its scientific status is
debatable.

\subsubsection{Definition}
In justifying multiverses, it is often stated that `all that can
occur, occurs' (or similarly). However that statement does not
adequately specify a multiverse. To define a multiverse properly
requires two steps \cite{ellstokir04}. First, one needs to specify
what is conceived of in the multiverse, by defining a
\textit{possibility space}: a space $\mathcal{M}$ of all possible
universes, each of which can be described in terms of a set of
states $s$ in a state space $\mathcal{S}$. Each universe $m$ in
$\mathcal{M}$ will be characterized by a set of distinguishing
parameters $p$, which are coordinates on $\mathcal{S}.$ Choices
are needed here. In geometrical terms, will it include only
Robertson--Walker models, or more general ones (e.g. Bianchi
models, or models without symmetries)? In gravitational terms,
will it include only General Relativity, or also brane theories,
models with varying G, loop quantum gravity models, string theory
models with their associated possibility `landscapes', and models
based on the wave function of the universe concept? Will it allow
only standard physics but with varying constants, or a much wider
spectrum of physical possibilities, e.g. universes without quantum
theory, some with five fundamental forces instead of four, and
others with Newtonian gravity?\ Defining the possibility space
means making some kind of assumptions about physics and geometry
that will then apply across the whole family of models considered
possible in the multiverse, and excluding all other possibilities.

Second, one needs to specify which of the possible universes are
physically realized in the multiverse, and how many times each one
occurs. \textit{A multiverse must be a physically realized
multiverse and not a hypothetical or conceptual one if it is to
have genuine explanatory power}. Thus one needs a distribution
function $f(m)$ specifying how many times each type of possible
universe $m$ in $\mathcal{M}$ is realised. The function $f(m)$
expresses the contingency in any actualization. Things could have
been different! Thus, $f(m)$ describes a specific \textit{ensemble
of universes} or \textit{multiverse} envisaged as being realised
out of the set of possibilities. For example, $f(m)$ might be
non-zero for all possible values of all the parameters $p$ (`all
that can happen, happens'); but it could be that $f$ describes a
multiverse where there are $10^{100}$ identical copies of one
particular universe (the realization process finds a particularly
successful recipe, and then endlessly replicates it).

Additionally we need a measure $d\pi$ that enables this function
to determine numbers and probabilities of various properties in
the multiverse: the number of universes corresponding to a set of
parameter increments will be $ \mathrm{d}N$ given by
\begin{equation}
\mathrm{d}N=f(m) d\pi  \label{dist1}
\end{equation}
for continuous parameters; for discrete parameters, we add in the
contribution from all allowed parameter values. The total number of
universes $N$ in the ensemble will be given by
\begin{equation}
N=\int_\mathcal{M} f(m)d\pi  \label{dist2}
\end{equation}
(which will often diverge), where the integral ranges over all
allowed values of the member parameters and we take it to include
all relevant discrete summations. The expectation value $P$ of a
quantity $p(m)$ defined on the set of universes will be given by
\begin{equation}
P=\int_\mathcal{M} p(m)f(m)d\pi.  \label{prob}
\end{equation}

These three elements (the possibility space, the measure, and the
distribution function) must all be clearly defined in order to give a proper
specification of a multiverse \cite{ellstokir04}. This is almost never done.

\subsubsection{Non-uniqueness:\ Possibilities}
There is non-uniqueness at both steps. Stating ``all that is
possible, happens" does not resolve what is possible. The concept
of multiverses is not well defined until the space of possible
universes has been fully characterized; it is quite unclear how to
do this uniquely. The issue of what is to be regarded as an
ensemble of `all possible' universes can be manipulated to produce
any result you want, by redefining what is meant by this phrase
--- standard physics and logic have no necessary sway over them:
what I envisage as `possible' in such an ensemble may be denied by
you. What super-ordinate principles are in operation to control
the possibilities in the multiverse, and why? A key point here is
that \textit{our understandings of the possibilities are always of
necessity arrived at by extrapolation from what we know,} and my
imagination may be more fertile than yours, and neither need
correspond to what really exists out there --- if indeed there is
anything there at all. Do we include only
\begin{itemize}
    \item \emph{Weak variation}:\ e.g. only the values of the constants of
physics are allowed to vary? This is an interesting exercise but
is certainly not an implementation of the idea `all that can
happen, happens'. It is an extremely constrained set of
variations.
    \item  \emph{Moderate variation}: different symmetry groups, or numbers
of dimensions, etc. We might for example consider the possibility
landscapes of string theory \cite{fre05} as realistic indications
of what may rule multiverses \cite{sus03,sus05}. But that is very
far indeed from `all that is possible', for that should certainly
include spacetimes not ruled by string theory.
    \item \emph{Strong variation}: different numbers and kinds of forces,
universes without quantum theory or in which relativity is untrue
(e.g. there is an aether), some in which string theory is a good
theory for quantum gravity and others where it is not, some with
quite different bases for the laws of physics (e.g. no variational
principles).
    \item \emph{Extreme variation}: universes where physics is not well
described by mathematics; with different logic; universes ruled by
local deities; allowing magic as in the Harry Potter series of
books; with no laws of physics at all? Without even mathematics or
logic?
\end{itemize}
\noindent Which is claimed to be the properties of the multiverse,
and why? We can express our dilemma here through the paradoxical
question: \textit{\ Are the laws of logic necessary in all
possible universes?}

\subsubsection{Non-uniqueness: existence and causation}
A specific multiverse is defined by specifying the distribution
function $f(m)$ of actually realized universes. It is unclear what
mechanism can underlie such a distribution, and any proposal for
such a mechanism is completely untestable. We need some indication
as to \textit{what determines existence within the possibilities
defined by the supposed possibility space}:\ What decides how many
times each one happens? Unless we understood the supposed
underlying mechanisms we can give no serious answer; and there is
no prospect whatever of testing any proposed mechanism. The
mechanisms supposed to underlie whatever regularities there are in
the multiverse must pre-exist the existence of not merely this
universe but also every other one. If one assumes a universe that
is connected in the large but is locally separated into causally
disconnected domains with different physical properties(as in
chaotic inflation), one attains a plausible picture of a creation
mechanism that can underlie an effective multiverse --- but at the
expense of supposing the validity of untested and perhaps
untestable physics. Because of this one does not obtain a
specification of a unique multiverse: the physics could be
different than what we assumed.

\subsubsection{Explanatory power}
What explanatory power do we get in return for these problems? It
has been suggested they explain the parameters of physics and of
cosmology and in particular the very problematic observed value of
the cosmological constant \cite{wei00,sus05}. The argument goes as
follows: assume a multiverse exists; observers can only exist in
one of the highly improbable biophilic outliers where the value of
the cosmological constant is very small \cite{har04}. A similar
argument has been proposed for neutrino masses \cite{tegetal03}.
If the multiverse has many varied locations with differing
properties, that may indeed help us understand the Anthropic
issue: some regions will allow life to exist, others will not
\cite{bartip84,les89}. This does provide a useful modicum of
explanatory power. However it is far from conclusive. Firstly, it
is unclear why the multiverse should have the restricted kinds of
variations of the cosmological constant assumed in the various
analyses mentioned. If we assume `all that can happen, happens'
the variations will not be of that restricted kind; those analyses
will not apply.

Secondly, ultimate issues remain: Why does this unique larger
whole have the properties it does? \emph{Why this multiverse
rather than any other one}?\ Why is it a multiverse that allows
life to exist? Many multiverses will not allow any life at all. To
solve this, we can propose an \emph{ensemble of ensembles of
universes}, with even greater explanatory power and even less
prospect of observational verification; and so on. The prospect of
an infinite regress looms. Indeed if we declare (as suggested at
the start of this article) that `the Universe' is the total of all
that physically exists, then when an ensemble of expanding
universe domains exists, whether causally connected or not, that
ensemble itself should be called `the Universe', for it is then
the totality of physically existing entities. All the foundational
problems for a single existing universe domain recur for the
multiverse --- because when properly considered, it is indeed the
Universe!

\subsubsection{Testability}
\label{sec:inftest} If an ensemble exists with members not connected
in any physical way to the observable universe, then we cannot
interact with them in any way nor observe them, so we can say
anything we like about them without fear of disproof.\footnote{But
there are counter arguments by Leibniz \cite{wil89} and Lewis
(\cite{lew86} section 2.4, pp. 108-115).} Thus any statements we
make about them can have no solid scientific or explanatory status;
they are totally vulnerable to anyone else who claims an ensemble
with different properties (for example claiming different kinds of
underlying logics are possible in their ensemble, or claiming many
physically effective gods and devils in many universes in their
ensemble).

\textbf{Thesis H1: Multiverse proposals are unprovable by
observation or experiment, but some self-consistency tests are
possible}. \textit{Direct observations cannot prove or disprove that
a multiverse exists, for the necessary causal relations allowing
observation or testing of their existence are absent. Their
existence cannot be predicted from known physics, because the
supposed causal or pre-causal processes are either unproven or
indeed untestable. However some self-consistency conditions for
specific multiverse models can be tested.}

\noindent Any proposed physics underlying a multiverse proposal,
such as Coleman-de Luccia tunneling \cite{coldel}, will be an
extrapolation of known physics; but the validity of that major
extrapolation to cosmology is untestable.

Attempts have been made to justify the existence of multiverses as
testable firstly via Rees' `slippery slope' argument \cite{ree02}.
This runs as follows: we can reasonably assume galaxies that we
cannot see exist outside the visual horizon (Sec.\ref{sec:epist});
why not extend this argument by small steps to show totally
disconnected universes exist? The problem is that this assumes a
continuity of existence that does not hold good. The domain outside
our horizon is assumed to exist with similar properties to those
inside because they are a continuous extension of it and have a
largely common causal origin; their nature can be inferred from what
we can see. Disconnected multiverse domains are assumed to have
quite different properties, and their nature cannot be inferred from
what we can see as there is no continuity or causal connection.

Secondly, several authors (Leslie \cite{les89}, Weinberg
\cite{wei00}, and Rees \cite{ree02} for example) have used
arguments based on the idea that the universe is no \emph{more}
special than it has to be; a form of ``speciality argument.''
According to Rees, if our universe turns out to be \textit{\ even
more specially} tuned than our presence requires, the existence of
a multiverse to explain such ``over-tuning'' would be refuted; but
the actual universe is not more special than this, so the
multiverse is not refuted.

In more detail: naive quantum physics predicts $\Lambda$ to be
very large. But our presence in the universe requires it to be
small enough that galaxies and stars can form, so $\Lambda$ must
obviously be below that galaxy-forming threshold. If our universe
belongs to an ensemble in which $\Lambda$ was equally likely to
take any value in the biophilic region (the uniform probability
assumption),\footnote{The probability distribution for $\Lambda$
will plausibly peak far away from the biophilic region, tailing
down to a low value that will be approximately constant in that
narrow region, cf.\cite{har04}.} then we would not expect it to be
too far below this threshold. This is because, if it's too far
below the threshold, the probability of randomly choosing that
universe in the ensemble becomes very small
--- there are very few universes with such small values of
$\Lambda$ in the biophilic subset of the ensemble. That is, it
would be more likely that any bio-friendly universe in the
ensemble would have a value of $\Lambda$ closer to the threshold
value. Present data on this value indicates that it is not too far
below the threshold. Thus, our universe is not markedly more
special that it needs to be as far as $\Lambda$ is concerned, and
so explaining its fine-tuning by existence of a multiverse is
legitimate.

Is this argument compelling? It is a reasonable test of
consistency for a multiverse that is known to exist, so that
probability considerations apply; but they do not apply if there
is no multiverse (Sec.\ref{sec:unique}). Additionally, probability
considerations cannot ever be \emph{conclusive}. Indeed,

\textbf{Thesis H2: Probability-based arguments cannot demonstrate the
existence of multiverses}. \textit{Probability arguments cannot be used to}
prove \textit{the existence of a multiverse,\ for they are only applicable
if a multiverse exists. Furthermore probability arguments can never prove
anything for certain, as it is not possible to violate any probability
predictions, and this is} a fortiori \textit{so when there is only one case
to consider, so that no statistical observations are possible}.

\noindent All one can say on the basis of probability arguments is
that some specific state is very improbable. But this does not
prove it is impossible, indeed if it is stated to have a low
probability, that is precisely a statement that it is possible.
Thus such arguments can at best only give plausibility indications
even when they are applicable. The assumption that probability
arguments can be conclusive is equivalent to the claim that the
universe is generic rather than special; but whether this is so or
not is precisely the issue under debate (see Thesis \textbf{D3}).
The argument is useful as a plausibility argument for a
multiverse, but is not \emph{proof} of its existence.

Finally, it has been proposed that the existence of multiverses is
an inevitable consequence of the universe having infinite space
sections \cite{teg03a,sei04}, because that leads to infinite spatial
repetition of conditions (cf. \cite{ellbru79}). But this supposed
spatial infinity is an untested philosophical assumption, which
certainly cannot be observationally proven to be correct. Apart from
the existence of horizons preventing confirmation of this
supposition, even if the entire universe were observable, proving it
correct would still not be possible because by definition counting
an infinite number of objects takes an infinite amount of time. This
is an untestable philosophical argument, not an empirically testable
one; furthermore, it can be argued to be implausible
(Sec.\ref{sec:infinity}). Indeed current data suggest it is not the
case; this is the one good consistency test one can use for some
multiverse proposals (Sec.\ref{sec:test}).

\subsubsection{Explanation vs Testability}
The argument that this infinite ensemble actually exists can be
claimed to have a certain explanatory economy, although others
would claim that Occam's razor has been completely abandoned in
favour of a profligate excess of existential multiplicity,
extravagantly hypothesized in order to explain the one universe
that we do know exists. Certainly the price is a lack of
testability through either observations or experiment --- which is
usually taken to be an essential element of any serious scientific
theory.\footnote{In \cite{stoetal04}, the framework and conditions
under which the multiverse hypothesis would be testable within a
retroductive framework, given the rigorous conditions formulated
in that paper; are indicated; these conditions are not fulfilled.}
It is not uniquely definable nor determinable, and there is a
complete loss of verifiability. There is no way to determine the
properties of any other universe in the multiverse if they do
indeed exist, for they are forever outside observational reach.
The point is that there is not just an issue of showing a
multiverse exists. If this is a scientific proposition one needs
to be able to show which specific multiverse exists; but there is
no observational way to do this. Indeed if you can't show
\textit{which particular} one exists, it is doubtful you have
shown \textit{any} one exists.

What does a claim for such existence mean in this context? Gardner
puts it this way: ``There is not the slightest shred of reliable
evidence that there is any universe other than the one we are in. No
multiverse theory has so far provided a prediction that can be
tested. As far as we can tell, universes are not even as plentiful
as even \textit{two} blackberries" \cite{gar03}.\footnote{This
contrasts strongly, for example, with Deutsch's and Lewis's defence
of the concept \cite{deu98,lew86}. Lewis defends the thesis of
``modal realism": that the world we are part of is but one of a
plurality of worlds.}

\textbf{Thesis H3: Multiverses are a philosophical rather than
scientific proposal}. \textit{The idea of a multiverse provides a
possible route for the explanation of fine tuning. But it is not
uniquely defined, is not scientifically testable apart from some
possible consistency tests, and in the end simply postpones the
ultimate metaphysical questions.}

\noindent The definitive consistency tests on some multiverse
proposals (Sec.\ref{sec:test}) are \emph{necessary} conditions for
those specific multiverse proposals, but are hardly by themselves
indications that the multiverse proposal is true. The drive to
believe this is the case comes from theoretical and philosophical
considerations (see e.g. \cite{sus05}) rather than from data. The
claim an ensemble physically exists\footnote{As opposed to
consideration of an explicitly hypothetical such ensemble, which can
indeed be useful, see Sec.\ref{sec:alternate}).} is problematic as a
proposal for scientific explanation, if science is taken to involve
testability. Indeed, adopting these explanations is a triumph of
theory over testability \cite{gar03}, but the theories being assumed
are not testable. It is therefore a metaphysical choice made for
philosophical reasons. That does not mean it is unreasonable (it can
be supported by quite persuasive plausibility arguments); but its
lack of scientific status should be made clear.

\subsubsection{Observations and disproof} \label{sec:test}
Despite the gloomy prognosis given above, there are some specific
cases where the existence of a chaotic inflation (multi-domain) type
scenario (Sec.\ref{sec:inflate}) can be \emph{dis}proved. These are
firstly when we live in a `small universe' where we have already
seen right round the universe (Sec.\ref{sec:small}), for then the
universe closes up on itself in a single FL-like domain, so that no
further such domains can exist that are causally connected to us in
a single connected spacetime. This `small universe' situation is
observationally testable (Sec.\ref{sec:small}); its confirmation
would disprove the usual chaotic inflationary scenario, but not a
truly `disconnected' multiverse proposal, for that cannot be shown
to be false by any observation. Neither can it be shown to be true.
Secondly, many versions of chaotic inflation, for example those
involving Coleman-de Luccia tunneling \cite{coldel} from a de Sitter
spacetime, demand $k=-1 \Leftrightarrow \Omega_0 <1$
\cite{fre05,sus05}. This requirement is currently marginally
disproved by the $2-\sigma$ bounds on $\Omega_0$ when WMAP
observations are combined with the best other available data
(Sec.\ref{sec:cbr}). The best current data is marginally consistent
with $k=-1$, but the value indicated most strongly by that data is
$k=+1$, indicating finite closed space sections rather than an
infinite multiverse such as that advocated by Susskind \emph{et al}
\cite{fre05,sus05}.

\subsubsection{Physical or biological paradigms --- Adaptive Evolution?}
Given that the multiverse idea must in the end be justified
philosophically rather than by scientific testing, is there a
philosophically preferable version of the idea? One can suggest
there is:\ greater explanatory power is potentially available by
introducing the major constructive principle of biology into
cosmology, namely adaptive evolution, which is the most powerful
process known that can produce ordered structure where none
pre-existed. This is realized in principle in Lee Smolin's idea
(Sec.\ref{sec:options}) of Darwinian adaptation when collapse to
black holes is followed by re-expansion, but with an alteration of
the constants of physics each time, so as to allow for
evolutionary selection towards those regions that produce the
maximum number of black holes. The idea needs development, but is
very intriguing:

\textbf{Thesis H4: The underlying physics paradigm of cosmology could be
extended to include biological insights}. \textit{The dominant paradigm in
cosmology is that of theoretical physics. It may be that it will attain
deeper explanatory power by embracing biological insights, and specifically
that of Darwinian evolution. The Smolin proposal for evolution of
populations of expanding universe domains \cite{smo92} is an example of this
kind of thinking}.

\noindent The result is different in important ways from standard
cosmological theory precisely because it embodies in one theory
three of the major ideas of the last century, namely (i) Darwinian
evolution of populations through competitive selection, (ii) the
evolution of the universe in the sense of major changes in its
structure associated with its expansion, and (iii) quantum theory,
underlying the only partly explicated mechanism supposed to cause
re-expansion out of collapse into a black hole. There is a great
contrast with the theoretical physics paradigm of dynamics governed
simply by variational principles shaped by symmetry considerations.
It seems worth pursuing as a very different route to the
understanding of the creation of structure.\footnote{cf. Chapter 13
of Susskind \cite{sus05}.}

\subsection{Issue I: Natures of Existence}
\label{sec:existence} Underlying all this is the issue of natures
of existence, which has a number of aspects, relating from the
purely physical to more metaphysical issues.

\subsubsection{Physical existence: kinds of matter}
\label{sec:darken}Unsolved key issues for physical cosmology relate
to what kind of matter and/or fields exist. While we understand
matter in the solar system quite well, at present we do not
understand most of what exists in the universe at large:\

\textbf{Thesis I1: We do not understand the dominant dynamical
matter components of the universe at early or late times}. \textit{A
key goal for physical cosmology is determining the nature of the
inflaton, of dark matter, and of dark energy. Until this is done,
the causal understanding of cosmology is incomplete, and in
particular the far future fate of the universe is unknown. }

\noindent This is the core activity of much work in cosmology at
present. Until they are all explicated, cosmology is not properly
linked to physics, and the nature of the matter that dominates the
dynamics of the universe is unknown. Its explication is surely one
of the key concerns of cosmology \cite{dur05}. A key requirement
is that even if we cannot experimentally verify the proposed
nature of the matter, at least it should be physically plausible.
This appears not to be the case for some current proposals, e.g.
so-called `phantom matter' which has negative kinetic energy
terms.

The far future fate of the universe depends crucially on the
effective equation of state for dark matter (`quintessence'). But
the problem is that even if we can determine these properties at
the present time (for one particular range of parameter values),
this does not necessarily guarantee what they will be in the far
future (for a quite different range of parameter values that are
probably outside the range of possible experimental test).
Furthermore adjusting a `dark energy' model to fit the supernova
data does not determine the underlying physics. One can fit any
monotonic evolution $S(t)$ with a suitable choice of the equation
of state function $p(\mu)$. Specifically, for any $S(t)$ and any
$k$ we define $\mu (t)$ and $p(t)$ by
\begin{equation}
\kappa \mu (t)=3\left[ \frac{\dot{S}^{2}(t)}{S^{2}(t)} +
\frac{k}{S^{2}(t)} \right] , \,\,\kappa p(t) =
\left[\frac{\dot{S}^{2}(t)}{S^{2}(t)} + \frac{k}{S^{2}(t)}\right]
-2\frac{\ddot{S}(t)}{S(t)}\,,  \label{p}
\end{equation}
then (\ref{Fried}), (\ref{Ray}) will be exactly satisfied, and we
have `solved' the field equations for this arbitrarily chosen
monotonic evolution $S(t)$. If we can observationally determine
the form of $S(t)$, for example from $(m,z)$--curves associated
with supernovae data, this is essentially how we can then
determine that some kind of `dark energy' or `quintessence' is
required to give that evolution, and we can find the equation of
state implied by eliminating $t$ between these two equations. This
is, however, not a \emph{physical} explanation until we have
either in some independent experimental test demonstrated that
matter of this form exists, or have theoretically shown why this
matter or field has the form it does in some more fundamental
terms than simply a phenomenological fit. If we assume the matter
is a scalar field, the kinetic energy term $\dot{\phi}^{2}$
implied by (\ref{scalar}), (\ref{p}) may be negative --- which is
the case for so-called `shadow matter' models proposed recently by
some worker. If normal physics criteria are applied, this is a
proof that this kind of matter is unphysical, rather than an
identification of the nature of the dark energy.

\subsubsection{Existence of Infinities}
\label{sec:infinity} The nature of existence is significantly
different if there is a finite amount of matter or objects in the
universe, as opposed to there being an infinite quantity in
existence. Some proposals claim there may be an infinite number of
universes in a multiverse and many cosmological models have
spatial sections that are infinite, implying an infinite number of
particles, stars, and galaxies. However, infinity is quite
different from a very large number! Following David Hilbert
\cite{hil64}, one can suggest these unverifiable proposals cannot
be true: the word `infinity' denotes a quantity or number that can
never be attained, and so will never occur in physical
reality.\footnote{An intriguing further issue is the dual
question: Does the quantity zero occur in physical reality? This
is related to the idea of physical existence of nothingness, as
contrasted with a vacuum \cite{sei00}. A vacuum is not nothing!
(cf. \cite{sus05}).} He states

\noindent \textit{``Our principal result is that the infinite is
nowhere to be found in reality. It neither exists in nature nor
provides a legitimate basis for rational thought . . . The role
that remains for the infinite to play is solely that of an idea .
. . which transcends all experience and which completes the
concrete as a totality . . ."} (\cite{hil64}, p. 151).

\noindent This suggests ``infinity'' cannot be arrived at, or
realized, in a concrete physical setting; on the contrary, the
concept itself implies its inability to be realized!\footnote{For a
contrasting view, see Bernadete \cite{ber64}.}

\textbf{Thesis I2: The often claimed physical existence of infinities is
questionable}. \textit{The claimed existence of physically realized
infinities in cosmology or multiverses raises problematic issues. One can
suggest they are unphysical; in any case such claims are certainly
unverifiable.}

\noindent This applies in principle to both small and large scales
in any single universe:
\begin{itemize}
    \item The existence of a physically existing spacetime continuum
    represented by a real (number) manifold at the
micro-level contrasts with quantum gravity claims of a discrete
spacetime structure at the Planck scale, which one might suppose
was a generic aspect of fully non-linear quantum gravity theories
\cite{rov04}. In terms of physical reality, this promises to get
rid of the uncountable infinities the real line continuum
engenders in all physical variables and fields.\footnote{To avoid
infinities entirely would require that nothing whatever is a
continuum in physical reality (since any continuum interval
contains an infinite number of points). Doing without that,
conceptually, would mean a complete rewrite of many things.
Considering how to do so in a way compatible with observation is
in my view a worthwhile project.} There is no experiment that can
\emph{prove} there is a physical continuum in time or space; all
we can do is test space-time structure on smaller and smaller
scales, but we cannot approach the Planck scale.
    \item Infinitely large space-sections at the macro-level raise problems
as indicated by Hilbert, and leads to the infinite duplication of
life and all events \cite{ellbru79}. We may assume space extends
forever in Euclidean geometry and in many cosmological models, but
we can never prove that any realised 3-space in the real universe
continues in this way --- it is an untestable concept, and the real
spatial geometry of the universe is almost certainly not Euclidean.
Thus Euclidean space is an abstraction that is probably not
physically real. The infinities supposed in chaotic inflationary
models derive from the presumption of pre-existing infinite
Euclidean space sections, and there is no reason why those should
necessarily exist. In the physical universe spatial infinities can
be avoided by compact spatial sections, resulting either from
positive spatial curvature, or from a choice of compact topologies
in universes that have zero or negative spatial curvature. Machian
considerations to do with the boundary conditions for physics
suggest this is highly preferable \cite{whe68}; and if one invokes
string theory as a fundamental basis for physics, then `dimensional
democracy' suggests the three large spatial dimensions should also
be compact, since the small (`compactified') dimensions are all
taken to be so. The best current data from CBR and other
observations (Sec.\ref{sec:cbr}) indeed suggest $k=+1$, implying
closed space sections for the best-fit FL model.
    \item The existence of an eternal universe implies that an
    infinite time actually
exists, which has its own problems: if an event happens at any
time $t_0$, one needs an explanation as to why it did not occur
before that time (as there was an infinite previous time available
for it to occur); and Poincar\'{e} eternal return (mentioned in
Sec.\ref{sec:start1}) will be possible if the universe is truly
cyclic. In any case it is not possible to \emph{prove} that the
universe as a whole, or even the part of the universe in which we
live, is past infinite; observations cannot do so, and the physics
required to guarantee this would happen (if initial conditions
were right) is untestable. Even attempting to prove it is future
infinite is problematic (we cannot for example guarantee the
properties of the vacuum into the infinite future --- it might
decay into a state corresponding to a negative effective
cosmological constant).
    \item It applies to the possible nature of a multiverse. Specifying the
geometry of a generic universe requires an infinite amount of
information because the quantities necessary to do so are fields
on spacetime, in general requiring specification at each point (or
equivalently, an infinite number of Fourier coefficients): they
will almost always not be algorithmically compressible. All
possible values of all these components in all possible
combinations will have to occur in a multiverse in which ``all
that can happen, does happen". There are also an infinite number
of topological possibilities. This greatly aggravates all the
problems regarding infinity and the ensemble. Only in highly
symmetric cases, like the FL solutions, does this data reduce to a
finite number of parameters, each of which would have to occur in
all possible values (which themselves are usually taken to span an
infinite set, namely the entire real line). Many universes in the
ensemble may themselves have infinite spatial extent and contain
an infinite amount of matter, with all the problems that entails.
To conceive of physical creation of an infinite set of universes
(most requiring an infinite amount of information for their
prescription, and many of which will themselves be spatially
infinite) is at least an order of magnitude more difficult than
specifying an existent infinitude of finitely specifiable objects.
\end{itemize}
One should note here particularly that problems arise in the
multiverse context from the continuum of values assigned by
classical theories to physical quantities. Suppose for example
that we identify corresponding times in the models in an ensemble
and then assume that \textit{all} values of the density parameter
and the cosmological constant occur at each spatial point at that
time. Because these values lie in the real number continuum, this
is a doubly uncountably infinite set of models. Assuming genuine
physical existence of such an uncountable infinitude of universes
is the antithesis of Occam's razor. But on the other hand, if the
set of realised models is either finite or countably infinite,
then almost all possible models are not realised. And in any case
this assumption is absurdly unprovable. We can't observationally
demonstrate a single other universe exists \cite{gar03}, let alone
an infinitude. The concept of infinity is used with gay abandon in
some multiverse discussions \cite{knoetal05}, without any concern
either for the philosophical problems associated with this
statement \cite{hil64}, or for its completely unverifiable
character. It is an extravagant claim that should be treated with
extreme caution.

\subsubsection{The Nature of the Laws of Physics}\label{sec:laws1}
Underlying all the above discussion is the basic concept of
ordered behaviour of matter, characterized by laws of physics of a
mathematical nature that are the same everywhere in the
universe.\footnote{The effective laws may vary from place to place
because for example the vacuum state varies \cite{sus05}; but the
fundamental laws that underlie this behaviour are themselves taken
to be invariant.} Three interlinked issues arise.

(i) \emph{What is the ontological nature of the laws of physics}:
descriptive, \emph{just characterizing the way things are, or}
prescriptive, \emph{enforcing them to be this way}? \cite{car04}.
If they are descriptive, the issue arising is, \emph{Why does all
matter have the same properties wherever it exists in the
universe}? Why are all electrons everywhere in the universe
identical, if the laws are only descriptive? If they are
prescriptive, then matter will necessarily be the same everywhere
(assuming the laws themselves are invariable); the issue arising
then is, \emph{In what way do laws of physics exist that enforce
themselves on the matter in the universe}? Do they for example
have an existence in some kind of Platonic space that controls the
nature of matter and existence? One can avoid talking about the
laws of physics \emph{per se} by instead considering the
\emph{space of possibilities} underlying what exists physically,
rigorously constraining the possible natures of what actually
comes into existence \cite{ell04}. This space is more or less
uniquely related to the underlying laws in the same way that the
space of solutions of differential equations is related to the
nature of the equations. This enables one to avoid the issue of
the ontology of the laws of physics, but does not solve it.

(ii) \emph{Why are the laws of physics so well explained by
mathematical descriptions}? If they are prescriptive, this deep
issue might be related to the suggested Platonic nature of the
space of mathematical reality \cite{pen04}. If they are
descriptive, then the mathematical expressions we use to
encapsulate them are just a convenient description but do not
reflect their ultimate nature. Many writings in physics and
cosmology seem to assume that their ultimate existential nature is
indeed mathematical --- perhaps a confusion of appearance and
reality (see Sec.\ref{sec:epist}).

(iii) \emph{Do they pre-exist the universe and control its coming
into being, or do they come into being with the universe}? This is
where this issue relates deeply to the nature of cosmology, and is
clearly related to the other two questions raised above. Many
theories of creation of the universe assume that all these laws,
or at least a basic subset, pre-exist the coming into being of the
physical universe, because they are presumed to underlie the
creation process, for example the entire apparatus of quantum
field theory is often taken for granted as pre-existing our
universe (Sec.\ref{sec:origins}). This is of course an unprovable
proposition

\textbf{Thesis I3: A deep issue underlying the nature of cosmology
is the nature of the laws of physics}. \textit{The nature of the
possibility space for physical existence is characterized by the
laws of physics. However it is unclear if these laws are
prescriptive or descriptive; whether they come into being with
space-time and matter, or pre-exist them.}

\subsubsection{`Ultimate Reality'}\label{sec:ultimate}
Philosophers have debated for millennia whether the ultimate
nature of existence is purely material, or embodies some form of
rationality (`Logos') and/or purpose (`Telos'). \emph{What in the
end underlies it all}? Is the ultimate nature of the universe
purely material, or does it in some way have an element of the
mental? (cf. Sec.\ref{sec:options}). That profound debate is
informed by physical cosmology, but cannot be resolved by the
physical sciences alone (Sec.\ref{sec:uncertain}). Here, I will
make just two comments on this deep issue.

Firstly, even in order to understand just the material world, it
can be claimed that one needs to consider forms of existence other
than the material only --- for example a Platonic world of
mathematics and a mental world, both of which can be claimed to
exist and be causally effective in terms of affecting the material
world \cite{ell04,pen04}. Our understanding of local causation
will be incomplete unless we take them into account.

Secondly, in examining these issues one needs to take into account
data about the natures of our existence that come from our daily
lives and the broad historical experience of humanity (our
experiences of ethics and aesthetics, for example), as well as
those discoveries attained by the scientific method. Many writings
claim there is no purpose in the universe: it is all just a
conglomerate of particles proceeding at a fundamental level in a
purposeless and meaningless algorithmic way. But I would reply,
the very fact that those writers engage in such discourse
undermines their own contention; they ignore the evidence provided
by their own actions. There is certainly meaning in the universe
to this degree: \emph{the fact they take the trouble to write such
contentions is proof that they consider it meaningful to argue
about such issues}; and this quality of existence has emerged out
of the nature of the physical universe (Sec.\ref{sec:new}). Indeed
the human mind is causally effective in the real physical world
precisely through many activities motivated by meanings perceived
by the human mind. Any attempt to relate physics and cosmology to
ultimate issues must take such real world experience seriously
\cite{ell05}, otherwise it will simply be ignoring a large body of
undeniable data. This data does not resolve the ultimate issues,
but does indicate dimensions of existence that indeed do occur.

\section{Conclusion}
\label{sec:conclude}

The physical scale of the Universe is enormous, and the images of
distant objects from which we obtain our information are extremely
faint. It is remarkable that we are able to understand the
Universe as well as we do. An intriguing feature is the way in
which the philosophy of cosmology is to a considerable degree
shaped by contingent aspects of the nature of the universe --- its
vast scale (Sec.\ref{sec:scale}), leading to the existence of
visual horizons (Sec.\ref{sec:horizon1}), and the occurrence of
extreme energies in the early universe (Sec.\ref{sec:energies}),
leading to the existence of physical horizons. Philosophical
issues arising in relation to cosmology (Sec.\ref{sec:philos})
would be quite different if its physical structure were very
different. Furthermore in order that philosophical analysis can
engage with cosmology in depth, the detailed nature of the
relation between observations and theory in cosmology
(Sec.\ref{sec:outline}) is relevant.

\subsection{Are there laws of cosmology?}
\label{sec:lawscosm} As we have discussed in detail, the
uniqueness of the universe implies the unique nature of cosmology.
We now return to the initial issue, \textit{Are there Laws of the
Universe}? (Sec.\ref{sec:unique}). At one level, the laws of the
cosmos are simply the local laws we know and love (e.g. Maxwell's
laws, Einstein's field equations) applied to the whole shebang. Of
course, there is the problem of extrapolation from the local to
the global. But although the extrapolation is bigger in cosmology,
it seems not to be different in kind from what we always do in
science. In that sense, there are no special laws for the
evolution of the universe. But that does not determine the
outcome: cosmology needs some prescription of boundary or initial
conditions as well, in order to determine the future. Is there a
true ``Cosmological principle", a law of initial conditions for
the universe, that determines this outcome?

The idea of ``Laws of initial conditions of the universe" seems
not to be a testable idea (Sec.\ref{sec:unique}). Scientifically,
one can only describe what occurred rather than relate it to
generic principles, for such principles cannot be tested. In fact
any description of boundary or initial conditions for the universe
seems to be just that: a description of these conditions, rather
than a testable prescription of how they must be. The
`Cosmological Principle' --- the universe is necessarily spatially
homogeneous and isotropic (Sec.\ref{sec:homog}) is of this kind: a
description of the way the initial data turned out, rather than a
fundamental reason for why this should be so. Justification of
this view was based by some workers on a \emph{Copernican
Principle} (the assumption we do not live in a privileged place in
the universe), strengthened to become a \emph{Cosmological
Principle} \cite{bon60,wei72,har81}; but this is a philosophical
assumption --- essentially, a claim that the universe must have
very special initial conditions --- which may or may not be true,
and does not attempt a physical explanation. This kind of argument
is out of fashion at present, because we now prefer generality to
speciality and physical argumentation to geometrical prescription;
but it was previously strongly proposed (e.g. \cite{wei72}, pp.
407-412). The tenor of philosophical argument has changed.

Nevertheless there is one kind of Law of the Universe one might
propose, following McCrea \cite{mac70}: namely an ``Uncertainty
principle in cosmology", dual to the uncertainty principle in
quantum theory. Uncertainty applies on the largest scale, as we
have discussed above in some detail, and also on the smallest,
where it is a profound feature of quantum theory. Its basis is
very different in the two cases, on the one hand (in quantum
theory)\ being ontological in nature, on the other (in cosmology)\
being epistemological in nature.\footnote{Assuming that quantum
uncertainty is indeed ontological rather than epistemological. One
should however keep an open mind on this: just because it is the
current dogma does not necessarily mean it is true.} Nevertheless
it is a key aspect of our relation to the cosmos, so that
(following McCrea) we might perhaps formalize it in order to
emphasize its centrality to the relation between cosmology and
philosophy:

\textbf{Thesis of Uncertainty: Ultimate uncertainty is a key aspect of
cosmology}. \textit{Scientific exploration can tell us much about the
universe but not about its ultimate nature, or even much about some of its
major geometrical and physical characteristics. Some of this uncertainty may
be resolved, but much will remain. Cosmological theory should acknowledge
this uncertainty.}

\subsection{What can we truly claim}
Cosmology considers questions of physical origins in the uniquely
existing physical universe (Sec.\ref{sec:origins}) which provides
the context of our existence (Sec.\ref{sec:background},
Sec.\ref{sec:anthropic}). These questions can be extended to
include ultimate issues if we so desire (Sec.\ref{sec:scope}), but
physical theory cannot resolve them (Sec.\ref{sec:uncertain}). In
the end, there are a variety of mysteries underlying the existence
and nature of the universe (Sec.\ref{sec:existence}). The
scientific study of cosmology can help illuminate their nature,
but cannot resolve them.

As well as celebrating the achievements of cosmology, one should
fully take into account the limits and problems considered in this
chapter, and not claim for scientific cosmology more than it can
actually achieve or more certainty than is in fact attainable.
Such claims will in the long term undermine cosmology's legitimate
status as a project with solid scientific achievements to its
name. That status can be vigorously defended as regards the
`Standard Model' of cosmology (Sec.\ref{sec:concord}), provided
this standard model is characterized in conservative terms so that
it is not threatened by relatively detailed shifts in theory or
data that do not in fact threaten the core business of cosmology.
Further, this defence must take adequate cognisance of the
difficult philosophical issues that arise if one pushes the
explanatory role of cosmological theory to its limits
(Sec.\ref{sec:origins}); for example one should not make too
strong \emph{scientific} claims in regard to the possible
existence of multiverses (Sec.\ref{sec:multiverse});
philosophically based plausibility arguments for them are fine, if
identified as such. Cosmology is not well served by claims that it
can achieve more explanatory power than is in fact attainable, or
by statements that its claims are verified when in fact the
requisite evidence is unavailable, and in some cases must forever
remain so.\\

\noindent\textbf{Acknowledgement}:

I thank Bill Stoeger, Martin Bojowald, Malcolm MacCallum, Jeremy
Butterfield, Henk van Elst, and John Earman
for useful comments that have improved this article.\\

\noindent\textbf{Abbreviations used:}

 HBB: Hot Big Bang

 CBR: Cosmic Blackbody Radiation

CDM: Cold Dark Matter

EFE: Einstein Field Equations

FL: Friedmann-Lema\^{\i}tre (universe models)

LSS: Last Scattering surface

RW: Robertson-Walker (geometry)

SAP: Strong Anthropic Principle

WAP: Weak Anthropic Principle

\newpage

\begin{center}
\textbf{Issues in the Philosophy of Cosmology}
\end{center}
\begin{center}
\textbf{SUMMARY TABLE OF ISSUES AND THESES}
\end{center}

{\small \noindent \textbf{Issue A: The uniqueness of the universe} }

{\small \emph{\textbf{Thesis A1}}: The universe itself cannot be subjected
to physical experimentation }

{\small \emph{\textbf{Thesis A2}}: The universe cannot be observationally
compared with other universes }

{\small \emph{\textbf{Thesis A3}}: The concept of `Laws of Physics' that
apply to only one object is questionable }

{\small \emph{\textbf{Thesis A4}}: The concept of probability is problematic
in the context of existence of only one object}

{\small \ }

{\small \noindent \textbf{Issue B: The large scale of the Universe in space
and time} }

{\small \emph{\textbf{Thesis B1}}: Astronomical observations are confined to
the past null cone, and fade with distance }

{\small \emph{\textbf{Thesis B2}}: `Geological' type observations
can probe the region near our past world line in the very distant
past}

{\small \emph{\textbf{Thesis B3}}: Establishing a Robertson-Walker geometry
relies on plausible philosophical assumptions }

{\small \emph{\textbf{Thesis B4}}: Interpreting cosmological observations
depends on astrophysical understanding }

{\small \emph{\textbf{Thesis B5}}: A key test for cosmology is that the age
of the universe must be greater than the ages of stars }

{\small \emph{\textbf{Thesis B6}}: Horizons limit our ability to
observationally determine the very large scale geometry of the universe }

{\small \emph{\textbf{Thesis B7}}: We have made great progress towards
observational completeness }

{\small \ }

{\small \noindent \textbf{Issue C: The unbound energies in the early
universe } }

{\small \emph{\textbf{Thesis C1}}: The Physics Horizon limits our knowledge
of physics relevant to the very early universe }

{\small \emph{\textbf{Thesis C2}}: The unknown nature of the inflaton means
inflationary universe proposals are incomplete }

{\small \ }

{\small \noindent \textbf{Issue D: Explaining the universe --- the
question of origins} }

{\small \emph{\textbf{Thesis D1}}: An initial singularity may or may not
have occurred }

{\small \emph{\textbf{Thesis D2}}: Testable physics cannot explain the
initial state and hence specific nature of the universe }

{\small \emph{\textbf{Thesis D3}}: The initial state of the universe may
have been special or general }

{\small \ }

{\small \noindent \textbf{Issue E: The Universe as the background for
existence} }

{\small \emph{\textbf{Thesis E1}}: Physical laws may depend on the nature of
the universe }

{\small \emph{\textbf{Thesis E2}}: We cannot take the nature of the laws of
physics for granted }

{\small \emph{\textbf{Thesis E3}}: Physical novelty emerges in the expanding
universe }

{\small \ }

{\small \noindent \textbf{Issue F: The explicit philosophical basis} }

{\small \emph{\textbf{Thesis F1}}: Philosophical choices necessarily underly
cosmological theory }

{\small \emph{\textbf{Thesis F2}}: Criteria for choice between theories
cannot be scientifically chosen or validated }

{\small \emph{\textbf{Thesis F3}}: Conflicts will inevitably arise in
applying criteria for satisfactory theories}

{\small \emph{\textbf{Thesis F4}}: The physical reason for believing in
inflation is its explanatory power re structure growth. }

{\small \emph{\textbf{Thesis F5}}: Cosmological theory can have a wide or
narrow scope of enquiry}

{\small \emph{\textbf{Thesis F6}}: Reality is not fully reflected in either
observations or theoretical models }

{\small \ }

{\small \noindent \textbf{Issue G: The Anthropic question:\ fine tuning for
life} }

{\small \emph{\textbf{Thesis G1}}: Life is possible because both the laws of
physics and initial conditions have a very special nature }

{\small \emph{\textbf{Thesis G2}}: Metaphysical uncertainty remains about
ultimate causation in cosmology }

{\small \ }

{\small \noindent \textbf{Issue H: The possible existence of multiverses} }

{\small \emph{\textbf{Thesis H1}}: The Multiverse proposal is unprovable by
observation or experiment }

{\small \emph{\textbf{Thesis H2}}: Probability-based arguments cannot
demonstrate the existence of multiverses }

{\small \emph{\textbf{Thesis H3}}: Multiverses are a philosophical rather
than scientific proposal }

{\small \emph{\textbf{Thesis H4}}: The underlying physics paradigm of
cosmology could be extended to include biological insights }

{\small \ }

{\small \noindent \textbf{Issue I: The natures of existence} }

{\small \emph{\textbf{Thesis I1}}: We do not understand the dominant
dynamical matter components of the universe at early or late times }

{\small \emph{\textbf{Thesis I2}}: The often claimed physical existence of
infinities is questionable }

{\small \emph{\textbf{Thesis I3}}: A deep issue underlying the nature of
cosmology is the nature of the laws of physics. }

{\small \ }

{\small \noindent \textbf{Thesis of Uncertainty}: Ultimate
uncertainty is one of the key aspects of cosmology }
\end{document}